\documentclass[pra,reprint,twocolumn,superscriptaddress,showpacs,floatfix]{revtex4-1}
\usepackage{changes}
\usepackage{amsmath}
\usepackage{mathrsfs}
\usepackage{txfonts}
\usepackage{amssymb}
\usepackage{graphicx}
\usepackage{hyperref}
\usepackage{ulem}
\usepackage{overpic}
\usepackage{psfrag}
\usepackage{tabularx}
\usepackage{multirow}
\usepackage{array}
\usepackage{placeins}
\newcommand{\PreserveBackslash}[1]{\let\temp=\\#1\let\\=\temp}
\usepackage{color}

\makeatletter

\begin{document}
	
\title{Fractals and spontaneous symmetry breaking with type-B Goldstone modes:\\ A perspective from entanglement}
\author{Huan-Qiang Zhou}
\affiliation{Centre for Modern Physics, Chongqing University, Chongqing 400044, The People's Republic of China}

\author{Qian-Qian Shi}
\affiliation{Centre for Modern Physics, Chongqing University, Chongqing 400044, The People's Republic of China}

\author{John O. Fj{\ae}restad}

\affiliation{Center for Quantum Spintronics, Department of Physics,
Norwegian University of Science and Technology, NO-7491 Trondheim, Norway}

\author{Ian P. McCulloch}
\affiliation{Department of Physics, National Tsing Hua University, Hsinchu 30013, Taiwan}
\affiliation{Frontier Center for Theory and Computation, National Tsing Hua University, Hsinchu 30013, Taiwan}
\affiliation{Centre for Modern Physics, Chongqing University, Chongqing 400044, The People's Republic of China}

\begin{abstract}
The one-dimensional spin-$s$ ${\rm SU}(2)$ ferromagnetic Heisenberg model, as a paradigmatic example for spontaneous symmetry breaking (SSB) with type-B Goldstone modes (GMs), is expected to exhibit an abstract fractal underlying the ground-state subspace. This intrinsic abstract fractal is here revealed from a systematic investigation into the entanglement entropy for a linear combination of factorized (unentangled) ground states on a fractal decomposable into a set of the Cantor sets. The entanglement entropy scales logarithmically with the block size, with the prefactor being half the fractal dimension of a fractal, as first noted by Castro-Alvaredo and Doyon for the spin-$1/2$ ferromagnetic Heisenberg model [Phys. Rev. Lett. \textbf{108}, 120401 (2012)], as long as the norm for the linear combination scales as the square root of the number of the self-similar building blocks kept at each step $k$ for a fractal, under an assumption that the maximum absolute value of the coefficients in the linear combination is chosen to be around one, and  the coefficients in the linear combination are almost constants within the building blocks.
Actually, the set of the fractal dimensions for all the Cantor sets forms a {\it dense} subset in the interval $[0,1]$. As a consequence, the ground-state subspace is separated into a disjoint union of countably infinitely many regions, each of which is labeled by a decomposable fractal.
Hence, the interpretation of the prefactor as half the fractal dimension is valid for any support beyond a fractal, which in turn leads to the identification of the fractal dimension with the number of type-B GMs for the orthonormal basis states.
Our argument may be extended to any quantum many-body systems undergoing SSB with type-B GMs.
\end{abstract}
\maketitle

\section{Introduction}
Spontaneous symmetry breaking (SSB) plays a fundamental role in classifying quantum states of matter and quantum phase transitions in condensed-matter physics.  As first put forward by Goldstone~\cite{goldstone}, if the symmetry group is continuous, then a gapless low-lying excitation mode emerges, which dominates the low-energy physics of a quantum many-body system undergoing SSB. However, not all the SSB patterns for continuous symmetry groups fall into the same category, as reflected in a debate between Anderson and Peierls~\cite{anderson}. A significant progress in this regard concerns the introduction of type-A and type-B (Goldstone modes) GMs~\cite{watanabe}, as a result of an earlier observation made by Nambu~\cite{nambu}. This in turn is relevant to the long term pursuit of a proper classification of GMs~\cite{nielsen,nambu,schafer, miransky, nicolis, brauner-watanabe, watanabe, NG}.  Accordingly, for the SSB pattern from the symmetry group $G$ to the residual symmetry group $H$,
a distinction between type-A and type-B GMs has to be made to understand the physics behind quantum many-body systems undergoing SSB.  To reveal the essential difference between them, it appears to be interesting to characterize SSB with type-A or type-B GMs from a quantum entanglement perspective, as already embodied in recent developments on this topic (see, e.g., Refs.~\cite{Metlitski, Rademaker, typeAmany1,typeAmany2,typeAmany3,typeAmany4} for type-A GMs and Refs.~\cite{FMGM,goldensu3,spinorbitalsu4,finitesize} for type-B GMs).

In particular, SSB with type-B GMs occurs in any spatial dimensions, in contrast with SSB with type-A GMs that is forbidden in one spatial dimension, as a result of the Mermin-Wagner-Coleman theorem~\cite{mwc}. Hence, one may choose to study one-dimensional quantum many-body systems for SSB with type-B GMs, although an extension to any spatial dimensions is possible~\cite{2dtypeb}. As it turns out, all the known quantum many-body systems undergoing SSB with type-B GMs are frustration-free~\cite{tasakibook}, thus rendering them to be exactly solvable, as far as their ground-state subspaces are concerned.   As argued in Ref.~\cite{FMGM}, the entanglement entropy scales logarithmically with the block size in the thermodynamic limit, if the system is partitioned into a block and an environment.  In fact, the prefactor in front of the logarithm is half the number of type-B GMs $N_B$ for a set of orthonormal basis states, which are constructed from the repeated action of the lowering operator(s) on the highest weight state. Further developments demonstrate that, if the system size is finite, then the block size in the logarithmic scaling relation, valid in the thermodynamic limit, is replaced by a universal finite system-size scaling function~\cite{finitesize}. As a remarkable feature, for a given quantum many-body system undergoing SSB with type-B GMs, the orthonormal basis states thus constructed admit an exact Schmidt decomposition, with the basis states in the system and its subsystems being self-similar. This implies that an abstract fractal underlying the ground-state subspace exists. However, although the self-similarities have been reflected in the ground-state degeneracies under both periodic boundary conditions (PBCs) and open boundary conditions (OBCs)~\cite{goldensu3,spinorbitalsu4}, a full characterization of such an intrinsic abstract fractal is still lacking.

To address this intriguing question, it appears necessary to clarify the nature of the self-similarities underlying the orthonormal basis states, as reflected in the observation that both the block and the environment, as a subsystem, share exactly the same  orthonormal basis states as the entire system, if a scale transformation is performed~\cite{FMGM}. Hence, the self-similarities manifest themselves in the real space via a scale transformation connecting the system with its subsystems. However,  the orthonormal basis states constitute a  countably infinite set even in the thermodynamic limit, they alone are thus not sufficient to describe an abstract fractal, given that generically a fractal contains uncountably infinitely many elements. In other words, there must be a {\it hidden} aspect of the self-similarities to be exposed beyond the real space itself.
As will be shown in this work, a key link is that the orthonormal basis states may be expressed as linear combinations of a set of  the overcomplete basis states that are factorized (unentangled) ground states, where the support of a linear combination is a subspace of the coset space $G/H$, with the dimension of the subspace being identical to the number of type-B GMs, thus opening up a possibility for an interpretation of the fractal dimension of the support of the linear combinations as the number of type-B GMs. This may be achieved by introducing {\it extrinsic} fractals on the coset space.  That is, the self-similarities manifest themselves in  a fractal as the support of a linear combination on the coset space.

It is worthwhile to stress that a set of the overcomplete basis states has been exploited in a pioneering work by Castro-Alvaredo and Doyon~\cite{doyon} for the  spin-$1/2$ ${\rm SU}(2)$ ferromagnetic Heisenberg model--a paradigmatic example for SSB with type-B GMs, with a SSB pattern from  ${\rm SU}(2)$ to  ${\rm U}(1)$.
A Cantor set $C[2,1/3; \{k\}]$  located on the great circle $S^1$ was chosen as a fractal, where $\{ k\}$ denotes the set of the natural numbers, with $k$ being the step number. We remark that it should be understood as the image of a Cantor  set under a mapping from the interval $[0,1]$ to a circle $S^1$ on the coset space $S^2$. The entanglement entropy for a linear combination of the overcomplete basis states on this particular fractal has been investigated~\cite{doyon}.  It was found that the entanglement entropy scales logarithmically with the block size, with the prefactor being half the fractal dimension of the Cantor set $C[2,1/3; \{k\}]$, when the coefficients in the linear combination are identical.
Although inspiring, their work is far from complete, even if one restricts attention to the  spin-$1/2$ ${\rm SU}(2)$ ferromagnetic Heisenberg model. In fact,  a few shortcomings in Ref.~\cite{doyon} need to be tackled. First, for a specific linear combination on a fractal, a proper characterization of  the coefficients is required, since a restriction must be imposed on the coefficients to ensure that the prefactor in front of the logarithm is half the fractal dimension of the fractal. Physically, this is due to a simple observation that the geometric information encoded in a fractal, e.g., a Cantor set, is simply washed away, if no restriction is imposed on the coefficients. Second, there is a loophole in the argument, as presented in Ref.~\cite{doyon}, which led to the conclusion that the same scaling relation for the entanglement entropy is valid for a linear combination of the overcomplete basis states on any support, defined as a subset of the coset space. Indeed, given that the set of the fractal dimensions for all the possible fractals is countably infinite, it is necessary to  show that a {\it minimal} set of fractals exists, whose fractal dimensions form a {\it dense} subset in the entire range. This is crucial if one attempts to interpret the prefactor in front of the logarithm as half the fractal dimension of the support of a linear combination on the coset space. Third, an arbitrary degenerate ground state can be expressed as apparently different linear combinations on uncountably infinitely many supports. Hence, the set of  all the supports itself remains to be elaborated on for a complete understanding of the scaling behaviors of the entanglement entropy.

In this work, we aim to expose an {\it intrinsic} abstract fractal underlying the ground-state subspace by introducing an {\it extrinsic} fractal and performing  a systematic investigation into the entanglement entropy for linear combinations on this fractal, which act as degenerate ground states arising from SSB with type-B GMs in a quantum many-body system. For this purpose, a conceptual framework needs to be developed,
with the following three notions  as the key ingredients: an equivalence class in the set of all the possible supports, an approximation of a generic support in terms of a fractal, and a decomposition of a fractal into a set of Cantor sets. These notions lie at the heart of our investigation into  the scaling behaviors of the entanglement entropy for highly degenerate ground states. Actually, one only needs to focus on one representative of the supports in a given equivalence class, because all the supports in the same class yield, by definition, exactly the same ground-state wave function, up to a local unitary operation induced from the symmetry group. Since such a generic support is not necessarily a well-defined fractal, an approximation of it in terms of a fractal is crucial for any further investigation. In addition, a decomposition of a fractal into a set of the Cantor sets makes it possible to reveal a {\it minimal} set of fractals--the set of all the Cantor sets.

As it turns out, the ground-state subspace is separated  into a disjoint union of countably infinitely many regions labeled by a fractal decomposable into a set of the Cantor sets.
It is argued that an appropriate restriction imposed on the coefficients in a linear combination is to require that
the norm for a linear combination scales as the square root of the number of the self-similar building blocks kept at each step $k$ for a fractal, under an assumption that the maximum absolute value of the coefficients in the linear combination is chosen to be around one, and  the coefficients in the linear combination are almost constants within the building blocks. This restriction leads to two specific realizations: the ratio between any two nonzero coefficients is either a random constant at each step $k$ or converges to a constant, as $k$ tends to infinity, subject to the condition that the number of zero coefficients scales polynomially with $k$. Here, by ``a restriction" we mean it is presumably a sufficient and necessary condition. In addition, we show that the set of the fractal dimensions for all the Cantor sets is {\it dense} in the interval $[0,1]$, thus offering a resolution to the loophole in Ref.~\cite{doyon}. As a result,
the entanglement entropy for a linear combination on any support scales logarithmically with the block size in the thermodynamic limit, with the prefactor being half the fractal dimension of the support, if the coefficients are subject to the restriction.

Our argument is applicable to any quantum many-body systems undergoing SSB with type-B GMs, with a SSB pattern from $G$ to $H$, as long as $G$ is a semisimple Lie group. Although the presentation is mainly unfolded for the  spin-$s$ ${\rm SU}(2)$ ferromagnetic Heisenberg model, a concise exposition  of the scaling behaviors of the entanglement entropy for degenerate ground states is also made for the ${\rm SO}(4)$  ferromagnetic spin-orbital model and the   ${\rm SU}(2s+1)$ ferromagnetic Heisenberg model (in the fundamental representation).  In addition, extensive numerical tests are performed for the  spin-$s$ ${\rm SU}(2)$  ferromagnetic Heisenberg model and the  ${\rm SU}(3)$ ferromagnetic Heisenberg model, thus confirming our theoretical predictions.
Here, we stress that the models under investigation always yield  permutation-invariant ground states.
However, highly degenerate ground states arising from SSB with type-B GMs are not necessarily permutation-invariant,  as demonstrated~\cite{TypeBtasaki}  for the  flat-band Tasaki model~\cite{tasaki} at quarter filling.

As a consequence of our analysis, we are led to the identification of the fractal dimension $d_f$ with the number of type-B GMs $N_B$ for quantum many-body systems undergoing SSB with type-B GMs, if one restricts to the orthonormal basis states that exhibit the self-similarities in the real space. Indeed, the orthonormal basis states generated from the repeated action of the lowering operator(s) on the highest weight state may be expressed in terms of
a linear combination of a set of the overcomplete basis states  in a subspace of the coset space, with the dimension being identical to  the number of type-B GMs $N_B$. In addition, the subspace  may be regarded as a limit of a sequence of decomposable fractals. This identification itself is a reflection of  the {\it intrinsic} abstract fractal underlying the ground-state subspace.

\section{The entanglement entropy and the orthonormal basis states for quantum many-body systems undergoing SSB with type-B GMs}~\label{eeobs}

For a quantum many-body system undergoing SSB with type-B GMs, highly degenerate ground states arise. Hence, it is legitimate to speak of the ground-state subspace. Here, we briefly recall some previous results for the spin-$s$ ${\rm SU}(2)$ ferromagnetic Heisenberg model, the ${\rm SO}(4)$ ferromagnetic spin-orbital model and the
 ${\rm SU}(2s+1)$ ferromagnetic Heisenberg model, in order to explain how to construct a set of orthonormal basis states for quantum many-body systems undergoing SSB with type-B GMs (for more details, we refer to Ref.~\cite{FMGM}).

Consider the spin-$s$ ${\rm SU}(2)$ ferromagnetic Heisenberg model, described by the Hamiltonian
\begin{equation}
	\mathscr{H}=-\sum_{j=1}^{L}\textbf{S}_j\cdot \textbf{S}_{j+1}, \label{su2ham}
\end{equation}
where $\textbf{S}_j=(S^x_j,S^y_j,S^z_j)$, with $S^x_j$, $S^y_j$, $S^z_j$ being the spin-$s$ operators at the $j$-th lattice site, and $L$ is the system size. Here, PBCs have been assumed in the  Hamiltonian (\ref{su2ham}).
The model is frustration-free, so its degenerate ground states are exactly solvable. It only becomes exactly solvable by means of the Bethe ansatz for $s=1/2$~\cite{faddeev,baxterbook}. The symmetry group ${\rm SU}(2)$ is generated by  $S^x=\sum_jS_{j}^x$, $S^y=\sum_jS_{j}^y$,  and $S^z=\sum_jS_{j}^z$, with
$S^z$ being the Cartan gernerator, and the lowering and raising operators defined as $S_{\mp}=S^x \mp iS^y$.
We remark that highly degenerate ground states arise from the SSB pattern: ${\rm SU}(2) \rightarrow {\rm U}(1)$. Hence, the number of type-B GMs is one: $N_B = 1$.

Highly degenerate ground states $|L,M\rangle$ may be constructed from the repeated action of the lowering operator $S_-$ on the highest weight state $|s\cdots s\rangle$:
 $|L,M\rangle=1/Z(L,M) S_-^M |s\cdots s\rangle$ ($M=0,1,2,\ldots,2sL$), which constitute a set of orthonormal basis states.

The orthonormal basis states $|L,M\rangle$ admit an exact Schmidt decomposition~\cite{FMGM}
\begin{equation}
	|L,M\rangle= \sum\limits_{\kappa}\lambda(L,n,\kappa,M)
	|n,\kappa\rangle|L-n,M-\kappa\rangle. \label{schmidt}
\end{equation}
Here, $\lambda(L,n,\kappa,M)$ denote the Schmidt coefficients, which take the form:
\begin{equation*}
	\lambda(L,n,\kappa,M)=\frac{\mu(L,n,\kappa,M)}{\nu(L,n,\kappa,M)},
\end{equation*}
with
\begin{equation*}
	\mu(L,n,\kappa,M)\!=\!\sqrt{\!{\sum}'_{n_{-\!s},\ldots,\; n_{s},\atop l_{-\!s},\ldots,\;l_{s}}\!\prod_{u,t=-s}^{s-1}\!\varepsilon(s,u)^{n_{u}}
		{C_{n\!-\!\sum_{m=-s}^{u-1}\!n_m}^{n_u}}\!\varepsilon(s,t)^{l_{t}}{C_{L\!-\!n-\!\sum_{m=-s}^{t-1}l_m}^{l_t}}},
\end{equation*}
and
\begin{equation*}
	\nu(L,n,\kappa,M)=\sqrt{{\sum}'_{N_{-s},\ldots,N_{s}}\prod_{u=-s}^{s-1}\varepsilon(s,u)^{N_{u}}
		{C_{L-\sum_{m=-s}^{u-1}N_m}^{N_u}}},
\end{equation*}
where $\sum'_{n_{-s},\ldots,\;n_{s}}$ is taken over all the possible values of $n_{-s}$,\ldots, $n_s$, subject to the constraints: $\sum_{m=-s}^s n_m=n$ and $\sum_{m=-s}^{s}(s-m)n_m=\kappa$, and $\sum'_{l_{-s},\ldots,\;l_{s}}$ is taken over all the possible values of $l_{-s}$,\ldots, $l_s$, subject to the constraints: $\sum_{m=-s}^s l_m=L-n$ and $\sum_{m=-s}^{s}(s-m)l_m=M-\kappa$.
Here, $C_{M}^{\kappa}$ is the binomial coefficients: $C_{M}^{\kappa}= M !/ (\kappa ! (M - \kappa)!)$ and $\varepsilon(s,u)$ is
\begin{equation*}
	\varepsilon(s,u)=\frac{\prod_{m=u+1}^{s}{(s+m)(s-m+1)}}{\prod_{m=u}^{s-1}(s-m)^2}. \label{epsilon}
\end{equation*}

The ${\rm SO}(4)$ spin-orbital model is described by the Hamiltonian~\cite{So4}
\begin{equation}
	\mathscr{H}=\pm\sum_{j=1}^L(\zeta+\mathbf{S}_j\cdot \mathbf{S}_{j+1})(\zeta+\mathbf{T}_j \cdot \mathbf{T}_{j+1}),\label{ham3}
\end{equation}
where  $\mathbf{S}_j=(S_{j}^{x},S_{j}^{y},S_{j}^{z})$ are the spin-$1/2$ operators and $\textbf{T}_j=(T_{j}^x,T_{j}^y,T_{j}^z)$ are the orbital pseudospin 1/2 operators at the  $j$-th lattice site. Here, PBCs have been assumed. The model (\ref{ham3}) shares the same ground-state subspace, consisting of the ferromagnetic ground states,  in the regime $(-\infty, -1/4)$ if a plus sign ``+" is taken in the Hamiltonian  (\ref{ham3}) and in the regime $(1/4,\infty)$ if a minus sign ``-" is taken in the Hamiltonian (\ref{ham3}). In particular, it consists of the two decoupled ${\rm SU(2)}$ spin-$1/2$ Heisenberg ferromagnetic models at $\zeta=-\infty$ if the plus sign ``+" is taken and at $\zeta=\infty$ if the minus sign ``-" is taken, so it is frustration-free. In addition, the  Hamiltonian (\ref{ham3})  in this ferromagnetic regime possesses  the  symmetry group ${\rm SO(4)}$, homomorphic to ${\rm SU(2)} \times {\rm (2)}$, with the generators of the two copies of  ${\rm SU(2)}$ being $S^x=\sum_jS_{j}^x$, $S^y=\sum_jS_{j}^y$,  and $S^z=\sum_jS_{j}^z$, and $T^x=\sum_jT_{j}^x$, $T^y=\sum_jT_{j}^y$ and $T^z=\sum_jT_{j}^z$, respectively.
The SSB pattern is from ${\rm SO(4)} $ to ${\rm U(1)}\times {\rm U(1)}$, with two type-B GMs: $N_B=2$. In passing, we remark that the Hamiltonian (\ref{ham3}) at  $\zeta=1/4$ is unitarily equivalent to the ${\rm SU(4)}$ ferromagnetic model (\ref{HsuNp1}) below and that the Hamiltonian (\ref{ham3}) at  $\zeta=-1/4$ is unitarily equivalent to the spin-$3/2$ staggered ${\rm SU(4)}$ ferromagnetic model~\cite{spinorbitalsu4}, which is exactly solvable by means of the Bethe ansatz~\cite{barber}, relevant to the Temperley-Lieb algebra~\cite{tla,martin}.

Highly degenerate ground states $|L,M_s,M_t\rangle$ ($M_s=0$, \ldots, $L$ and $M_t=0$, \ldots, $L$) may be constructed from
the repeated action of the lowering operators $S_-$ and $T_-$  on the highest weight state $|\uparrow_s\uparrow_t\ldots\uparrow_s\uparrow_t\rangle$, with $S_-=S^x-iS^y$ and $T_-=T^x-iT^y$:
\begin{equation*}
	|L,M_s,M_t\rangle=\frac{1}{Z(L,M_s,M_t)}S_-^{M_s}T_-^{M_t}|\uparrow_s\uparrow_t\ldots\uparrow_s\uparrow_t\rangle,
	\label{lmsmt}
\end{equation*}
where $Z(L,M_s,M_t)$ is introduced to ensure that $|L,M_s,M_t\rangle$ is normalized,
\begin{equation*}
	Z(L,M_s,M_t)=M_s!M_t!\sqrt{C_{L}^{M_s}C_{L}^{M_t}}.\label{zm1m2m3}
\end{equation*}
Hence, the orthonormal basis states $|L,M_s,M_t\rangle$ span an irreducible representation of the symmetry group ${\rm SU(2)} \times {\rm SU(2)}$, with the dimension being $(L+1)^2$. Actually, the orthonormal basis states $|L,M_s,M_t\rangle$ may be factorized as  $|L,M_s,M_t\rangle = |L,M_s\rangle|L,M_t\rangle$.

The orthonormal basis states $|L,M_s,M_t\rangle$ admit an exact Schmidt decomposition:
\begin{align}
	|L,M_s,M_t\rangle=&\sum_{k_s,k_t=0}^{n}\lambda(L,n,k_s,k_t,M_s,M_t)\times\nonumber\\
	&|n,k_s,k_t\rangle|L-n,M_s-k_s,M_t-k_t\rangle,	\label{lmschmiditso}
\end{align}
where the Schmidt coefficients $\lambda(L,n,k_s,k_t,M_s,M_t)$ take the form
\begin{equation*}
	\lambda(L,n,k_s,k_t,M_s,M_t)=\sqrt{\frac{C_{n}^{k_s}C_{n}^{k_t}C_{L-n}^{M_s-k_s}C_{L-n}^{M_t-k_t}} {C_{L}^{M_s}C_{L}^{M_t}}}.
\end{equation*}

We now turn to the  ${\rm SU}(2s+1)$ ferromagnetic Heisenberg model, described by the Hamiltonian
\begin{equation}
	\mathscr{H}=-\sum_{j=1}^LP_{j\;j+1}. \label{HsuNp1}
\end{equation}
Here, PBCs have been assumed, and $P$ is the permutation operator, which may be realized in terms of the spin-$s$ operators $\textbf{S}=(S_x,S_y,S_z)$:
\begin{equation*}
	P=\sum_{t=0}^{2s}(-1)^{2s+t} \prod_{m\neq t}^{2s} \frac{2(\textbf{S}\otimes\textbf{S})-m(m+1)+2s(s+1)}{t(t+1)-m(m+1)}.
\end{equation*}
Note that the model is frustration-free and exactly solvable by means of the Bethe ansatz~\cite{sutherland}. Indeed, the Hamiltonian (\ref{HsuNp1}) possesses the symmetry group $\rm{SU(2s+1)}$, with the local Hilbert space being the fundamental representation space of $\rm{SU(2s+1)}$ at each lattice site $j$, spanned by the (local) orthonormal basis states $|1\rangle_j, \ldots, |2s+1\rangle_j$.
The symmetry group $\rm{SU(2s+1)}$ is spontaneously broken into $\rm{U(1)} \times\rm{SU(2s+1)}$, the number of type-B GMs is thus $2s$.
For the symmetry group $\rm{SU(2s+1)}$, one may choose the Cartan generators $H_\alpha=\sum_jH_{\alpha,j}$  as
$H_{\alpha,j}=|1 \rangle_{jj} \langle 1| - |\alpha+1\rangle_{jj} \langle \alpha+1|$
($\alpha=1$, \ldots, $2s$). Accordingly,
for each $H_{\alpha}$, the lowering operator $F_{\alpha,j}$ and the raising operator $E_{\alpha,j}$ may be chosen as: $F_\alpha=\sum_jF_{\alpha,j}$ $E_\alpha=\sum_jE_{\alpha,j}$, with $F_{\alpha,j}=  |\alpha+1 \rangle_{jj} \langle 1|$ and $E_{\alpha,j}=|1 \rangle_{jj} \langle \alpha+1|$, satisfying
$[H_\alpha,E_\alpha]=2E_\alpha$, $[E_\alpha,F_\alpha]=H_\alpha$ and $[F_\alpha,H_\alpha]=2F_\alpha$.

Highly degenerate ground states $|L,M_1,\ldots,M_{2s}\rangle$ are generated from the repeated action of the lowering operators $F_{\alpha}$ on the highest weight state $|1\ldots1\rangle$:
\begin{equation*}
	|L,M_1,\ldots,M_{2s}\rangle
	=\frac{1}{Z(L,M_1,\ldots,M_{2s})}\prod_{\alpha=1}^{{2s}}F_\alpha^{\,\,M_\alpha}|1\ldots1\rangle,
	\label{gsineq}
\end{equation*}
which are orthonormal basis states and span an irreducible representation of the symmetry group $\rm{SU(2s+1)}$, with the dimension being the binomial coefficients  $C_{L+2s}^{2s}$.  Here, we note that
there is a sophisticated way to evaluate the dimension of the irreducible representation of the symmetry group $\rm{SU(2s+1)}$ by means of the standard Young tableau~\cite{mila}, in which the matrix of the Hamiltonian (\ref{HsuNp1}) takes a simple form.

Here,  $Z(L,M_1,\ldots,M_{2s})$ is introduced to ensure that $	|L,M_1,\ldots,M_{2s}\rangle$ is normalized,
which takes the form
\begin{equation*}
	Z(L,M_1,\ldots,M_{2s})=\prod_{\alpha=1}^{2s} M_\alpha!\sqrt{C_{L-\sum_{\beta=1}^{\alpha-1}{M_\beta}}^{M_{\alpha}}} \;.
	\label{zlmsuN}
\end{equation*}

\begin{widetext}
	The orthonormal basis states $|L,M_1,\ldots,M_{2s}\rangle$ admit an exact Schmidt decomposition:
	\begin{align}
		|L,M_1,\ldots,M_{2s}\rangle=
		\prod_{\alpha=1}^{2s}\sum\limits_{k_\alpha=0}^{\min(M_\alpha,n)}\lambda(L,n,k_1,\ldots,k_{2s},M_1,\ldots,M_{2s})
		|n,k_1,\ldots,k_{2s}\rangle|L-n,\!M_1-k_1,.\ldots,M_{2s}-k_{2s}\rangle,\label{lmschmidit}
	\end{align}
	where the Schmidt coefficients $\lambda(L,n,k_1,\ldots,k_{2s},M_1,\ldots,M_{2s})$ take the form
	\begin{align*}
		\lambda(L,n,k_1,\ldots,k_{2s},M_1,\ldots,M_{2s})=
		\sqrt{\frac{\prod_{\alpha=1}^{2s} {C_{n-\sum_{\beta=1}^{\alpha-1}{k_\beta}}^{k_{\alpha}}\prod_{\gamma=1}^{2s} C_{L-n-\sum_{\beta=1}^{\gamma-1}{(M_\beta-k_\beta)}}^{M_\gamma-k_{\gamma}}}}
			{\prod_{\alpha=1}^{2s} {C_{L-\sum_{\beta=1}^{\alpha-1}{M_\beta}}^{M_{\alpha}}}}} \;.
	\end{align*}
	
\end{widetext}

The orthonormal basis states $|L,M\rangle$, $|L,M_s,M_t\rangle$ and $|L,M_1,\ldots,M_{2s}\rangle$ admit an exact Schmidt decomposition, as seen in Eqs.(\ref{schmidt}), (\ref{lmschmiditso}) and (\ref{lmschmidit}). This fact reflects the self-similarities of the abstract fractals underlying the ground-state subspaces for the spin-$s$  ${\rm SU}(2)$ ferromagnetic Heisenberg model, the ${\rm SO}(4)$ spin-orbital model and the  ${\rm SU}(2s+1)$ ferromagnetic Heisenberg  model. Here,  the system is partitioned into a block consisting of $n$ lattice sites and an environment consisting of $L-n$ lattice sites.
We remark that, in each case, there are three sets of the orthonormal basis states, defined for the block, the environment and the system, respectively.  That is, they are similar to each other, in the sense that they are identical after performing a scale transformation connecting the block, the environment and the system. This requires that the number of the basis states for both the entire system and the subsystems must match to each other, if a proper scale transformation is taken into account.
Hence, the self-similarity manifests itself in the real space via a scale transformation connecting the system and the subsystems.
However,  the orthonormal basis states constitute a  countably infinite set even in the thermodynamic limit, they alone are thus not sufficient to describe an abstract fractal, given that a fractal contains uncountably infinitely many elements. In other words, there must be a {\it hidden} aspect of the self-similarities to be exposed.

To proceed, we introduce the reduced density matrix $\rho_L(n)$, which is determined from tracing out the degrees of freedom in the environment ${\cal E}$:  $\rho_L(n) = {\rm Tr}_{\cal E} \; \rho_L$, where $\rho_L$ denotes a density matrix. Hence, the entanglement entropy $S(L,n)$ is defined as $S(L,n)= - {\rm Tr} \rho_L(n) \log_2 \rho_L(n)$.
Generically, the eigenvalues of the reduced density matrix are the Schmidt coefficients squared.
For the orthonormal basis states $|L,M\rangle$,
the eigenvalues $\Lambda(L,n,\kappa,M)$ of the reduced density matrix $\rho (L,n,M)$ are $\Lambda(L,n,\kappa,M)=[\lambda(L,n,\kappa,M)]^2$.
For the orthonormal basis states $|L,M_s,M_t\rangle$,
the eigenvalues $\Lambda(L,n,k_s,k_{t},M_s,M_t)$ of the reduced density matrix $\rho (L,n,M_s,M_t)$ are  $\Lambda(L,n,k_s,k_t,M_s,M_t)=[\lambda(L,n,k_s,k_t,M_s,M_t)]^2$.
For the orthonormal basis states $|L,M_1,\ldots,M_{2s}\rangle$,
the eigenvalues $\Lambda(L,n,k_1,\ldots,k_{2s},M_1,\ldots,M_{2s})$ of the reduced density matrix $\rho (L,n,M_1,\ldots,M_{2s})$ are  $\Lambda(L,n,k_1,\ldots,k_{2s},M_1,\ldots,M_{2s})=[\lambda(L,n,k_1,\ldots,k_{2s},M_1,\ldots,M_{2s})]^2$.
In particular, the entanglement entropy $S_f(n)$ in the thermodynamic limit $L \rightarrow \infty$  may be defined for fixed filling $f$:  $S_f(n) = \lim _{L \rightarrow \infty} S(L,n,M)$. For the spin-$s$ ${\rm SU}(2)$ ferromagnetic Heisenberg model, we have $f=M/L$. For the ${\rm SO}(4)$ spin-orbital model,  we have  $f=(f_s,f_t)$, where $f_s=M_s/L$ and $f_t=M_t/L$. For the  ${\rm SU}(2s+1)$ ferromagnetic Heisenberg model, we have  $f=(f_1,\ldots,f_{2s})$, where $f_\alpha=M_\alpha/L$ ($\alpha=1$, \ldots, $2s$).

As argued in Ref.~\cite{FMGM}, for the orthonormal basis states in the ground-state subspace for a quantum many-body system undergoing SSB with type-B GMs, the entanglement entropy $S_f(n)$ scales as follows
\begin{equation}
	S_f(n)= \frac{N_B}{2}\log_2 n+S_{f0}, \label{srtlim}
\end{equation}
where $S_{f0}$ denotes a nonuniversal additive constant.  Here, $N_B =1$ for the spin-$s$ ${\rm SU}(2)$ ferromagnetic Heisenberg model, $N_B =2$ for the ${\rm SO}(4)$ spin-orbital model in the ferromagnetic regime and $N_B =2s$ for the  ${\rm SU}(2s+1)$ ferromagnetic Heisenberg model. Note that this scaling relation is consistent with an analytical analysis of the entanglement entropy for permutation-invariant states in Ref.~\cite{popkov}. In fact, the logarithmic scaling relation is also valid for the Renyi entropy~\cite{2dtypeb,doyon}. In passing, we emphasize that highly degenerate ground states arising from SSB with type-B GMs are not always permutation-invariant, although this is the case for $|L,M\rangle$, $|L,M_s,M_t\rangle$ and $|L,M_1,\ldots,M_{2s}\rangle$.

An important question that remains to be addressed is whether or not the entanglement entropy, as a physical observable~\cite{klich}, is able to reflect the abstract fractal underlying the ground-state subspace. The answer to the question demands to expose an abstract fractal that is characterized by the fractal dimension $d_f$. If so,
then one may anticipate to see a connection between the fractal dimension $d_f$ and the number of type-B GMs $N_B$ for the orthonormal basis states $|L,M\rangle$, $|L,M_s,M_t\rangle$ and $|L,M_1,\ldots,M_{2s}\rangle$, respectively.

\section{The number of type-B GMs and the fractal dimension: the spin-$s$ ${\rm SU}(2)$ ferromagnetic Heisenberg model}

 For clarity, we  shall mainly focus on the  spin-$s$ ${\rm SU}(2)$ ferromagnetic Heisenberg model (\ref{su2ham}) in this Section, and then extend to the ${\rm SO}(4)$ ferromagnetic spin-orbital model and the  ${\rm SU}(2s+1)$ ferromagnetic Heisenberg  model in the next Section.

\subsection{A  set of overcomplete basis states in the ground-state subspace}

To begin with, we stress that the  orthonormal basis states $|L,M\rangle$ are essentially {\it unique}, in the sense that they are the {\it only} degenerate ground states exhibiting the self-similarities in the real space, up to a rotation in the spin space induced from the symmetry group ${\rm SU}(2)$.  Mathematically, this stems from an observation that the self-similarities in the real space, reflected in the Schmidt decomposition (\ref{schmidt}) for the transformed basis states $U_S|L,M\rangle$,  impose a strict constraint on a unitary transformation $U_S$ acting on the orthonormal basis states $|L,M\rangle$. As a result, this implies that $U_S$ must be factorized into its counterparts  $U_B$ and  $U_E$ in the block and the environment according to the bipartition. As a result, $U_S$ is uniform. That is, $U_S$ is translation-invariant under PBCs or invariant under the permutation operation that exchanges $j$ with $j+1$ for $j=1,\ldots,L-1$ and $L$ with $1$ under OBCs. Hence, $U_S$ is an element of the symmetry group ${\rm SU}(2)$, since $U_S|L,M\rangle$ is a degenerate ground state.

However,  the orthonormal basis states $|L,M\rangle$ {\it only} constitute a  countably infinite set even in the thermodynamic limit, they alone are thus not sufficient to describe an abstract fractal, because a fractal generically contains uncountably infinitely many elements. In contrast, the highest weight state $|s\cdots s\rangle$, in combination with a rotation in the spin space induced from the symmetry group ${\rm SU}(2)$,  offers a set of factorized (unentangled) ground states. Indeed, this set is uncountably infinite, thus constituting a set of the overcomplete basis states.  But they are still subject to the self-similarities in the real space. Hence, one may expect that this set helps to reveal a hidden aspect of the self-similarities that manifest themselves in  a fractal as the support of a linear combination on the coset space $S^2$. In this sense,  a fractal is introduced as an {\it extrinsic} reference to reveal an {\it intrinsic} abstract fractal underlying the ground state subspace. To this end,  it appears to be beneficial to adopt a set of the overcomplete basis states in the ground state subspace, which plays a crucial role in a complete characterization of the scaling behaviors of the entanglement entropy for highly degenerate ground states arising from SSB with type-B GMs.

Here, we restrict ourselves to  linear combinations on a fractal which contains, by definition, uncountably infinitely many elements. That amounts to excluding linear combinations on a set of finite elements,  which may fall into two types: one type consists of finite but fixed elements so that the number of the elements does not scale with the system size $L$; the other type consists of finite elements so that the number of the elements scales with  the system size  $L$. Obviously, the first type consists of
finite elements even in the thermodynamic limit. However, the second type is quite subtle,  which points towards the interplay between the system size $L$ in the real space and the support of a linear combination on the coset space.

Actually, such an alternative approach to the entanglement entropy of a linear combination on a fractal has been developed by Castro-Alvaredo and  Doyon~\cite{doyon} for the  spin-$1/2$ ${\rm SU}(2)$ ferromagnetic Heisenberg model. For this model, a set of the overcomplete basis states, as degenerate factorized (unentangled) ground states, are nothing but the spin coherent states~\cite{shankar}.
In fact, the spin coherent states $|\psi(\theta,\phi)\rangle$  are  expressed in terms of the two spherical coordinates $\theta \in [0,\pi]$ and $\phi \in [0,2\pi]$ on the sphere $S^2$,
\begin{equation*}
	|\psi(\theta,\phi)\rangle=|v(\theta,\phi)\rangle_1 \cdots |v(\theta,\phi)\rangle_j \cdots |v(\theta,\phi)\rangle_L,
\end{equation*}
with
\begin{equation*}
	|v(\theta,\phi)\rangle_j=\exp(i\phi S^z_{j})\exp(i\theta S^y_{j})\;|s\rangle_j,
\end{equation*}
where $|s\rangle_j$ represents the eigenvector of $S_j^z$ with the eigenvalue being $s$ at a lattice site $j$. Indeed, one may form a linear combination of a set of the overcomplete basis states $|\psi(\theta,\phi)\rangle$ on a fractal $C$,  e.g., a Cantor set $C[N,r;\{k\}]$ (for a brief summary about the Cantor sets, cf. Appendix A).
Here, we adopt a convention that  $C[N,r;\{k\}]$ should be understood as the image on the sphere $S^2$ under the mapping $\phi: [0,1] \rightarrow S^1$, defined as $\phi (\xi) = 2\pi \xi$, or under the mapping $\theta: [0,1] \rightarrow S^1$, defined as $\theta (\xi) = \pi/2 \xi$, where $\xi \in [0,1]$. Owing to a symmetric consideration, the two mappings defined above are sufficient to meet our needs. From now on, we do not make any distinction between the image of a fractal under the mapping $\phi$ or $\theta$ and  a fractal itself for brevity.

Now we are ready to introduce a linear combination on a fractal $C$. For a Cantor set $C[N,r;\{k\}]$ under the mapping $\phi$, we have
\begin{equation}
	|\Phi_C(\theta)\rangle=\frac{1}{Z_C}\sum_{\phi_\gamma \in C} c(\phi_\gamma) |\psi(\theta,\phi_\gamma)\rangle, \label{lcfractal}
\end{equation}
where $c(\phi_\gamma)$ ($\gamma =1,2,\ldots,|C|$) are complex numbers, with $|C|=N^k$ being the number of the subintervals in  the Cantor set $C[N,r;\{k\}]$, and ${Z_C}$ is a normalization factor to ensure that $|\Phi_C(\theta)\rangle$ has been normalized, and the sum over $\phi_\gamma$ is carried out for all the subintervals at the step $k$.  One may also define a linear combination $|\Phi_C(\phi)\rangle$ on a Cantor set $C[N,r;\{k\}]$ under the mapping $\theta$, with the roles of $\theta$ and $\phi$ being swapped. Later on, our detailed discussion is restricted to $|\Phi_C(\theta)\rangle$ and its coefficients $c(\phi_\gamma)$, but may be carried forward to $|\Phi_C(\phi)\rangle$ and its coefficients $c(\theta_\gamma)$. Here, we have assumed that the maximum absolute value of the coefficients in the linear combination is chosen to be around one so that it fixes the normalization factor ${Z_C}$, in the sense that it makes sense to speak of the scaling behaviors of ${Z_C}$ with the step number $k$.

It is anticipated~\cite{doyon} that, in the thermodynamic limit $L \rightarrow \infty$, the entanglement entropy $S(n)$ for a linear combination  $|\Phi_C(\theta)\rangle$ on a fractal $C$  scales as follows
\begin{equation}
	S(n)= \frac{d_f}{2}\log_2 n+S_0, \label{srf}
\end{equation}
where $d_f$ is the fractal dimension of the fractal $C$, which is  $-\ln N/\ln r$ for the Cantor set $C[N,r;\{k\}]$  and $S_0$ is a nonuniversal additive constant. This scaling relation may be extended to the Renyi entropy~\cite{doyon}, with the prefactor being independent of the Renyi index.

We remark that the scaling relation   (\ref{srf}) was argued to be valid for the linear combination with all the coefficients being identical on the Cantor set  $C[2,1/3;\{k\}]$~\cite{doyon}. However, it was incorrectly claimed~\cite{doyon} that it also works for a linear combination with nonzero coefficients.  Instead, a proper characterization of linear combinations on a fractal requires imposing a restriction on the coefficients in a linear combination on a fractal.

Alternatively, the scaling relation (\ref{srf}) may be justified from  a scaling argument, as presented in Ref.~\cite{FMGM}, under the supposition that the entanglement entropy of a linear combination on a fractal $C$ is scale-invariant. The supposition itself may be attributed to the presence of the abstract fractal underlying the ground-state subspace. As a result, the prefactor is universal, in the sense that it is model-independent. However, the interpretation of the prefactor as half the fractal dimension of the fractal $C$ requires clarifying the physical meaning of the  self-similarities underlying the linear combinations at different steps, which leads us back to the restriction on the coefficients in a linear combination on a fractal $C$.

\subsection{A characterization of linear combinations on a fractal}~\label{clinearcombination}

The set of the overcomplete basis states exhibits self-similarities in the real space, since they are unitarily equivalent to the highest weight state  $|s\cdots s\rangle$. Indeed, the highest weight state  $|s\cdots s\rangle$ is self-similar in the real space, as already mentioned in Sec.~\ref{eeobs}. In addition, if one introduces a degenerate ground state as a linear combination of the overcomplete basis states on a fractal $C$ at each step, then degenerate ground states at different steps are self-similar, if the coefficients are chosen in a proper way.   That is, a key step is to look for an appropriate restriction imposed on linear combinations on a fractal $C$. For brevity, we restrict ourselves to the Cantor set  $C[N,r;\{k\}]$, but an extension to other types of fractals is straightforward.

To proceed, let us emphasize that the coefficients  $c(\phi_\gamma)$ in a linear combination [cf. Eq.~(\ref{lcfractal})] cannot be arbitrary, if one attempts to keep track of the geometric information encoded in the Cantor set  $C[N,r;\{k\}]$.   Physically, that amounts to stating that the geometric information encoded in the Cantor set  $C[N,r;\{k\}]$ is simply washed away, if no restriction is imposed on  the coefficients  $c(\phi_\gamma)$. This is due to the fact that the set of all the linear combinations on the Cantor set  $C[N,r;\{k\}]$ span the ground-state subspace, if the coefficients  $c(\phi_\gamma)$ are arbitrary. Mathematically, the dimension of the ground-state subspace is $2sL+1$, so it is much less than the number of the subintervals, i.e., $N^k$ in the Cantor set  $C[N,r;\{k\}]$ at the step $k$, if $k$ is large enough. Hence, only $2sL+1$ subintervals are needed to construct a set of linearly independent states that is sufficient to exhaust the entire ground-state subspace, as long as their coefficients are arbitrary.

This implies that it is necessary to impose a proper restriction on the coefficients $c(\phi_\gamma)$ in a linear combination on the Cantor set  $C[N,r;\{k\}]$. Given the geometric information encoded in the Cantor set  $C[N,r;\{k\}]$ must be kept,  the restriction imposed on the coefficients $c(\phi_\gamma)$ in the linear combination (\ref{lcfractal})  demands that the norm ${Z_C}$ should scale
as the square root of the number of subintervals, i.e., $N^{k/2}$, kept at each step $k$ for the Cantor set  $C[N,r;\{k\}]$
and that  the coefficients $c(\phi_\gamma)$ in the linear combination must remain to be almost constants within the subintervals at the step $k$, if $k$ is large enough.  On the one hand, the necessity for the requirement on the norm ${Z_C}$ originates from the fact that a linear combination, with all the coefficients being identical, must be a legitimate choice. We remark that the norm ${Z_C}$ for this particular linear combination is approximately the square root of the number of subintervals kept at each step $k$, as long as the size $L$ is large enough, since the overcomplete basis states $|\psi(\theta,\phi)\rangle$ are asymptotically orthogonal in the thermodynamic limit. This requirement ensures the information about the  number of the subintervals not to be lost at each step $k$.  On the other hand, the requirement on  the sizes and locations of the subintervals amounts to demanding that the coefficient $c(\phi_\gamma)$ does not vary abruptly as $\phi_\gamma$ varies from one endpoint to the other within a specific subinterval, thus ensuring that the information on the sizes and locations of the subintervals are well kept at each step $k$.

The restriction on the coefficients $c(\phi_\gamma)$ in a linear combination of the overcomplete basis states $|\psi(\theta,\phi)\rangle$ on the Cantor set  $C[N,r;\{k\}]$ ensures that degenerate ground states at different steps are self-similar to each other, with the two specific realizations:
the ratio between any two nonzero coefficients either is a random constant at each step $k$ or converges to any nonzero value, as the step number $k$ tends to infinity,  subject to the condition that the number of zero coefficients scales polynomially with $k$. Indeed, it is readily seen that the two  realizations satisfy all the requirements for  the norm ${Z_C}$, the sizes and locations of the subintervals and the self-similarities at different steps.
In particular, we emphasize that the presence of zero coefficients in the linear combination, with the number being polynomial in $k$, does not affect the exponential scaling of  the norm ${Z_C}$ with $k$.

Hence, we are led to conclude that the prefactor in the scaling relation (\ref{srf}) must be a function of $N$ and $r$, which does not depend on $k$.
Physically, this follows from the fact that the prefactor is universal~\cite{FMGM}, so it does not depend on the specifics of the coefficients in the linear combination.  In particular, the prefactor should be the same  if $N$ and $r$ are replaced by $N^q$ and $r^q$, respectively, with $q$ being a positive integer, as  a result of the self-similarities at different steps. The prefactor is thus only a function of $N$ and $r$ through the fractal dimension $-\ln N/ \ln r$ for the Cantor set  $C[N,r;\{k\}]$. Now it is readily seen that the prefactor is identical to half the fractal dimension $-\ln N/ \ln r$, if a detailed evaluation of the entanglement entropy for  the linear combination with all the coefficients being identical on the Cantor set  $C[2,1/3;\{k\}]$ is performed, as done in Ref.~\cite{doyon} (for a numerical test, cf. Sec.~\ref{numetext}). Actually, one may directly resort to
the results for the orthonormal basis states $|L,M \rangle$ in the spin-$1/2$ ${\rm SU}(2)$ ferromagnetic Heisenberg model, combining with that for the orthonormal basis states $|L, M_s, M_t \rangle$ in the ${\rm SO}(4)$ spin-orbital model, to establish that the prefactor is half the fractal dimension of the Cantor set $C[N,r;\{k\}]$ (for the details, see Sec.~\ref{identification}).

The above discussion is applicable to any other types of fractals, with the subintervals in the Cantor set  $C[N,r;\{k\}]$ replaced by the self-similar building blocks adapted to specific types of fractals. As an example,  the self-similar building blocks of the Sierpinski carpet and the Sierpinski triangle are squares and triangles.
Note that the set of all the fractals is still countably infinite. However, it is certainly not convenient if one has to study  all different types of fractals. In this sense, one may ask what types of fractals should be taken into account, in order to reveal the intrinsic abstract fractal underlying the ground-state subspace. For this purpose, we demonstrate that there exists a {\it minimal} set of fractals, whose fractal dimensions form a {\it dense} subset in the entire range. This is also crucial if one attempts to interpret the prefactor in front of the logarithm as half the fractal dimension of the support of a linear combination on the coset space, if the support is not a well-defined fractal.

\subsection{The set of the fractal dimensions for all the Cantor sets  is dense in the interval $[0,1]$}
Now we are ready to establish that the prefactor in front of the logarithm [cf. Eq.~(\ref{srf})] is half the fractal dimension $d_f$ of the support of linear combinations. However, for this statement to be valid, it is necessary to show that there exists a subset of fractals, whose fractal dimensions are densely distributed in the entire range, given that even the set of all the possible fractals we are able to construct is countably infinite. Indeed, for the set of all the Cantor sets $C[N,r;\{k\}]$, the set of the fractal dimensions constitutes a {\it dense} subset in the interval $[0,1]$, as we shall show below.

Given two arbitrary Cantor sets  $C[N_1,r_1;\{k\}]$ and  $C[N_2,r_2;\{k\}]$ at the step $k$, we assume that their fractal dimensions satisfy $d_{f1} < d_{f2}$, with $d_{f1} = -\ln N_1/\ln r_1$ and $d_{f2} = -\ln N_2/\ln r_2$. Before proceeding, let us make an observation that  $C[N^{q_1}_1,r^{q_1}_1;\{k\}]$ and $C[N^{q_2}_2,r^{q_2}_2;\{k\}]$ share the same fractal dimensions as $C[N_1,r_1;\{k\}]$ and  $C[N_2,r_2;\{k\}]$, respectively, as long as $q_1$ and $q_2$ are positive integers. Now we have to show that there is always a Cantor set  $C[N,r;\{k\}]$, with the fractal dimension $d_f$ satisfying $d_{f1} < d_f < d_{f2}$. Note that $(d_{f1} + d_{f2})/2$ does not take the form $-\ln N/\ln r$. However, as follows from the above observation,  one may demand that $N =[\exp(q_1 \;q_2 \;(N_1 \; \ln r_2+N_2 \; \ln r_1))]$ and $r^{-1}=[\exp (- 2q_1 \;q_2 \;(\ln r_1 \; \ln r_2))]$, where $[\xi]$ indicates the integer part of a real number $\xi$, thus  ensuring that $d_f =-\ln N/\ln r$ is  infinitesimally close to $(d_{f1} + d_{f2})/2$, as long as $q_1$ and/or $q_2$ are sufficiently large.

The ramifications from this fact are far-reaching. However, we have to further develop a conceptual framework, necessary for a deep understanding of the nature of linear combinations of the overcomplete basis states on a generic support.

\subsection{An equivalence class in the set of all the supports, an approximation of a fractal to a support, and a decomposition of a fractal to a set of the Cantor sets}\label{concept}

For our purpose, a  support  may be regarded as a subset of the coset space, on which a linear combination is formed, with the coefficients subject to the restriction stated above. Then each element in the subset labels one of the overcomplete basis states. Actually, a specific degenerate ground state admits apparently different representations on uncountably infinitely many supports.
It is therefore necessary to introduce an important notion - an equivalence class in the set of all the supports. That is, for any two supports on which two linear combinations  are formed, then they are equivalent, if and only if the two linear combinations represent the same ground state, up to a local unitary operation induced from the symmetry group. Here we remark that the two sets of the coefficients in the two linear combinations are not necessarily identical.  As such, we have defined a binary relation that is reflexive, symmetric and transitive. Mathematically, the set of all the supports are nothing but the set of all the subsets of the coset space, modulo linear combinations on each subset. Accordingly, we refer to the fractal dimension of a support as that of a subset of the coset space. We may thus resort to a mathematical theorem~\cite{settheory}, stating that the cardinality of the set of all subsets of a given set is greater than the cardinality of this given set, to account for the occurrence of apparently different but equivalent representations for a degenerate ground state on uncountably infinitely many supports.

To proceed, we introduce two other notions: an approximation of a generic support in terms of a fractal and a decomposition of a fractal into a set of the Cantor sets. For each equivalence class, there is at least one support on which a specific linear combination is formed. Generically, this support is not necessarily a well-defined fractal, then we have to introduce a fractal to approximate it by demanding that a linear combination on a fractal is infinitesimally close to a linear combination on the support. Here, both of the two linear combinations are  degenerate ground states, and the closeness is mathematically measured in terms of the norm for the difference between the two states or the fidelity between them. More precisely, if the norm is infinitesimally close to zero, or the fidelity is infinitesimally close to 1, then the support,  as a representative of the supports in an equivalent class, is well approximated in terms of a fractal. In addition, a fractal in a two-dimensional or even  higher dimensional setting may be decomposed into a set of the Cantor sets, although not all fractals admit such a decomposition.   As an example, the Cantor teepees, the Sierpinski carpet does not admit such a decomposition, but a variant of  the  Sierpinski carpet, which is defined in  a two-dimensional setting, is decomposed into two Cantor sets (for a brief summary about the Cantor teepees and the variants of  the  Sierpinski carpet, cf. Appendixes A and B).  Note that such a decomposition is very much like a decomposition of a torus $S^1 \times S^1$ into two circles $S^1$'s.

The implications of the three notions introduced above are far-reaching, in combination with the fact that the set of the fractal dimensions for all the Cantor sets is {\it dense} in the interval $[0,1]$. First, the fractal dimension $d_f$ is now well-defined for {\it any} support on which a linear combination is formed, since one is able to find a fractal to approximate this support as closely as possible. Therefore, the scaling relation (\ref{srf})
is now valid for a linear combination on {\it any}  support, as long as the coefficients are subject to the restriction. Actually, the closeness, measured in terms of the norm or the fidelity, for an approximation to a support in terms of a fractal, may be replaced by the closeness in the fractal dimensions for a fractal and the support, as justified in the next section.
Second,  one only needs to focus on fractals that admit a decomposition into a set of the Cantor sets. That is, the set of all the Cantor sets constitutes the {\it minimal} set of fractals for describing the scaling behaviors of the entanglement entropy.
Indeed, other types of fractals, such as the Sierpinski carpet or the Sierpinski triangle, are not necessary, because they may be well approximated in terms of two Cantor sets.  In fact, two Cantor sets are needed to approximate them, as follows from the fact that the Sierpinski carpet and the Sierpinski triangle are in the two-dimensional setting, whereas the Cantor sets are in a one-dimensional setting. Note that the fractal dimensions for the Sierpinski carpet and the Sierpinski triangle are $\ln 8/\ln3$ and $\ln3/\ln2$, respectively, whereas the fractal dimension of the Cantor set  $C[N,r;\{k\}]$ is  $-\ln N /\ln r$.
In other words,  we only need to consider fractals decomposable into a set of the Cantor sets,  with the fractal dimension of such a fractal being the sum of the Cantor sets contained in the decomposition.
Third, all the Cantor sets  $C[N,r;\{k\}]$, with the same $N$ and $r$, are in the same equivalence class. Here, we remark that the notation  $C[N,r;\{k\}]$ does not fix a Cantor set uniquely for generic $N$ and $r$, since the way to keep $N$ subintervals among $1/r$ subintervals is not specified.  Mathematically, this is due to the fact that a linear combination on the Cantor set  $C[N,r;\{k\}]$ at step $k$ may be expanded in terms of the orthonormal basis states $|L,M\rangle$, with the number of the coefficients being $2sL+1$. Therefore, we are led to a set of equations from equating the coefficients in the expansions of the two linear combinations of the overcomplete basis states on the two Cantor sets, which are expressed in terms of  the orthonormal basis states $|L,M\rangle$. As long as $k$ is large enough, the number of equations, i.e., $2sL+1$, is much less than the number of the coefficients in the linear combinations of the overcomplete basis states on the two Cantor sets. As a result, if one set of the coefficients is fixed, then there is always a solution to the set of equations to yield the other set of the coefficients, and vice versa.
In passing, we emphasize that the Cantor set $C[N^q,r^q;\{k\}]$ at step $k$ is identical to the Cantor set $C[N,r; \{qk\}]$
at step $q\; k$, with $q$ being a positive integer. That is, $C[N^q,r^q;\{k\}]$ is identical to $C[N,r; \{qk\}]$, both of which share the same fractal dimension.  Hence, the geometric information encoded in the Cantor set $C[N,r;\{k\}]$ is simply compressed into the fractal dimension, as far as the entanglement entropy is concerned.

As we shall see below, the three notions introduced above constitute the key ingredients in a conceptual framework, which is necessary to formalize  detailed theoretical predictions for the scaling behaviors of the entanglement entropy for highly degenerate ground states in a quantum many-body system undergoing SSB with type-B GMs.

\subsection{Closeness in the norm or the fidelity versus closeness in the fractal dimension}

The closeness, measured in terms of the norm or the fidelity, for an approximation to a support in terms of a fractal, may be replaced by the closeness in the fractal dimensions for a fractal and the support.

Consider two degenerate ground states, as linear combinations on the fractal and the support, respectively. Both of them may be expanded in terms of the orthonormal basis states $|L,M\rangle$ ($M=0,1,2,\ldots,2sL$), with the coefficients, denoted as $w_M$ and $w'_M$, respectively.  The closeness in the norm of the difference between the two states implies that  $w_M$ and $w'_M$ are close to each other.  On the other hand, the Schmidt coefficients for such a state is continuous as  a function of  $w_M$. This implies that
both $d_f$ and $S_0$, which appear in the scaling relation (\ref{srf}) of the entanglement entropy, are close to each other for the two states.
Conversely, if the fractal dimensions for the fractal and the support are close to each other, then one may form two linear combinations of the overcomplete basis states, one on the fractal and the other on the support. Afterwards, it is possible to adjust one of the two sets of the coefficients in the two linear combinations of the overcomplete basis states to ensure that {\it not only} the fractal dimension $d_f$, {\it but  also} $S_0$ are close to each other for the two linear combinations. Indeed, the same argument also works for the Renyi entropy. This in turn implies that the Schmidt coefficients are close to each other, given that the Schmidt coefficients and the Renyi entropy contain the same information on quantum entanglement. Hence,  $w_M$ and $w'_M$ must be close to each other, thus implying that the two states are close to each other  (measured in terms of the norm or fidelity), as long as the difference between $w_M$ and $w'_M$ vanishes faster than the square root of the system size $1/L$. Here, we remark that there might be a local  unitary operation induced from the symmetry group between the two states, since such a unitary operation does not affect the Schmidt coefficients.

The equivalence between the closeness in the norm or the fidelity and the closeness in the fractal dimension provides insights into
a characterization of the ground-state subspace from a perspective of linear combinations on a fractal $C$ that itself is decomposable into a set of the Cantor sets.  As a result, one is led to a notion--a region consisting of all the linear combinations on the fractal $C$, with the coefficients subject to the restriction.

\subsection{Separation of the ground-state subspace into a disjoint union of countably infinitely many regions}

Mathematically, a full characterization of linear combinations on a fractal $C$ decomposable into a set of the Cantor sets leads to the separation of the  ground-state subspace into a disjoint union of countably infinitely many regions, each of which consists of uncountably infinitely many linear combinations, with the coefficients subject to the restriction, so that the geometric information encoded in the Cantor set  $C[N,r;\{k\}]$ is kept. As already argued in Sec.~\ref{concept}, the geometric information is compressed into the fractal dimension $d_f$,
which is the sum of the fractal dimensions of  the Cantor sets contained in the decomposition of the fractal $C$.

Generically, as  one moves from one region to another,  many coefficients of a linear combination in one region vanish in another region, where the number of vanishing coefficients scales exponentially with the step number $k$, and vice versa. In this sense, any two distinct regions are well separated.
We stress that there is a one-to-one correspondence between the set of countably infinitely many regions and the set of fractals that may be decomposed into a set of  the Cantor sets. In particular, the number of the Cantor sets  is up to two, when the number of type-B GMs is one, as in the  spin-$s$ ${\rm SU}(2)$ ferromagnetic Heisenberg model. In other words, a region is always labeled by a fractal that is decomposed into a set of the Cantor sets.
Hence, all the regions constitute a countably infinite set, which is dense in the ground-state subspace. As a result, no clear-cut boundary exists between any two distinct regions, since there is always a region in between, no matter how close they are.

Actually, there are uncountably infinitely many linear combinations in each region, but each of them leads to the same prefactor in the logarithmic scaling relation  (\ref{srf}) of the entanglement entropy. This is consistent with the fact that the prefactor is universal~\cite{FMGM}, so it does not depend on any specifics of the coefficients in the linear combinations, as long as the coefficients are subject to the restriction. In contrast, the nonuniversal additive constant $S_0$ is continuously varying with the coefficients.
Here, we emphasize that the set of all the decomposable fractals and the set of the fractal dimensions $d_f$ are countably infinite, in contrast with the set of the linear combinations in each region and the set of the values of $S_0$ that are uncountably infinite.

Accordingly, a region is characterized by the fractal dimension of a fractal $C$, on which a linear combination is formed. Hence, it offers a physical interpretation of the prefactor in the logarithmic scaling relation of the entanglement entropy for the linear combinations.

\subsection{The physical interpretation of the prefactor in the logarithmic scaling relation of the entanglement entropy}~\label{interpretation}

The physical interpretation of the prefactor in the logarithmic scaling relation (\ref{srf}) of the entanglement entropy for a linear combination is necessary for a proper understanding of the physics underlying quantum many-body systems undergoing SSB with type-B GMs.

For a nonzero value of the fractal dimension $d_f$, the support, such as the Cantor set  $C[N,r;\{k\}]$, must contain uncountably infinitely many elements, modulo an exceptional case that the support contains countably infinitely many elements (cf. a brief discussion at the end of Sec.~\ref{identification}).  This in turn implies that the entanglement entropy is saturated for a linear combination on a support consisting of finite elements, given that the fractal dimension $d_f$ is zero for such a support.

Taking into account the decomposition of a fractal, as a subset of the coset space $S^2$, into a set of the Cantor sets, we have only up to two Cantor sets for the spin-$s$ ${\rm SU}(2)$ ferromagnetic Heisenberg model. Indeed, they are associated with the two spherical coordinates $\theta$ and $\phi$ on the coset space $S^2$: one is associated with $\phi$ and the other is associated with $\theta$.
Hence the maximum number of the Cantor sets in the decomposition is two. As a result,
the fractal dimension $d_f$ of the support may be re-interpreted  as the sum of the fractal dimensions of the Cantor sets contained in the decomposition of the support: $d_f= \sum _C \eta_C$,  where $\eta_C$ represents the fractal dimension of the Cantor set $C[N,r;\{k\}]$ contained in the decomposition.  In particular, if we restrict to a fractal decomposable into  $N_B^C$ pairs of the Cantor sets that consist of the same Cantor set $C[N,r;\{k\}]$,  with one pair located on the coset space $S^2$, then we have $N_B^C =1$ and 0, depending on whether or not the support is able to sense the presence of one type-B GM. As a result, we have $d_f = 2 N_B^C \eta_C$. Physically, $N_B^C =0$ means that the support consists of only finite elements, so it does not sense the presence of one type-B GM. In other words, for a linear combination on a specific support, the prefactor  reflects {\it not only} the information encoded in the intrinsic abstract fractal underlying the ground-state subspace, {\it but also} the information encoded in a Cantor set  contained in the decomposition of the support.
Here, we emphasize that the number of  type-B GMs must be an integer, in contrast with the claim in Ref.~\cite{doyon} that it could be a noninteger. Otherwise, a contradiction with the counting rule~\cite{nielsen,nambu,schafer, miransky, nicolis, brauner-watanabe, watanabe, NG} arises.

There is a discontinuous singularity in the prefactor for the highest weight state $|s\cdots s\rangle$ and the lowest weight state $|-s\cdots -s\rangle$. Indeed, they correspond to  $|L,M\rangle$ with $M=0$ and $M=2sL$ [cf. Eq.(\ref{lmexp})]. Hence, we have $d_f =1$ if one chooses $\theta \neq 0$, since the support is $S^1$. In contrast,  $d_f=0$ if one chooses $\theta = 0$ or $\theta = \pi$, since the support is one point. This singularity has already been noticed in a different guise that the fractal dimension $d_f$ is identical to the number of type-B GMs for the orthonormal basis states $|L,M\rangle$ in the thermodynamic limit, if filling $f$ is nonzero but not full~\cite{FMGM}, whereas the fractal dimension $d_f$ is zero  if filling $f$ is either zero or full: $f=0$ or $f=2s$.

\subsection{Identification of the fractal dimension with the number of type-B GMs for the orthonormal basis states}~\label{identification}

Now we turn to the question as to whether or not there is any connection between the fractal dimension $d_f$ and the number of type-B GMs $N_B$.

The orthonormal basis states $|L,M\rangle$  ($M=0$, \ldots, $2sL$) span the ground-state subspace.  Hence, $|\psi(\theta,\phi)\rangle$ may be expanded into a linear combination
in terms of $|L,M\rangle$: $|\psi(\theta,\phi)\rangle=\sum_{M=0}^{2sL}a_{LM}(\theta,\phi)|L,M\rangle$, where $a_{LM}$ are complex numbers, which are formally equal to $a_{LM}(\theta,\phi)= \langle L,M |\psi(\theta,\phi)\rangle$. As it turns out, $a_{LM}(\theta,\phi)=b_{LM}(\theta)\exp(i\phi(sL-M))$. However, the explicit expressions for $b_{LM}(\theta)$ are very complicated for arbitrary $s$. Here, we present the explicit expression for $b_{LM}(\theta)$ when $s=1/2$
\begin{equation*}
	b_{LM}(\theta)=(-1)^M\sqrt{C_L^M} \cos(\frac{\theta}{2})^{L-M}\sin(\frac{\theta}{2})^M.
\end{equation*}
Taking advantage of $1/(2\pi)\int_{0}^{2\pi} d\phi  \exp(i\phi(sL-M)) = \delta_{sL\;M} $, we are able to express $|L,M\rangle$  ($M=0$, \ldots, $2sL$) in terms of $|\psi(\theta,\phi)\rangle$ on a circle $S^1$ with fixed $\theta$ ($0<\theta<\pi$),
\begin{equation}
	|L,M\rangle=\frac {1} {b_{LM}(\theta)}\int_{0}^{2\pi} d\phi \; \exp \left(-i\phi(sL-M)\right)|\psi(\theta,\phi)\rangle.
	\label{lmexp}
\end{equation}
Obviously, this representation may be regarded as a linear combination in the overcomplete basis states $|\psi(\theta,\phi)\rangle$, with the coefficients only involving a phase factor.  Keeping this representation in mind and  taking into account the fact that the fractal dimension $d_f$ of the support of this linear combination is equal to one: $d_f=1$, we are led from the scaling relations (\ref{srtlim}) and (\ref{srf}) to conclude that the fractal dimension $d_f$ may be identified with the number of type-B GMs for the orthonormal basis states $|L,M\rangle$: $d_f=N_B$.
Hence, the fact that the fractal dimension $d_f$ is equal to one for the orthonormal basis states $|L,M\rangle$ reflects the {\it intrinsic} abstract fractal underlying the ground-state subspace.

In addition to the fact that $|L,M\rangle$ ($M=0,1,\ldots,2sL$) are expressed in terms of $|\psi(\theta,\phi)\rangle$ on a circle $S^1$ for fixed $\theta$ [cf. Eq.(\ref{lmexp})],  it is also possible to express them as a linear combination on an interval $[0, \phi_{\rm max}]$,  with $\phi_{\rm max}$ being a chosen value of $\phi$ less than $2\pi$. The construction thus provides an illustrative example for the notion of an equivalence class in the set of all the supports. The proof goes as follows.

Note that the interval $[0, \phi_{\rm max}]$ should be regarded as a subset of $S^1$ for fixed $\theta$. Suppose we choose a set of $2sL+1$ values of $\phi$, i.e., $\{ \phi_1,\phi_2, \ldots,\phi_{2sL+1}\}$, with $0<\phi_1 <\phi_2< \ldots < \phi_{2sL+1}$ and $\phi_{2sL+1} = \phi_{\rm max}$. Taking into account that  $|\psi(\theta,\phi)\rangle$ may be expressed in terms of $|L,M\rangle$ ($M=0,1,\ldots,2sL$):
$|\psi(\theta,\phi)\rangle=\sum_{M=0}^{2sL}a_{LM}(\theta,\phi)|L,M\rangle$ and integrating  the expression for $|\psi(\theta,\phi)\rangle$  over $\phi$ from 0 to $\phi_\delta$ ($\delta = 1,2,\ldots, 2sL+1$), we are able to establish a set of  linear equations, which express the integrals of $|\psi(\theta,\phi)\rangle$ over $\phi$ from 0 to $\phi_\delta$ ($\delta = 1,2,\ldots, 2sL+1$) in terms of $|L,M\rangle$ ($M=0,1,\ldots,2sL$). Solving this set of linear equations yields the desired result, with the coefficients being discontinuous at $ \phi =\phi_2, \ldots,\phi_{2sL}$. In other words, $|L,M\rangle$ ($M=0,1,\ldots,2sL$) are a linear combination in the overcomplete basis states $|\psi(\theta,\phi)\rangle$, with the fractal dimension $d_f$ of the support being one. Note that the number of discontinuous points scales as $L$, which becomes countably infinite in the thermodynamic limit $L \rightarrow \infty$. Hence, the coefficients in the linear combination are (piece-wisely) continuous as a function of $\phi$.

Here, we remark that a circle $S^1$ or an interval $[0, \phi_{\rm max}]$ on $S^1$ in the coset space $S^2$ may be regarded as a limit of a sequence of the  Cantor sets, as long as both $N$ and $1/r$ tend to infinity in proportionality: $1/r = g\;N$, with $g$ being a fixed integer. Indeed, the fractal dimensions of the  Cantor sets in the sequence tend to one, as $N$ tends to infinity.
In this sense, a Cantor set is introduced as an {\it extrinsic} fractal to reveal an {\it intrinsic} abstract fractal underlying the ground-state subspace, spanned by the orthonormal basis states $|L,M\rangle$.
The subtleties are reflected in the fact that a certain number of subintervals are discarded during the construction of the Cantor sets, in contrast with a circle $S^1$ or an interval $[0, \phi_{\rm max}]$ on the coset space $S^2$. In other words, discarding subintervals during the construction of the Cantor sets draws a demarcation line between the intrinsic and extrinsic nature of the fractals involved. However, this is not the only way to reveal the intrinsic abstract fractal underlying the ground-state subspace from  a linear combination on a set of uncountably infinitely many elements. Indeed, there is another way to expose this abstract fractal from a linear combination on a set of countably infinitely many elements, which is the thermodynamic limit of a set of finite elements, with the number of the elements scaling with  the system size $L$.  In fact, the fractal dimension of such a countably infinitely many set in the thermodynamic limit is either zero or one, depending on whether or not this set is dense. This in turn is relevant to whether this set is able to sense the presence of one GM in the spin-$s$ ${\rm SU}(2)$ ferromagnetic Heisenberg model. In order words, the intrinsic abstract fractal underlying the ground-state subspace may also be exposed from a linear combination on a support, a subset in the coset space consisting of countably infinite many elements.  A detailed investigation into this problem will be carried out in a forthcoming presentation.

Actually, the above discussions offer an alternative means to justify that the prefactor in the logarithmic scaling relation (\ref{srf}) for a linear combination on the  Cantor set $C[N,r; \{k\}]$ must be half the fractal dimension $-\ln N / \ln r$,  if one takes advantage of
the representation of a linear combination in the overcomplete basis states $|\psi(\theta,\phi)\rangle$	for $|L, M \rangle$ [cf. Eq.(\ref{lmexp})]
and its counterpart for the orthonormal basis states $|L, M_s, M_t \rangle$ in the ${\rm SO}(4)$ spin-orbital model, as presented in Sec.~\ref{extension}. The justification goes as follows.
As already mentioned in Sec.~\ref{clinearcombination}, the prefactor is a function of the fractal dimension $-\ln N/\ln r$ for the Cantor set $C[N,r; \{k\}]$.
In fact, the results for $|L, M_s, M_t \rangle$ imply that the prefactor must be proportional to the fractal dimension $d_f$, with some unknown proportionality constant to be determined yet,  as follows from the fact that for a fractal decomposable into two Cantor sets, with one on each of the two factor spaces $S^2$'s of the coset space $S^2 \times S^2$, the fractal dimension is the sum of the fractal dimensions of the Cantor sets contained in the decomposition. Meanwhile, a generic degenerate ground state wave function is the tensor product of linear combinations on the two Cantor sets that appear in the decomposition, so the entanglement entropy is additive (see a detailed discussion in Sec.~\ref{extension}).   Moreover, the results for  $|L, M \rangle$ imply that the proportionality constant  must be $1/2$, because $d_f =1$ for $M \neq 0$ and $M \neq 2s$ (see also the discussion in Sec.~\ref{interpretation}).

\section{Extension to other quantum many-body systems undergoing SSB with type-B GMs}~\label{extension}

Our argument  may be extended to any quantum many-body systems undergoing SSB with type-B GMs, with a SSB pattern from $G$ to $H$, as long as $G$ is a semisimple Lie group. The number of GMs is equal to the rank of the symmetry group $G$~\cite{FMGM}.  Indeed, the whole machinery developed for the spin-$s$ ${\rm SU}(2)$ ferromagnetic Heisenberg model works for any quantum many-body systems undergoing SSB with type-B GMs.

As a generic remark, the entanglement entropy for a linear combination on a fractal decomposable into a set of the Cantor sets scales logarithmically with the block size $n$, with the prefactor being half the fractal dimension of the fractal, as long as the norm for the linear combination scales as
the square root of the number of the self-similar building blocks kept at each step $k$ and  the coefficients in the linear combination are almost constants within the building blocks, under an assumption that the maximum absolute value of the coefficients in the linear combination is chosen to be around one, as already discussed in Sec.~\ref{clinearcombination}. This restriction leads to two specific realizations: the ratio between any two nonzero coefficients either is a random constant at each step $k$ or converges to any non-zero value, as the step number $k$ tends to infinity,  subject to the condition that the number of zero coefficients scales polynomially with $k$. Note that the number of the Cantor sets contained in a decomposable  fractal is up to twice the number of type-B GMs $N_B$.

Here, we restrict our discussions to the ${\rm SO}(4)$ spin-orbital model in the ferromagnetic regime and the  ${\rm SU}(2s+1)$ ferromagnetic Heisenberg model. However, an extension to other quantum many-body systems undergoing SSB with type-B GMs is straightforward.

\subsection{The ${\rm SO}(4)$ spin-orbital model in the ferromagnetic regime}~\label{sofour}

For the ${\rm SO}(4)$ spin-orbital model, SSB occurs from ${\rm SO}(4)$ to ${\rm U}(1) \times {\rm U}(1)$ in the ferromagnetic regime. We remark that the symmetry group ${\rm SO}(4)$, as a semisimple Lie group, is isomorphic to ${\rm SU}(2) \times {\rm SU}(2)$. As a consequence, two type-B GMs emerge, with the coset space being $S^2 \times S^2$. Note that the ground-state subspace is factorized into the tensor product of two copies of the ground-state subspace of the spin-$1/2$ ${\rm SU}(2)$ ferromagnetic Heisenberg model, thus implying that the two type-B GMs are completely independent from each other. As a result,  the evaluation of the entanglement entropy for the orthonormal basis states  $|L,M_s,M_t\rangle$ ($M_s=0$, \ldots, $L$ and $M_t=0$, \ldots, $L$)  is a trivial mathematical problem. Hence, the scaling relation (\ref{srtlim}) is valid, with $N_B=2$ for  $|L,M_s,M_t\rangle$.

Now we move to the entanglement entropy for a linear combination of a set of the overcomplete basis states on a generic support. Here, the overcomplete basis states $|\psi(\theta_s,\phi_s; \theta_t,\phi_t)\rangle$ are factorized into two copies of the counterpart $|\psi(\theta,\phi)\rangle$ of the spin-$1/2$ ${\rm SU}(2)$ ferromagnetic Heisenberg model: $|\psi(\theta_s,\phi_s; \theta_t,\phi_t)\rangle=|\psi(\theta_s,\phi_s)\rangle  |\psi(\theta_t,\phi_t)\rangle$, where two copies of the spherical coordinates $\theta_\beta \in [0,\pi]$ and $\phi_\beta \in [0,2\pi]$ ($\beta = s $ and $t$) on the coset space $S^2 \times S^2$  have been introduced, labeled by the subscripts $s$ and $t$.

An important lesson we have learned from the scaling behaviors of the entanglement entropy for the spin-$s$ ${\rm SU}(2)$ ferromagnetic Heisenberg model is that
a support may be well approximated in terms of a fractal, which may be decomposed into a set of the Cantor sets. Hence, it is sufficient to restrict to a decomposable fractal. Recall that a Cantor set is located on an interval as a subset of one circle $S^1$, which in turn is a submanifold embedded into the coset space $S^2 \times S^2$. In particular,  a generic linear combination of $|\psi(\theta_s,\phi_s; \theta_t,\phi_t)\rangle$ on a support may be well approximated in terms of
a linear combination on a decomposable fractal, which in turn is the tensor product of two linear combination on two  subfractals, each of which is  decomposed into a set of the Cantor sets located on one of the two factor spaces $S^2$'s. We remark that the fractal dimension of the fractal is the sum of the fractal dimensions of the two subfractals. This stems from the observation that a linear combination of  the overcomplete basis states $|\psi(\theta_s,\phi_s; \theta_t,\phi_t)\rangle$ on a decomposable fractal, located on the coset space $S^2 \times S^2$, may be represented as  the tensor product of two linear combinations of $|\psi(\theta_s,\phi_s)\rangle$ and $  |\psi(\theta_t,\phi_t)\rangle$  on two decomposable subfractals, located on the two factor spaces $S^2$'s, because the linear combination of  the overcomplete basis states $|\psi(\theta_s,\phi_s; \theta_t,\phi_t)\rangle$ may be expanded in terms of the orthonormal basis states  $|L,M_s,M_t\rangle$,
whereas  the two linear combinations of $|\psi(\theta_s,\phi_s)\rangle$ and $  |\psi(\theta_t,\phi_t)\rangle$ may be expanded in terms of the orthonormal basis states $|L,M_s\rangle$ and $|L,M_t\rangle$, respectively. Mathematically, this amounts to solving a set of $(L+1)^2$ equations, with $N_1^k+N_2^k$ unknown coefficients for the two linear combinations of $|\psi(\theta_s,\phi_s)\rangle$ and $  |\psi(\theta_t,\phi_t)\rangle$, given that $N_1^k\times N_2^k$ coefficients for the linear combination of $|\psi(\theta_s,\phi_s; \theta_t,\phi_t)\rangle$ are known at the step $k$. The existence of the solution to the set of equations is guaranteed, as long as $k$ is large enough. Here, $N_1^k$, $N_2^k$ and $N_1^k \times N_2^k$ represent the number of the self-similar building blocks for the two subfractals and the decomposable fractal at the step $k$, respectively. In fact, there is an obvious solution if we demand that all the coefficients in each of the three linear combinations on the two subfractals and the decomposable fractal are equal to one.

Hence, the scaling relation (\ref{srf}) is valid for a generic linear combination of $|\psi(\theta_s,\phi_s; \theta_t,\phi_t)\rangle$ on a support,
where the prefactor in front of the logarithm is half the fractal dimension of the support. That is, the fractal dimension $d_f$ of the support may be re-interpreted
as the sum of the fractal dimensions of the Cantor sets contained in a decomposition of a fractal that well approximates the support: $d_f= \sum _C \eta_C$,  where $\eta_C$ represents the fractal dimension of a Cantor set contained in the decomposition.  In particular, if we restrict to a fractal  decomposable into  $N_B^C$ pairs of the Cantor sets that consist of the same Cantor set $C[N,r;\{k\}]$,  with each pair located on one factor space $S^2$, then we have $N_B^C =2,1,$ and $0$, depending on whether or not the support is able to partially sense the presence of the two type-B GMs. As a result, we have  $d_f= 2 N_B^C \eta_C$.   Physically, $N_B^C =0$ means that the support consists of only finite elements, so it does not sense the presence of the two type-B GMs and $N_B^C =1$  means that the support consists of one pair of the Cantor sets located on one factor space $S^2$, so it {\it only} senses the presence of one of the two type-B GMs.
Note that the number of type-B GMs, the presence of which is partially sensed in a specific fractal, is always an integer.

We turn to the connection between the fractal dimension $d_f$ and the number of type-B GMs $N_B$ for the orthonormal basis states  $|L,M_s,M_t\rangle$ ($M_s=0$, \ldots, $L$ and $M_t=0$, \ldots, $L$), which span the ground-state subspace.  Hence, $|\psi(\theta_s,\phi_s; \theta_t,\phi_t)\rangle$ may be expanded into a linear combination
in terms of $|L,M_s,M_t\rangle$: $|\psi(\theta_s,\phi_s; \theta_t,\phi_t)\rangle=\sum_{M_s,M_t=0}^{L}a_{LM_s,M_t}(\theta_s,\phi_s; \theta_t,\phi_t)|L,M_s,M_t\rangle$, where $a_{LM_sM_t}$ are complex numbers, which are formally equal to $a_{LM_sM_t}(\theta_s,\phi_s; \theta_t,\phi_t)= \langle L,M_s,M_t |\psi(\theta_s,\phi_s; \theta_t,\phi_t)\rangle$. It is readily seen that $a_{LM_sM_t}(\theta_s,\phi_s; \theta_t,\phi_t)$ is proportional to $\exp(i\phi_s(L/2-M_s))\exp(i\phi_t(L/2-M_t))$. Taking advantage of $1/(2\pi)\int_{0}^{2\pi} d\phi_\beta \exp(i\phi_\beta (L/2-M_\beta)) = \delta_{L/2\;M_\beta}$, we are able to express $|L,M_s,M_t\rangle$ in terms of $|\psi_s(\theta_s,\phi_s; \theta_t,\phi_t)\rangle$ on  $S^1 \times S^1$ with fixed $\theta_s$ and $\theta_t$,
\begin{align}
	|L,M_s,M_t\rangle & \propto \int_{0}^{2\pi} d\phi_s  \int_{0}^{2\pi} d\phi_t \; \exp \left(-i\phi_s(L/2-M_s)\right) \times \nonumber\\
	&\exp \left(-i\phi_t(L/2-M_t)\right) |\psi(\theta_s,\phi_s; \theta_t,\phi_t)\rangle.  \nonumber
\end{align}
This representation may be regarded as a linear combination in the overcomplete basis states $|\psi(\theta_s,\phi_s; \theta_t,\phi_t)\rangle$, with the coefficients only involving a phase factor. In addition, it is also possible to express the orthonormal basis states  $|L,M_s,M_t\rangle$ as a linear combination on a support $[0, \phi_{\rm s, max}] \times [0, \phi_{\rm t, max}]$,  with $\phi_{\rm \beta, max}$ being a chosen value of $\phi _\beta $ less than $2\pi$  ($\beta =s,t$), as an extension of the construction for the counterpart in the spin-$s$ ${\rm SU}(2)$ ferromagnetic Heisenberg model in Sec.~\ref{identification}. The two apparently different representations for the orthonormal basis states  $|L,M_s,M_t\rangle$ provide again an illustrative example for the notion of an equivalence class in the set of all the supports.
Hence, the fractal dimension $d_f$ of the support of this linear combination is equal to two: $d_f=2$, identical to the number of type-B GMs $N_B$ for the orthonormal basis states $|L,M_s,M_t\rangle$: $d_f=N_B$.

The presence of the two factor spaces $S^2$'s in the coset space $S^2 \times S^2$ implies that it is possible to choose a support consisting of two Cantor teepees, if the apex point of each Cantor teepee is located at the north pole of each factor space $S^2$,  for the ${\rm SO}(4)$ spin-orbital model in the ferromagnetic regime.

\subsection{The ${\rm SU}(2s+1)$ ferromagnetic Heisenberg model}

For the  ${\rm SU}(2s+1)$ ferromagnetic Heisenberg model, the SSB pattern is from ${\rm SU}(2s+1)$ to ${\rm U}(1) \times {\rm SU}(2s)$, with the number of type-B GMs being $2s$. Hence,
the coset space may be identified as the complex projective space $CP^{2s}$.  The orthonormal basis states $|L,M_1,\ldots,M_{2s}\rangle$ span the ground-state subspace, with the dimension being $C_{L+2s}^{2s}$.

To reveal the intrinsic abstract fractal underlying the ground-state subspace, we move to a set of the overcomplete basis states.
For the symmetry group ${\rm SU}(2s+1)$, we introduce $2s$  ${\rm SU}(2)$ subgroups, each of which is associated with one of $2s$ type-B GMs.  
We denote the generators for a ${\rm SU}(2)$  subgroup as $\Sigma_{\alpha}^x$, $\Sigma_{\alpha}^y$ and $\Sigma_{\alpha}^z$ ($\alpha=1,\ldots,2s$), which are defined as $\Sigma_{\alpha,j}^x=(E_{\alpha,j}+F_{\alpha,j})/2$, $\Sigma_{\alpha,j}^y=-i(E_{\alpha,j}-F_{\alpha,j})/2$ and  $\Sigma_{\alpha,j}^z=H_{\alpha,j}/2$, satisfying $[\Sigma_{\alpha}^{x},\Sigma_{\alpha}^{y}]=i\Sigma_{\alpha}^{z}$, $[\Sigma_{\alpha}^{y},\Sigma_{\alpha}^{z}]=i\Sigma_{\alpha}^{x}$ and $[\Sigma_{\alpha}^{z},\Sigma_{\alpha}^{x}]=i\Sigma_{\alpha}^{y}$, respectively.
The set of the overcomplete basis states are thus factorized (unentangled) ground states  $|\psi(\theta_1,\phi_1;\ldots;\theta_{2s},\phi_{2s})\rangle$, which are parametrized in terms of  $\theta_\alpha \in [0,\pi]$ and $\phi_\alpha \in [0,2\pi]$ ($\alpha=1,\ldots,2s$),
\begin{widetext}
	\begin{equation*}
		|\psi(\theta_1,\phi_1;\ldots;\theta_{2s},\phi_{2s})\rangle=|v(\theta_1,\phi_1;\ldots;\theta_{2s},\phi_{2s})\rangle_1 \cdots |v(\theta_1,\phi_1;\ldots;\theta_{2s},\phi_{2s})\rangle_j \cdots |v(\theta_1,\phi_1;\ldots;\theta_{2s},\phi_{2s})\rangle_L,
	\end{equation*}
	with
	\begin{equation*}
		|v(\theta_1,\phi_1;\ldots;\theta_{2s},\phi_{2s})\rangle_j=\exp(i\phi_{2s} \Sigma_{2s,j}^{z})\exp(i\theta_{2s} \Sigma_{2s,j}^{y})\ldots \exp(i\phi_1 \Sigma_{1,j}^{z})\exp(i\theta_1 \Sigma_{1,j}^{y})\;|s\rangle_j,
	\end{equation*}
where $|s\rangle_j$ represents the eigenvector of $S_j^z$ with the eigenvalue being $s$ at a lattice site $j$. Indeed, $|\psi(\theta_1,\phi_1;\ldots;\theta_{2s},\phi_{2s})\rangle$ may be regarded as an extension of the spin coherent states from  ${\rm SU}(2)$ to ${\rm SU}(2s+1)$.  We note that there are other choices for the  spin coherent states, see, e.g., Refs.~\cite{shankar,perelomov}. Actually, this stems from the fact that the standard coherent states may be defined in different but equivalent ways~\cite{shankar,Glauber}. However, these different ways do not lead to equivalent definitions of the spin coherent states for ${\rm SU}(2)$ and other Lie groups.
\end{widetext}

Generically, an arbitrary degenerate ground state, represented as a linear combination of $|\psi(\theta_1,\phi_1;\ldots;\theta_{2s},\phi_{2s})\rangle$ on a support,  yields a value of the fractal dimension $d_f$ in the range $0 \le d_f \le 4s$.  This follows from the observation that
the support is well approximated in terms of a fractal that may be decomposed into a set of the Cantor sets, with the number of the Cantor sets being up to twice the number of type-B GMs, i.e., $4s$.  Actually,
the fractal dimension $d_f$ of the support may be re-interpreted
as the sum of the fractal dimensions of the Cantor sets that appear in a decomposition of the support: $d_f= \sum _C \eta_C$,  where $\eta_C$ represents the fractal dimension of the Cantor sets contained in the decomposition. In particular, if we restrict to a fractal decomposable into
$N_B^C$ pairs of the Cantor sets that consist of the same Cantor set $C[N,r;\{k\}]$,  with each pair located on a two-dimensional subspace parametrized in terms of  $\theta_\alpha$ and $\phi_\alpha$ ($\alpha=1,\ldots,2s$),
then we have  $N_B^C=N_B,\ldots, 1, 0$, depending on whether or not the support is able to partially sense the presence of the $2s$ type-B GMs. As a result, we have  $d_f= 2N_B^C \eta_C$.  Physically, $N_B^C =0$ means that the support consists of only finite elements, so it does not sense the presence of any type-B GMs. Meanwhile, $N_B^C =1$  means that the support  consists of two Cantor sets located on a two-dimensional subspace parametrized in terms of  $\theta_\alpha$ and $\phi_\alpha$, so it {\it only} senses the presence of one of the $2s$ type-B GMs. Moreover,
$N_B^C =2$  means that the support  consists of four Cantor sets that are located on two pairs of $\theta_\alpha$ and $\phi_\alpha$, so it senses the presence of two of the $2s$ type-B GMs. This scenario is valid for any $N_B^C$, up to $N_B^C=N_B$.
Note that the number of type-B GMs, which  are partially sensed by the support,  is always an integer.

We turn to the connection between the fractal dimension $d_f$ and the number of type-B GMs $N_B$ for the orthonormal basis states $|L,M_1,\ldots,M_{2s}\rangle$.  Recall that $|L,M_1,\ldots,M_{2s}\rangle$ ($M_1\cdots+M_{2s}=L$) span the ground-state subspace, the overcomplete basis states $|\psi(\theta_1,\phi_1;\;\ldots;\;\theta_{2s},\phi_{2s})\rangle$ may be expressed as linear combinations in terms of the orthonormal basis states $|L,M_1,\ldots,M_{2s}\rangle$: $|\psi(\theta_1,\phi_1;\;\ldots;\;\theta_{2s},\phi_{2s})\rangle=\sum_{M_{\alpha},\alpha\in\{1,\ldots,2s\}}a_{LM_1\cdots M_{2s}}(\theta_1,\phi_1;\;\ldots;\;\theta_{2s},\phi_{2s})|L,M_1,\ldots,M_{2s}\rangle$, where $a_{LM_1\cdots M_{2s}}(\theta_1,\phi_1;\;\ldots;\;\theta_{2s},\phi_{2s})$ are complex numbers, which are formally equal to $a_{LM_1\cdots M_{2s}}(\theta_1,\phi_1;\;\ldots;\;\theta_{2s},\phi_{2s})= \langle L,M_1,\ldots,M_{2s}|\psi(\theta_1,\phi_1;\;\ldots;\;\theta_{2s},\phi_{2s})\rangle$. It is readily seen that $a_{LM_1\cdots M_{2s}}(\theta_1,\phi_{1};\;\ldots;\;\theta_{2s},\phi_{2s})$ is proportional to $\exp(i\phi_1(L/2-M_1))\exp(i\phi_2(L/2-M_1/2-M_2))\cdots \exp(i\phi_{2s}(L/2-\sum_{\alpha=1}^{2s-1}M_\alpha/2-M_{2s}))$. Taking advantage of $1/(2\pi)\int_{0}^{2\pi} d\phi_\alpha \exp(i\phi_{\alpha}(L/2-\sum_{\beta=1}^{\alpha-1}M_\beta/2-M_{\alpha})) = \delta_{L/2\;\sum_{\beta=1}^{\alpha-1}M_\beta/2+M_{\alpha}}$, we are able to express $|L,M_1,\ldots,M_{2s}\rangle$ in terms of $|\psi(\theta_1,\phi_1;\;\ldots;\;\theta_{2s},\phi_{2s})\rangle$ with fixed $\theta_1$,\ldots,$\theta_{2s}$,
\begin{align*}
 &	|L,M_1,\ldots,M_{2s}\rangle \propto \int_{0}^{2\pi} d\phi_1 \cdots \int_{0}^{2\pi} d\phi_{2s}  \\ \nonumber
	&\prod_\alpha \exp \left(-i\phi_{\alpha}(L/2-\sum_{\beta=1}^{\alpha-1}M_\beta/2-M_{\alpha})\right)  |\psi(\theta_1,\phi_1;\;\ldots;\;\theta_{2s},\phi_{2s})\rangle.  \nonumber
\end{align*}

This representation may be regarded as a linear combination in the overcomplete basis states, with the coefficients only involving a phase factor.
In addition, it is also possible to express the orthonormal basis states  $|L,M_1,\ldots,M_{2s}\rangle$ as a linear combination on a support $[0, \phi_{1,{\rm max}}]  \times \ldots \times [0, \phi_{2s, {\rm max}}]$,  with $\phi_{{1,\rm \beta_1, max}}, \ldots, \phi_{{2s,\rm \beta_{2s}, max}}$ being some chosen values of $\phi _1, \ldots, \phi_{2s} $ less than $2\pi$, as an extension of the construction for the counterpart in the spin-$s$ ${\rm SU}(2)$ ferromagnetic Heisenberg model in Sec.~\ref{identification}. The two apparently different representations for $|L,M_1,\ldots,M_{2s}\rangle$ provide again an illustrative example for the notion of an equivalence class in the set of all the supports.
Hence, the support of a linear combination is a $2s$-dimensional subspace of the coset space $CP^{2s}$, with the fractal dimension $d_f$ being $2s$.  We are thus able to identify the fractal dimension $d_f$ with the number of type-B GMs $N_B$ for the orthonormal basis states  $|L,M_1,\ldots,M_{2s}\rangle$: $d_f = N_B$.

Mathematically, this amounts to introducing a sequence of decomposable fractals in the $2s$-dimensional subspace, which itself may in turn be regarded as a limit of a sequence of decomposable fractals. In particular, for the ${\rm SU}(3)$ ferromagnetic Heisenberg model,  the support of such a linear combination is a $2$-dimensional subspace of the coset space $CP^{2}$, which may in turn be regarded as a limit of a sequence of the variants of the Sierpinski carpet, which are subject to a decomposition into a set of the Cantor sets, with the number of the Cantor sets being two.  In fact, it is possible to consider two
${\rm SU}(2)$ subgroups, each of which is associated with one of the two type-B GMs. Now if we consider a support consisting of two identical copies of a Cantor set on two copies of a circle $S^1$, then it is anticipated that the fractal dimension of the support is twice the fractal dimension of  the Cantor set, although the two type-B GMs are not independent to each other, in contrast with  the ${\rm SO}(4)$ spin-orbital model. Meanwhile, for the ${\rm SU}(4)$ ferromagnetic Heisenberg model, the support of such a linear combination is a $3$-dimensional subspace of the coset space $CP^{3}$, which may in turn be regarded as a limit of a sequence of analogs of the variants of the Sierpinski carpet in a three-dimensional setting, which may be decomposed into a set of the Cantor sets on three copies of a circle $S^1$, with the number of the Cantor sets being three (for a brief discussion on analogs of the variants of the Sierpinski carpet, cf. Appendix B). Generically, for the  ${\rm SU}(2s+1)$ ferromagnetic Heisenberg model, the support of a linear combination is a $2s$-dimensional subspace of the coset space $CP^{2s}$, which may in turn be regarded as a limit of a sequence of analogs of the variants of the Sierpinski carpet in a $2s$-dimensional setting, decomposable into a set of the Cantor sets on $2s$ copies of a circle $S^1$, with the number of the Cantor sets being $2s$.

Specifically,  we may consider a linear combination on a fractal decomposable into $2s$ Cantor sets $C[N_\alpha,r_\alpha;\{k\}]$ ($\alpha=1$,\ldots, $2s$) located on
the $2s$ circles with fixed  $\theta_\alpha$'s,
\begin{widetext}
	\begin{equation*}
		|\Phi_C(\theta_1,\ldots,\theta_{2s})\rangle=\frac{1}{Z_{C}}\sum_{\phi_{\alpha,\gamma_\alpha} \in C_\alpha,\;,\alpha\in\{1,\ldots,2s\}} c(\phi_{1,\gamma_1},\;\ldots,\;\phi_{2s,\gamma_{2s}}) |\psi(\theta_1,\phi_{1,\gamma_1};\;\ldots;\;\theta_{2s},\phi_{2s,\gamma_{2s}})\rangle, \label{lcsun}
	\end{equation*}
where the coefficients $c(\phi_{1,\gamma_1},\;\ldots,\;\phi_{2s,\gamma_{2s}})$ ($\gamma_\alpha =1,2,\ldots,N_\alpha^k$) are complex numbers, and ${Z_C}$ is a normalization factor to ensure that $|\Phi_C(\theta_1,\ldots,\theta_{2s})\rangle$ has been normalized. In particular, one may restrict to the
factorized coefficients $c(\phi_{1,\gamma_1},\;\ldots,\;\phi_{2s,\gamma_{2s}}) = c(\phi_{1,\gamma_1}) \ldots c(\phi_{2s,\gamma_{2s}})$, as follows from a similar argument to the ${\rm SO}(4)$ spin-orbital model in the ferromagnetic regime in Sec.~\ref{sofour}. Hence, both the linear combinations with  $c(\phi_{1,\gamma_1},\;\ldots,\;\phi_{2s,\gamma_{2s}})$ and $c(\phi_{1,\gamma_1}) \ldots c(\phi_{2s,\gamma_{2s}})$ as the coefficients yield the same fractal dimension, as long as they are subject to the restriction stated earlier. We stress that such a factorization generically does not yield a tensor product structure in degenerate ground state wave functions, in contrast with the ${\rm SO}(4)$ spin-orbital model in the ferromagnetic regime. Here, by a tensor product structure we mean that a linear combination on a fractal decomposable into two subfractals is expressed as the tensor product of two linear combinations on the two subfractals, with all the three linear combinations involved as degenerate ground-state wave functions. Physically, this is due to the fact that the $2s$ type-B GMs are generically not independent to each other for the  ${\rm SU}(2s+1)$ ferromagnetic  Heisenberg model.
\end{widetext}

However, it is still possible to have a tensor product structure in degenerate ground-state wave functions if we only concern about a subspace in the ground-state subspace. Mathematically, this amounts to asking what semisimple Lie subgroups the symmetry group ${\rm SU}(2s+1)$ contains. This in turn is relevant to the question regarding a submanifold embedded into the coset space $CP^{2s}$. According to the fundamental theorem in number theory, $2s+1$, as a positive integer, may be factorized into a product of prime numbers uniquely. For brevity, we consider the simplest case as an illustrative example: $2s+1 = p_1 p_2$, with $p_1$ and $p_2$ being the two prime numbers. It is convenient to set  $p_1 = 2s_1+1$ and $p_2=2s_2+1$, with both $s_1$ and $s_2$ being an integer or $1/2$, given that all the prime numbers are odd, except for 2.  The group ${\rm SU}(2s_1+1) \times {\rm SU}(2s_2+1)$ is thus contained in the symmetry group ${\rm SU}(2s+1)$ as a semisimple Lie subgroup. Accordingly, the coset space $CP^{2s}$ accommodates $CP^{2s_1} \times CP^{2s_2}$ as a submanifold.
As a consequence, a linear combination on a decomposable fractal located on the submanifold $CP^{2s_1} \times CP^{2s_2}$ is the tensor product of linear combinations on two decomposable subfractals  located on the submanifolds $CP^{2s_1}$ and $CP^{2s_2}$, respectively, as a result of an extension of the argument  for a linear combination on the coset space $S^2 \times S^2$ for the ${\rm SO}(4)$ ferromagnetic spin-orbital model in Sec.~\ref{sofour}. Note that $CP^1$ is diffeomorphic to $S^2$. Physically, this stems from the fact that the $2s_1$ type-B GMs from the subgroup ${\rm SU}(2s_1+1)$ are completely independent to the $2s_2$ type-B GMs from the subgroup ${\rm SU}(2s_2+1)$, with the total number of type-B GMs involved being $2s_1+2s_2$, which is less than the number of type-B GMs $N_B=2s$: $2s_1+2s_2 < 2s$. In particular, if $s=3/2$, then we have $p_1 = p_2 =2$,  we are thus led to $s_1=s_2=1/2$. Hence, the coset space $CP^3$ accommodates a submanifold $CP^1 \times CP^1$, diffeomorphic to $S^2 \times S^2$. This explains the physics underlying the ${\rm SO}(4)$ spin-orbital model as it evolves from deep inside the ferromagnetic regime to the ${\rm SU(4)}$ symmetric point (see the Hamiltonian (\ref{ham3}) at  $\zeta=1/4$, which is unitarily equivalent to the ${\rm SU(4)}$ ferromagnetic model (\ref{HsuNp1})).  Moreover, there is another embedding of $S^2 \times S^2$ into the coset space $CP^3$ for
the spin-$3/2$ staggered ${\rm SU(4)}$ ferromagnetic model, unitarily equivalent to the  staggered ${\rm SU}(4)$ point in the spin-orbital model (\ref{ham3}) at  $\zeta=-1/4$~\cite{spinorbitalsu4}, with extra complications arising from the staggered nature of the symmetry group ${\rm SU(4)}$.
An extension to more general cases is straightforward.

Another relevant question concerns how many Cantor teepees the coset space $CP^{2s}$ is able to accommodate  for the ${\rm SU}(2s+1)$ ferromagnetic Heisenberg model, under the condition that any two Cantor teepees are separated from each other, in the sense that one factor space $S^2$ accommodates one Cantor teepee, with the apex point located at the north pole of the factor space $S^2$. A simple answer is up to $\sigma$ Cantor teepees, with $\sigma = [[\log_2 (2s+1)]]$, where $[[\mu]]$ indicates the closest integer to but not greater than $\mu$. Mathematically, this follows from the fact that  the coset space $CP^{2s}$ accommodates $S^2 \times \cdots \times S^2$ as a submanifold, with the number of $S^2$'s being up to $\sigma$. As a result of such a decomposition, a tensor product structure in a degenerate ground-state wave function exists on the submanifold $S^2 \times \cdots \times S^2$.

The usefulness of such a tensor product structure in a degenerate ground-state wave function lies in the fact that the entanglement entropy for a linear combination on a decomposable fractal is the addition of the counterparts for linear combinations on the two subfractals. In other words, the entanglement entropy becomes additive for such a decomposable fractal, as a result of the independence between the two sets of type-B GMs the two subfractals are able to sense.

\section{A finite system-size scaling analysis of the entanglement entropy: A numerical test} \label{numetext}
Now we turn to a systematic finite system-size scaling analysis of the entanglement entropy $S(L,n)$ for a linear combination  $|\Phi_C(\theta)\rangle$ on a fractal.
If the system size $L$ is finite, then the block size $n$ in the scaling relation (\ref{srf}) is replaced by a universal finite system-size scaling function~\cite{finitesize}. As a result, the entanglement entropy $S(L,n)$ for a linear combination  $|\Phi_C(\theta)\rangle$ on a decomposable fractal $C$ scales as,
\begin{equation}
	S(L,n)= \frac{d_f}{2}\log_2 \left(\frac{n(L-n)}{L}\right)+S_0, \label{srfs}
\end{equation}
where $d_f$ is the fractal dimension of the decomposable fractal $C$. It is readily seen that the finite system-size scaling relation (\ref{srfs}) reduces to the scaling relation  (\ref{srf}) as the thermodynamic limit $L \rightarrow \infty$ is approached.

It is proper to mention that there is a subtle interplay between the system size $L$ and the step number $k$, if one tries to reach the thermodynamic limit from a finite system-size approach. A plausible estimate from the evaluation of the norm for  a linear combination  $|\Phi_C(\theta)\rangle$ on the Cantor set  $C[N,r;\{k\}]$ leads to a criterion that $k$ should be much less than $\ln L$ such that the cross terms from the overcomplete basis states located at different subintervals are negligible, in order to ensure that the thermodynamic limit is reached. It is therefore challenging to reach the thermodynamic limit in a practical implementation of the evaluation of the entanglement entropy of a linear combination on a fractal.
In this sense,  it is necessary to perform a finite system-size scaling analysis for the entanglement entropy.
In this case, we are able to keep the contributions from the dominant cross terms that do not violate the requirement on the norm ${Z_C}$ discussed in Sec.~\ref{clinearcombination}. It appears to be sufficient that $k$ is roughly tracking with  $\ln L$. This quantifies an intuitive picture that $k$ increases with the system size $L$, when one attempts to perform a finite system-size scaling analysis for a quantum many-body system undergoing SSB with type-B GMs.

As already demonstrated in Sec.~\ref{concept}, it is sufficient to restrict to  a fractal decomposable into a set of the Cantor sets, as far as the scaling behaviors of the entanglement entropy for degenerate ground states as linear combinations on a generic support are concerned.  This observation is practically valuable, since a numerical implementation of the evaluation of the entanglement entropy for a linear combination on a fractal  is computationally demanding in a higher dimensional setting. However, if a sphere $S^2$  appears as a submanifold in the coset space $G/H$, then a Cantor teepee $C_{\rm tp}[N,r;\{k\}]$, a fractal in a two-dimensional setting,  may be efficiently computed, although it may be well approximated in terms of two Cantor sets. Note that  the Cantor teepees are defined in a two-dimensional setting,  in contrast with the Cantor sets defined in a one-dimensional setting (for a brief summary about the Cantor teepees, cf. Appendix A). This is particularly so, if we consider a linear combination $|\Phi_{T}\rangle$ on a Cantor teepee $C_{\rm tp}[N,r;\{k\}]$, defined as follows,
\begin{equation}
	|\Phi_{T}\rangle=\frac{1}{Z_T}\sum_{\phi_\gamma \in C}\int_{0}^{\pi/2} d\theta \;c(\phi_\gamma) |\psi(\theta,\phi_\gamma)\rangle ,
	\label{lcteepee}
\end{equation}
where $c(\phi_\gamma)$ ($\alpha =1,2,\ldots,|C|$) are complex numbers, which are assumed to be independent of $\theta$, and the sum over $\theta$ is replaced by an integral on the line segment connecting  the Cantor set on the great circle $S^1$ to the apex point located at the north pole. Here, the two spherical coordinates $\theta \in [0,\pi]$ and $\phi \in [0,2\pi]$ are used to parametrize the sphere $S^2$,  and $Z_T$ is a normalization factor to ensure that $|\Phi_{T}\rangle$ has been normalized. We remark that one is able to accommodate one  Cantor teepee $C_{\rm tp}[N,r;\{k\}]$ on one sphere $S^2$, as occurs in the spin-$s$ ${\rm SU}(2)$ ferromagnetic Heisenberg model.
Hence, our numerical tests are also carried out for a linear combination $|\Phi_{T}\rangle$ on the Cantor teepee $C_{\rm tp}[N,r;\{k\}]$.

Here we present our numerical results for a finite system-size scaling analysis of the entanglement entropy $S(L,n)$ for the spin-$s$ ${\rm SU}(2)$ ferromagnetic Heisenberg model and the ${\rm SU}(3)$ ferromagnetic Heisenberg model. Before delving into the details, let us mention that, in our numerical implementation to evaluate the entanglement entropy of a linear combination on a decomposable fractal, it is convenient to reexpress it as
a linear combination of the orthonormal basis states.

\subsection{The spin-$s$ ${\rm SU}(2)$ ferromagnetic Heisenberg model}

 Our aim is to numerically check that $d_f$ is equal to the fractal dimension $-\ln N/\ln r$ for a degenerate ground state constructed as a linear combination on the Cantor set  $C[N,r;\{k\}]$ for $0 <d_f <1$ and on the Cantor teepee  $C_{\rm tp}[N,r;\{k\}]$ for $1< d_f < 2$. To keep consistency with the restriction imposed on the coefficients $c(\phi_\gamma)$ in the linear combinations (\ref{lcfractal}) and (\ref{lcteepee}),
 we have two distinct types of realizations, in addition to a linear combination with equal coefficients: one is that both the amplitudes and the phase angles of the coefficients $c(\phi_\gamma)$ in a given linear combination
 are randomly chosen from their respective finite sets, with equal probability; the other is that both the amplitudes and the phase angles of the coefficients $c(\phi_\gamma)$ in a given linear combination are predetermined from a (piece-wisely) continuous function. In both cases, it is always allowed to set some coefficients to be zero, as long as the number of zero coefficients scales polynomially with $k$ at each step $k$. Indeed, the zero coefficients may be chosen randomly, because the scaling of the norm with $k$ is not affected by the way one chooses them.

 For the first type of realization,  at each step $k$, for each of the $N^k$ subintervals, one picks a value of the coefficient randomly, which is uniform on this subinterval in both the amplitude and the phase angle. More precisely, the values of both the amplitude and the phase angle are chosen from the two prescribed finite sets of the values, with equal probability for each value.
 Specifically,
 we have implemented that the amplitudes are chosen randomly from the set $\{0.1,\;0.2,\;0.3,\;0.4,\;0.5,\;0.6,\;0.7,\;0.8,\;0.9,\;1\}$ and the phase angles  are chosen from the set
 $\{\pi/10,\pi/5,3\pi/10,2\pi/5,\pi/2,\;3\pi/5,\;7\pi/10,\;4\pi/5,\;9\pi/10,\;\pi\}$, respectively.
 For the second type of realization, we have set the amplitudes to be a function $\cos(2\phi_\gamma+1)+2$ and  the phase angles to be  a function $\phi_\gamma^2/(2\pi)$. Note that the functions have been chosen to ensure that they are the same for $\phi =0$ and $\phi =2\pi$.

Here, we only consider a linear combination on the Cantor set $C[N,r;\{k\}]$ on the great circle $S^1$ or the Cantor teepee $C_{\rm tp}[N,r;\{k\}]$, with its base on the great circle $S^1$ and the apex point located at the north pole. That is, we  always set $\theta=\pi/2$ for the Cantor set  $C[N,r;\{k\}]$, and demand that $\theta$ ranges from 0 to $\pi$ for the Cantor teepee $C_{\rm tp}[N,r;\{k\}]$.

 \begin{figure}[h!]
 	\centering
 	\includegraphics[width=0.35\textwidth]{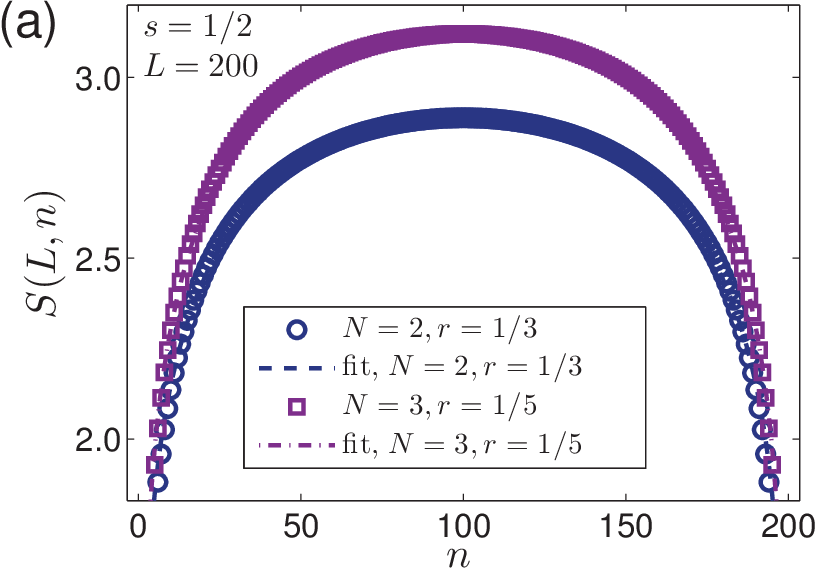}
 	\includegraphics[width=0.35\textwidth]{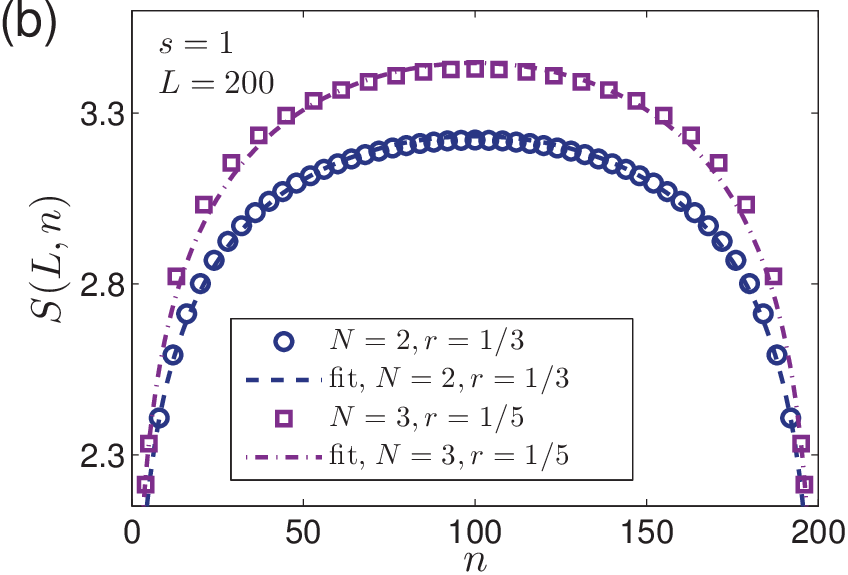}
 	\includegraphics[width=0.35\textwidth]{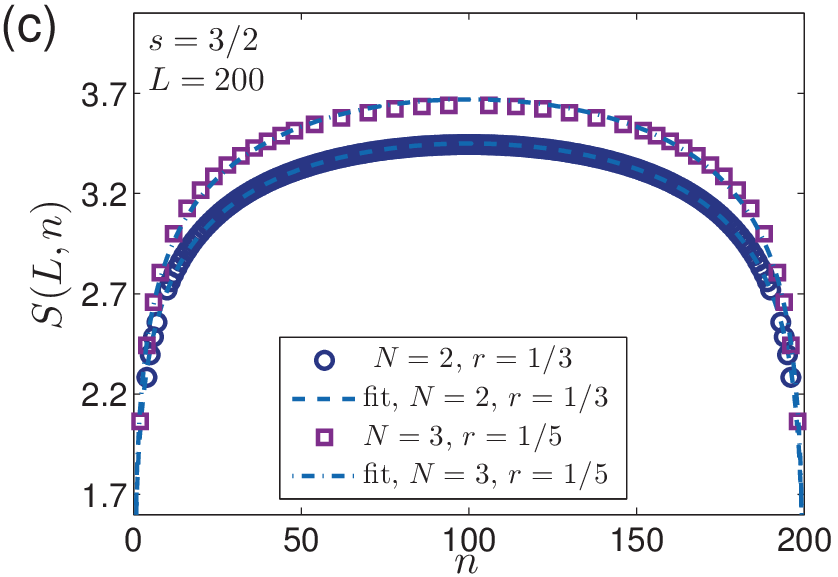}
 	\includegraphics[width=0.35\textwidth]{Fig1d.eps}
 	\caption{The entanglement entropy $S(L,n)$ versus $n$ for a degenerate ground state $|\Phi_C(\theta)\rangle$ on the Cantor set $C[N,r;\{k\}]$ [cf. Eq.~(\ref{lcfractal}), with $\theta=\pi/2$] in the spin-$s$ ${\rm SU}(2)$ ferromagnetic Heisenberg model: (a) $s=1/2$; (b) $s=1$; (c) $s=3/2$ and (d) $s=2$. Here, $N=2$ and $r=1/3$, and  $N=3$ and $r=1/5$ have been chosen,  respectively, when $L=200$ and $k=20$. The prefactor is half the fractal dimension $d_f$ of the Cantor set $C[N,r;\{k\}]$: $d_f=-\ln N/\ln r$, with a relative error being less than 3$\%$.}
 	\label{spinsg3g5}
 \end{figure}

In Fig.~\ref{spinsg3g5}, we plot the entanglement entropy $S(L,n)$  versus $n$  for a degenerate ground state $|\Phi_C(\theta)\rangle$, as a linear combination with equal coefficients, on a Cantor set [cf. Eq.~(\ref{lcfractal})] with $\theta=\pi/2$), where $s=1/2$, $s=1$, $s=3/2$ and $s=2$, respectively. The Cantor sets are taken to be $C[N,r; \{k\}]$, with $N=2$ and $r=1/3$, and $N=3$ and $r=1/5$, at the step $k=20$.

\begin{figure}[h!]
	\centering
	\includegraphics[height=4.3cm]{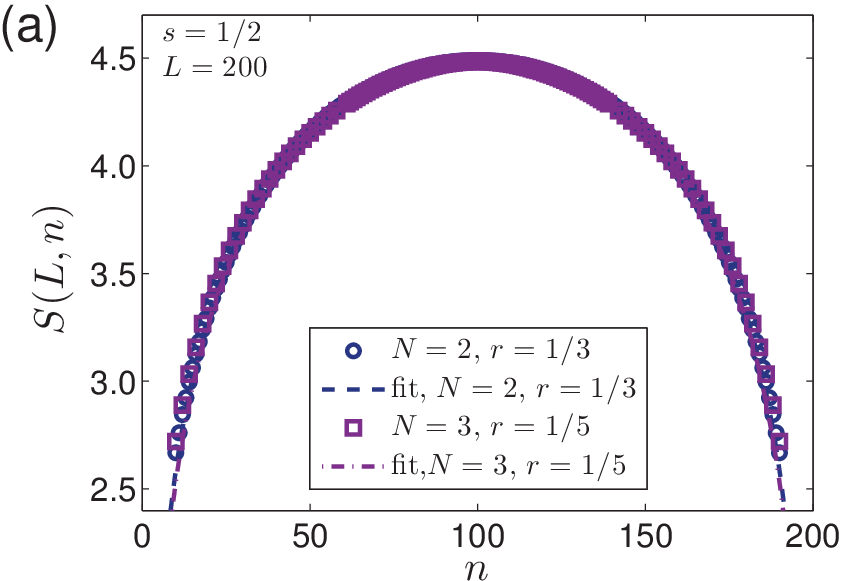}
	\includegraphics[height=4.3cm]{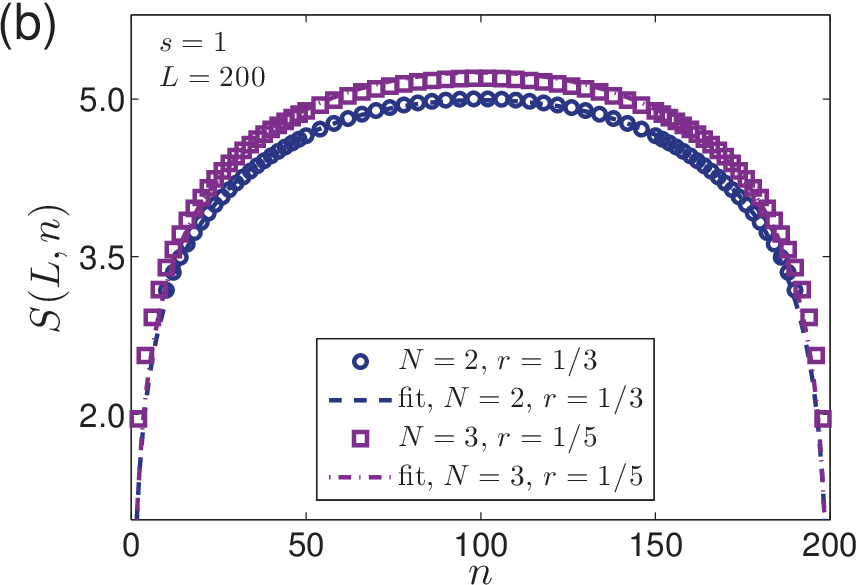}
  	\caption{The entanglement entropy $S(L,n)$ versus $n$  for  a degenerate ground state $|\Phi_T\rangle$ on the Cantor teepee  $C_{\rm tp}[N,r;\{k\}]$ [cf. Eq.~(\ref{lcteepee})] in the  spin-$s$ ${\rm SU}(2)$ ferromagnetic Heisenberg model: (a)  $s=1/2$ and (b) $s=1$, when $L=200$.  Here, $N=2$ and $r=1/3$ with $k=20$, and $N=3$ and $r=1/5$ with $k=18$ have been chosen. The prefactor is half the fractal dimension  of the Cantor teepee  $C_{\rm tp}[N,r;\{k\}]$: $1-\ln N/\ln r$, with a relative error being less than 2$\%$.}
	\label{spinsg3g5teepee}
\end{figure}

In Fig.~\ref{spinsg3g5teepee}, we plot the entanglement entropy $S(L,n)$ versus $n$ for a degenerate ground state $|\Phi_T\rangle$, as a linear combination with equal coefficients,  on a Cantor teepee [cf. Eq.~(\ref{lcteepee})], where $s=1/2$ and $s=1$, respectively.
The Cantor teepees are taken to be $C_{\rm tp}[N,r;\{k\}]$, where  $N=2$ and $r=1/3$, with $k=20$, and $N=3$ and $r=1/5$, with $k=18$.

\begin{figure}[ht!]
	\centering
	\includegraphics[height=4.3cm]{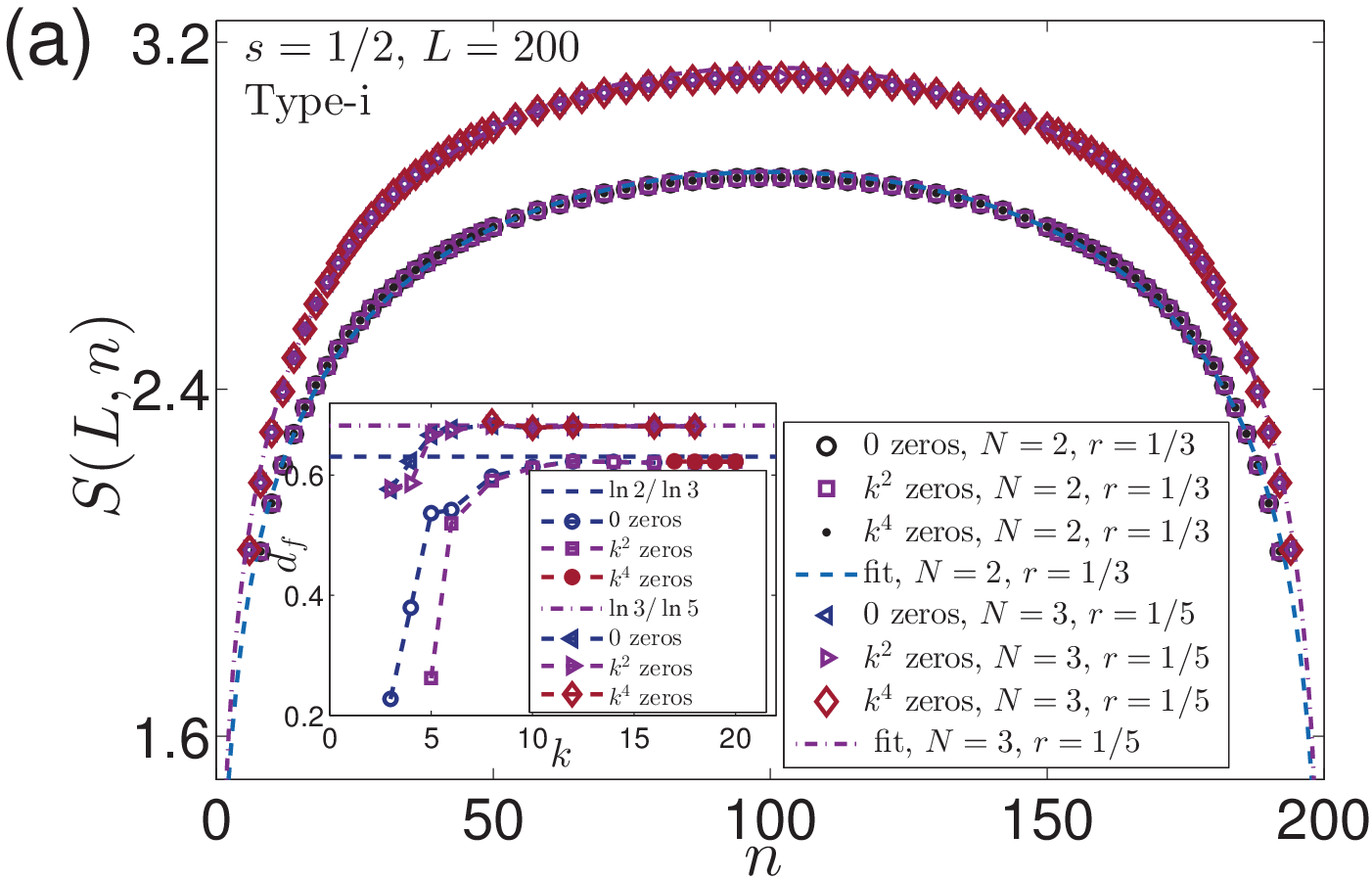}
	\includegraphics[height=4.3cm]{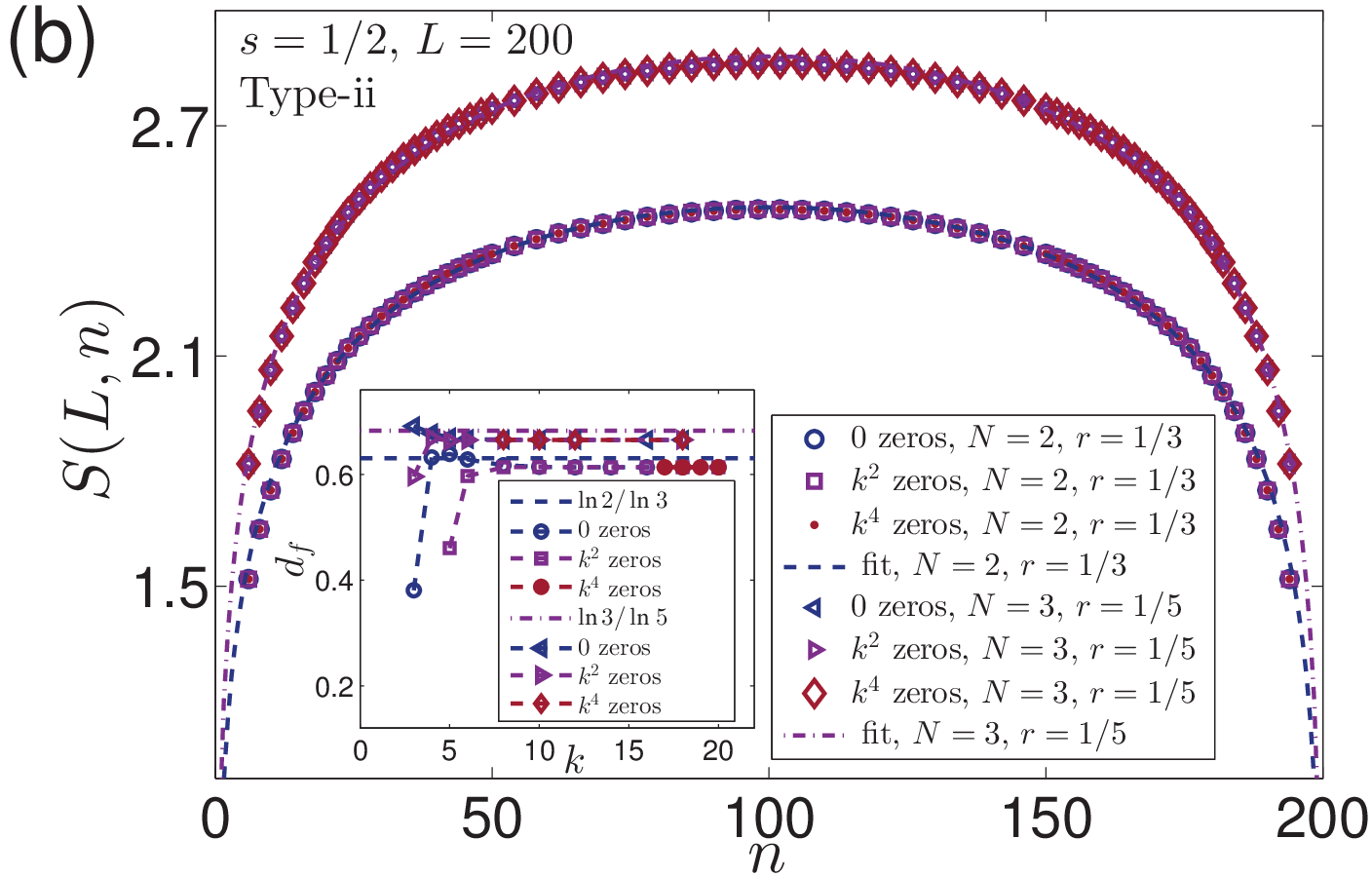}
	\includegraphics[height=4.3cm]{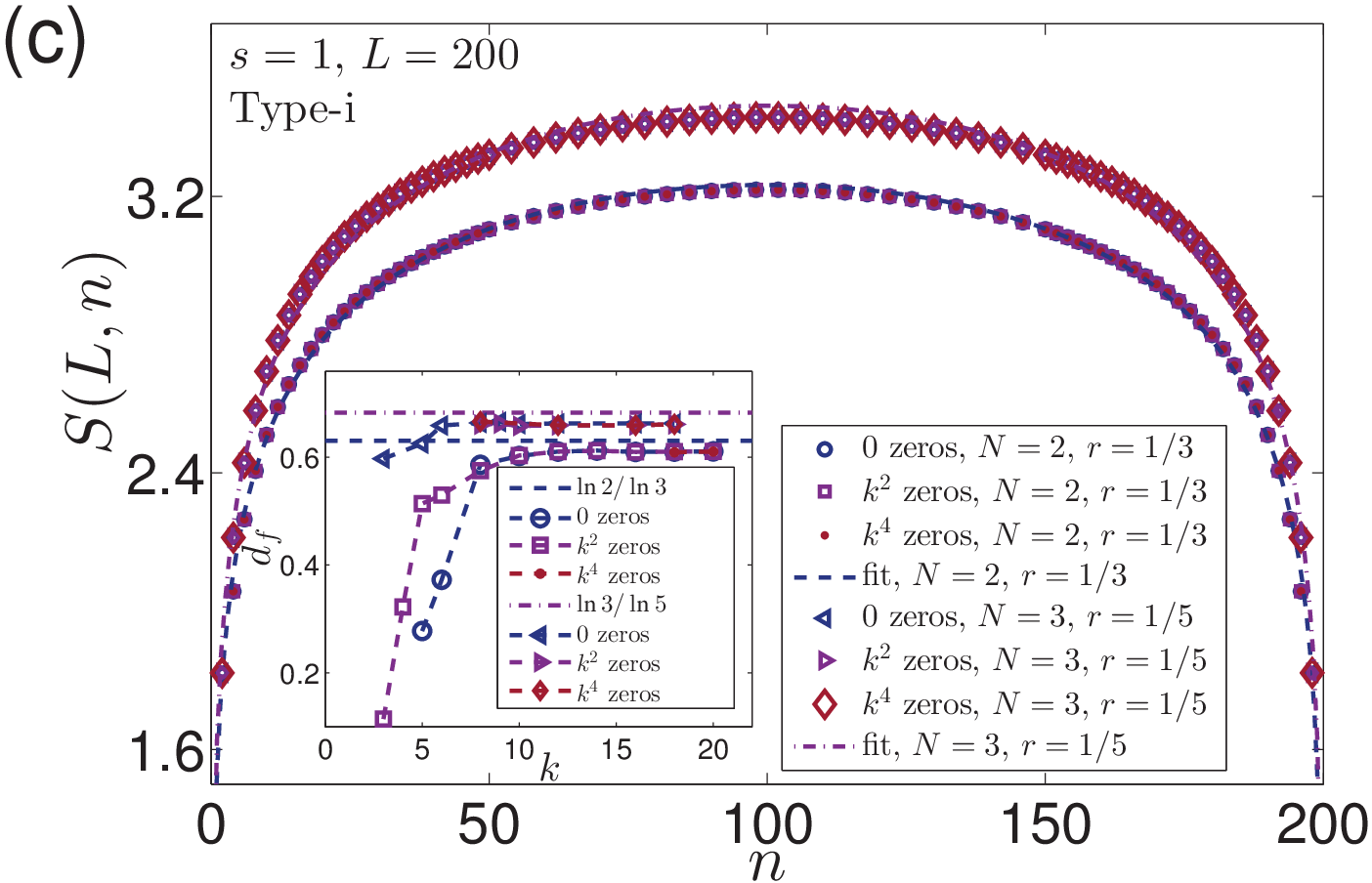}
	\includegraphics[height=4.3cm]{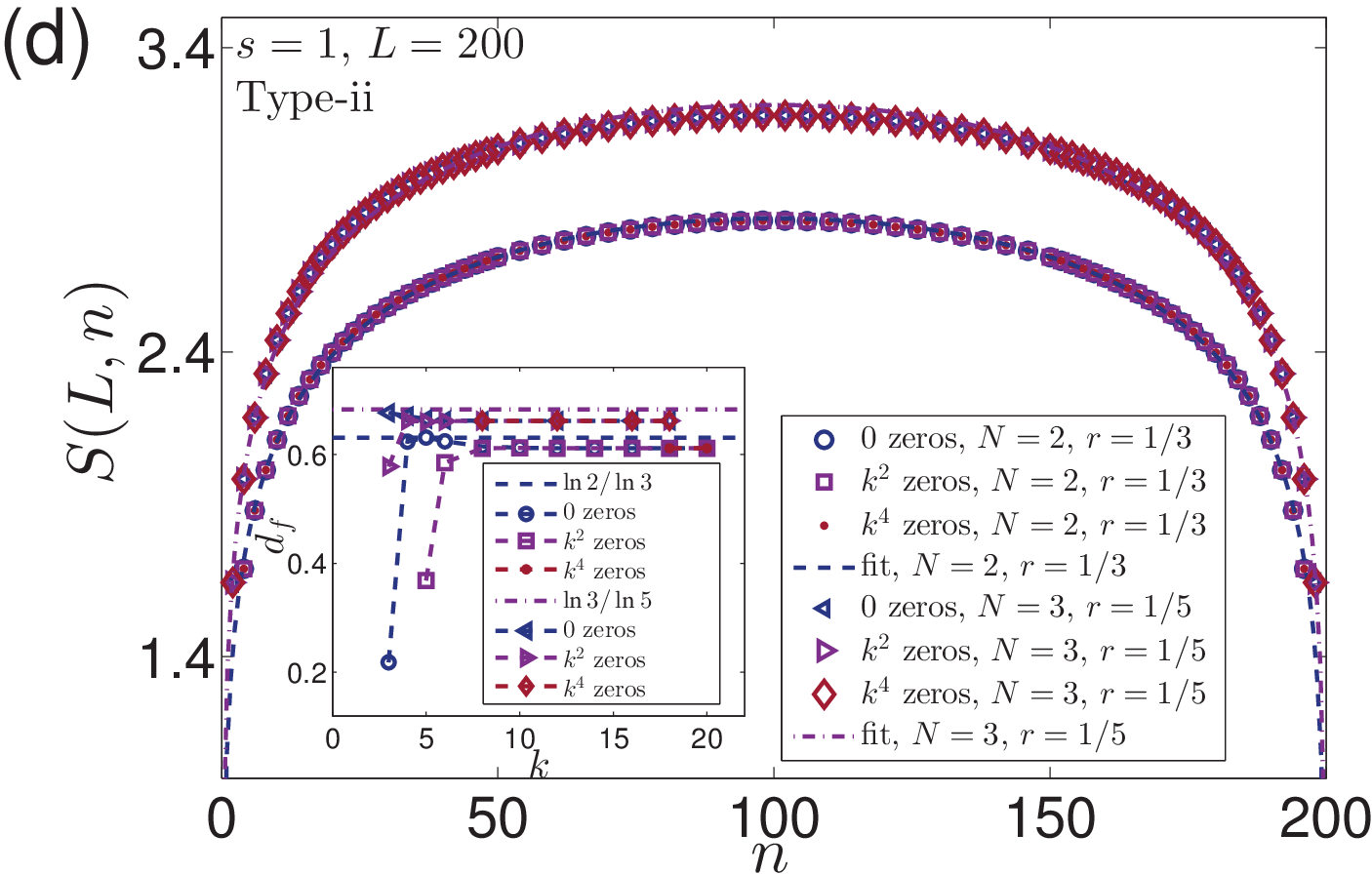}
		\caption{The entanglement entropy $S(L,n)$ versus $n$ for a degenerate ground state $|\Phi_C(\theta)\rangle$ on the Cantor set  $C[N,r;\{k\}]$ (cf. Eq.~(\ref{lcfractal}), with $\theta=\pi/2$) in the spin-$s$ ${\rm SU}(2)$ ferromagnetic Heisenberg model: (a) $s=1/2$ for the first type of realization (labeled as type-i); (b) $s=1/2$ for the second type of realization (labeled as type-ii); (c) $s=1$ for the first type of realization; (d) $s=1$ for  the second type of realization. Here, $N=2$ and $r=1/3$ with $k=20$, and  $N=3$ and $r=1/5$ with $k=18$ have been chosen,  respectively, when $L=200$. Insets: The  fractal dimensions $d_f(k)$  for different values of  $k$, where a dashed line and a dash-dotted line denote $\ln 2/\ln 3$ and $\ln 3/\ln 5$, respectively. The prefactor is half the fractal dimension $d_f$ of the Cantor set  $C[N,r;\{k\}]$, when $k$ is large enough: $d_f=-\ln N/\ln r$, with a relative error being less than 3$\%$ for $s=1/2$ and 3.5$\%$ for $s=1$ when $L=200$.
	}
	\label{spins1s12g3g5typeItypeII}
\end{figure}
\begin{figure}[ht!]
	\centering
	\includegraphics[width=0.35\textwidth]{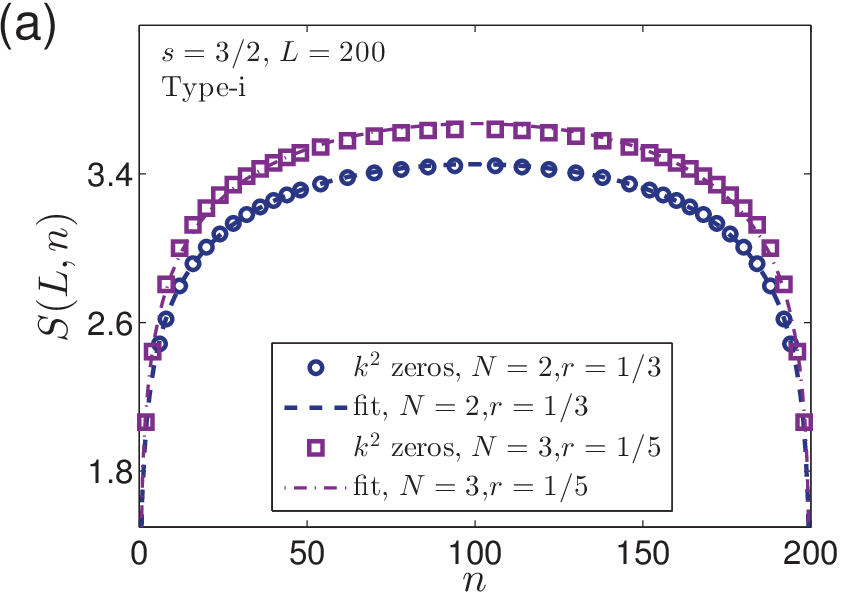}
	\includegraphics[width=0.35\textwidth]{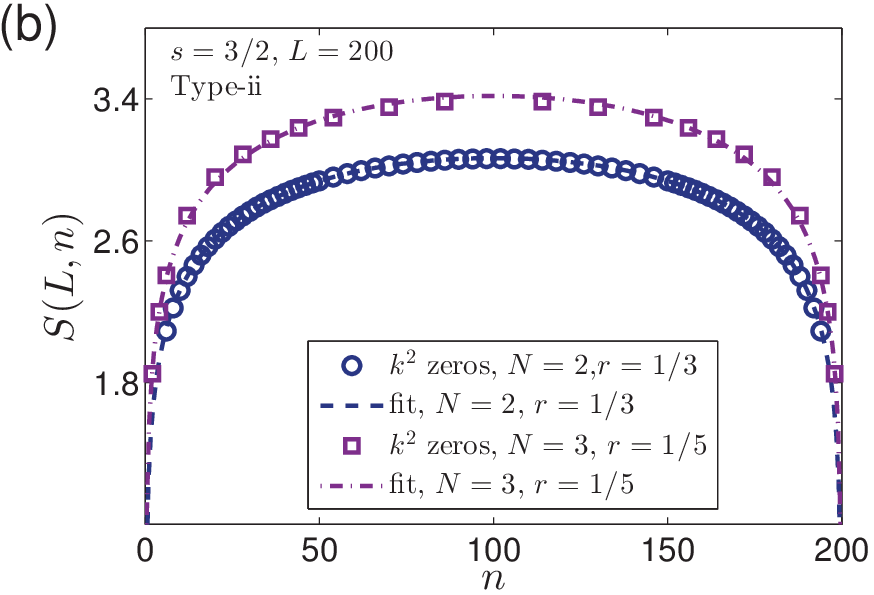}
	\includegraphics[width=0.35\textwidth]{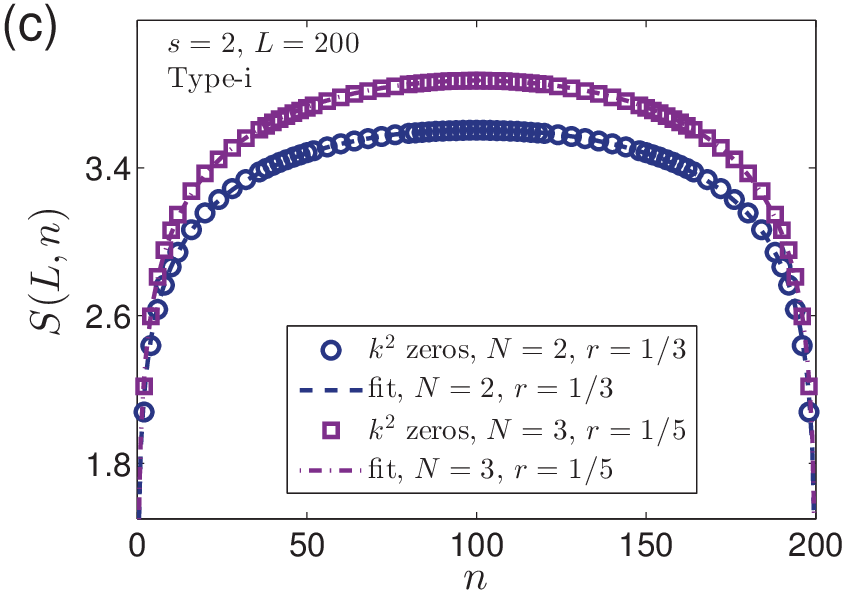}
	\includegraphics[width=0.35\textwidth]{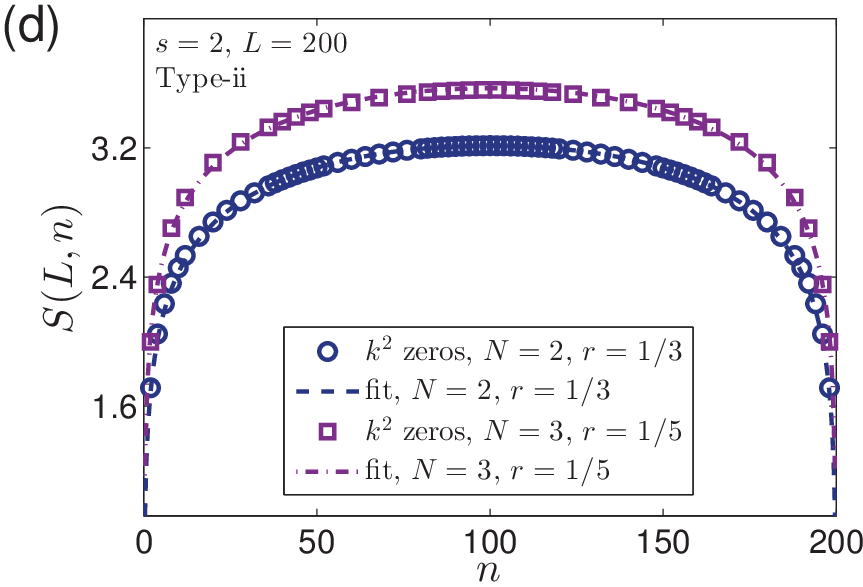}
	\caption{The entanglement entropy $S(L,n)$ versus $n$  for  a degenerate ground state $|\Phi_C(\theta)\rangle$ on the Cantor set  $C[N,r;\{k\}]$ [cf. Eq.~(\ref{lcfractal}), with $\theta=\pi/2$] in the spin-$s$ ${\rm SU}(2)$ ferromagnetic Heisenberg model: (a) $s=3/2$ for the first type of realization (labeled as type-i); (b) $s=3/2$ for the second type of realization (labeled as type-ii); (c) $s=2$ for the first type of realization; (d) $s=2$ for  the second type of realization. Here, $N=2$ and $r=1/3$ with $k=20$, and  $N=3$ and $r=1/5$ with $k=18$ have been chosen,  respectively, when $L=200$. The prefactor is half the fractal dimension $d_f$ of the Cantor set  $C[N,r;\{k\}]$, when $k$ is a large enough: $d_f=-\ln N/\ln r$, with a relative error being less than 3.5$\%$ for $s=3/2$ and less than  $3.5\%$ for $s=2$ when $L=200$.
	}
	\label{spins23s2g3g5typeItypeII}
\end{figure}

In Fig.~\ref{spins1s12g3g5typeItypeII}, we plot the entanglement entropy $S(L,n)$ versus $n$ for a degenerate ground state $|\Phi_C(\theta)\rangle$, as a linear combination on a Cantor set [cf. Eq.~(\ref{lcfractal}) with $\theta=\pi/2$], where  $s=1/2$ and $s=1$, when $L=200$.
The Cantor sets are taken to be $C[N,r;\{k\}]$, with $N=2$ and $r=1/3$, $N=3$ and $r=1/5$.
We have implemented the first type of realization (labeled as type-i) in Fig.~\ref{spins1s12g3g5typeItypeII} (a)  with $s=1/2$ and the second type of realization (labeled as type-ii) in Fig.~\ref{spins1s12g3g5typeItypeII} (b) with $s=1/2$, the first type of realization in Fig.~\ref{spins1s12g3g5typeItypeII} (c)  with $s=1$ and the second type of realization in Fig.~\ref{spins1s12g3g5typeItypeII} (d) with $s=1$.
Here, the number of zero coefficients has been taken to be 0, $k^2$ and $k^4$, respectively.
Our numerical results indicate that, when $k=20$, a relative difference for the values of the entanglement entropy $S(L,n)$ is less than $0.1\%$ for the two linear combinations: in one linear combination there is no  zero coefficients and in the other linear combination the number of zero coefficients is $k^4$.

In Fig.~\ref{spins23s2g3g5typeItypeII}, we plot the entanglement entropy $S(L,n)$  versus $n$  for a degenerate ground state $|\Phi_C(\theta)\rangle$,  as a linear combination on a Cantor set [cf. Eq.~(\ref{lcfractal}) with $\theta=\pi/2$], where $s=3/2$, and $s=2$ when $L=200$.
We have implemented the first type of realization (labeled as type-i) in Fig.~\ref{spins23s2g3g5typeItypeII} (a)  with $s=3/2$ and the second type of realization (labeled as type-ii) in Fig.~\ref{spins23s2g3g5typeItypeII} (b) with $s=3/2$, the first type of realization in Fig.~\ref{spins23s2g3g5typeItypeII} (c)  with $s=2$ and the second type of realization in Fig.~\ref{spins23s2g3g5typeItypeII} (d) with $s=2$.
The Cantor sets are taken to be $C[N,r;\{k\}]$, with $N=2$ and $r=1/3$ at the step $k=20$, and $N=3$ and $r=1/5$, at the step $k=18$.
Here the number of zero coefficients is chosen to be $k^2$.

\begin{figure}[ht!]
	\centering
	\includegraphics[width=0.35\textwidth]{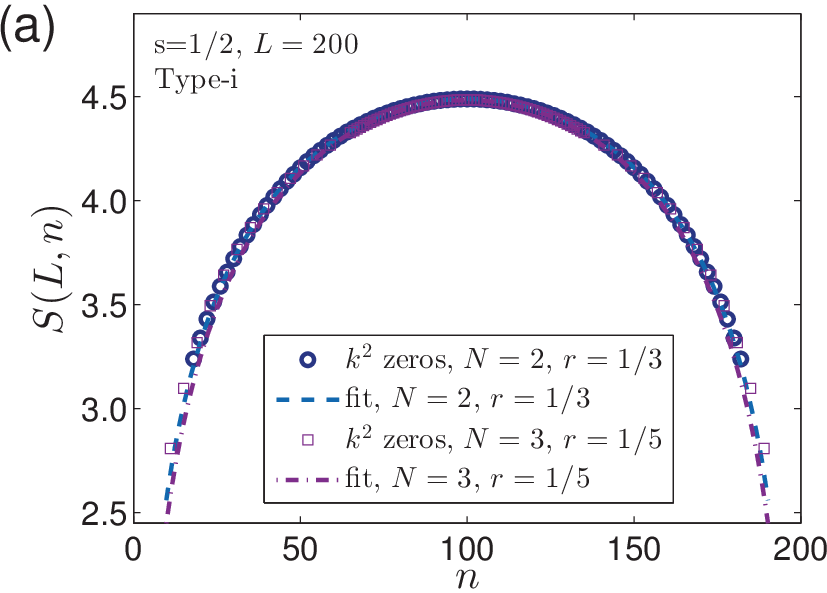}
	\includegraphics[width=0.35\textwidth]{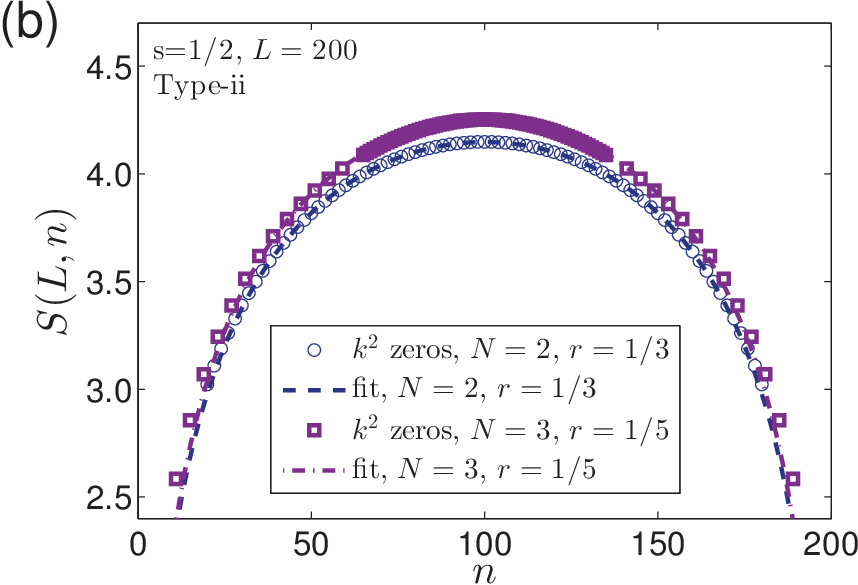}
	\includegraphics[width=0.35\textwidth]{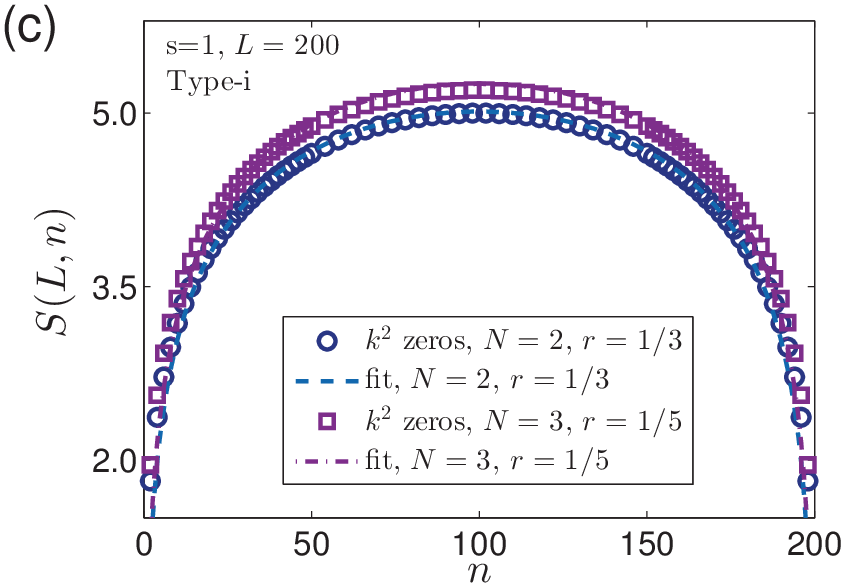}
	\includegraphics[width=0.35\textwidth]{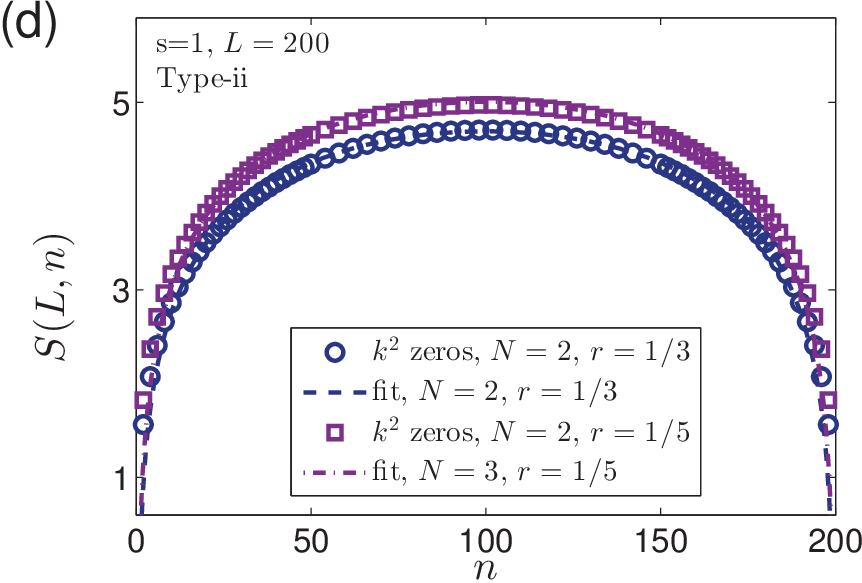}
	\caption{The entanglement entropy $S(L,n)$ versus $n$ for a degenerate ground state $|\Phi_T\rangle$ on the Cantor teepee $C_{\rm tp}[N,r;\{k\}]$ [cf. Eq.~(\ref{lcteepee})] in the  spin-$s$ ${\rm SU}(2)$ ferromagnetic Heisenberg model: (a) $s=1/2$ for the first type of realization (labeled as type-i); (b) $s=1/2$ for the second type of realization (labeled as type-ii); (c) $s=1$ for the first type of realization; (d) $s=1$ for  the second type of realization, when $L=200$.  Here, $N=2$ and $r=1/3$ with $k=20$, and $N=3$ and $r=1/5$ with $k=18$ have been chosen. The prefactor is half the fractal dimension $d_f$ of the Cantor teepee $C_{\rm tp}[N,r;\{k\}]$, when $k$ is large enough: $d_f = 1-\ln N/\ln r$, with a relative error being less than 2$\%$.
	}
	\label{spinsg3g5teepeetypeItypeII}
\end{figure}

In Fig.~\ref{spinsg3g5teepeetypeItypeII}, we plot the entanglement entropy $S(L,n)$ versus $n$  for  a degenerate ground state $|\Phi_T\rangle$, as a linear combination  on a Cantor teepee (cf. Eq.~(\ref{lcteepee})), where $s=1/2$ and $s=1$, respectively.
We have implemented the first type of realization (labeled as type-i) in Fig.~\ref{spinsg3g5teepeetypeItypeII} (a)  with $s=1/2$ and the second type of realization (labeled as type-ii) in Fig.~\ref{spinsg3g5teepeetypeItypeII} (b) with $s=1/2$, the first type of realization in Fig.~\ref{spinsg3g5teepeetypeItypeII} (c)  with $s=1$ and the second type of realization in Fig.~\ref{spinsg3g5teepeetypeItypeII} (d) with $s=1$.
The Cantor teepees are taken to be $C_{\rm tp}[N,r;\{k\}]$, with  $N=2$ and $r=1/3$, and $N=3$ and $r=1/5$, at the step $k=20$.
Here the number of zero coefficients is chosen to be $k^2$.

In our numerical implementation, we have chosen a sequence of  the step numbers to extract a sequence of  the fractal dimensions, labeled as $d_f(k)$, in order to see if the fractal dimensions $d_f(k)$ thus extracted converge to the exact values,  as $k$ gets large enough.  As shown in the insets of Fig.~\ref{spins1s12g3g5typeItypeII} (a), (b), (c) and (d) for $s=1/2$ and $s=1$,  the  fractal dimensions $d_f(k)$ approach $-\ln N/\ln r$ for the Cantor set  $C[N,r;\{k\}]$. We also observe the fractal dimensions $d_f(k)$ approach $1-\ln N/\ln r$ for the Cantor teepee $C_{\rm tp}[N,r;\{k\}]$, if $k$ is large enough (but not shown here).

The same conclusion may be drawn from a numerical test for any other circle on the sphere $S^2$ by setting $\theta$ to be any value in the interval $(0,\pi/2)$ for the Cantor set  $C[N,r;\{k\}]$ and by demanding $\theta$ to  range from 0 to $\theta_{\rm max}$, with $\theta_{\rm max}$ being a value less than $\pi/2$, for the Cantor teepee $C_{\rm tp}[N,r;\{k\}]$. However,  for the Cantor set  $C[N,r;\{k\}]$ with $0<\theta<\pi/2$, a larger $L$ is required to reach the same accuracy in the fractal dimension than that for  the Cantor set  $C[N,r;\{k\}]$ with $\theta=\pi/2$. For the Cantor teepee $C_{\rm tp}[N,r;\{k\}]$, if $\theta$ ranges from 0 to $\theta_{\rm max}$, with $\theta_{\rm max}$ being a value less than $\pi/2$, then the minimum $L$ to achieve the same accuracy is much larger, compared to that if one demands $\theta$ to range from 0 to $\pi/2$.   Therefore, the computational costs are much higher to achieve the convergence of the values of the fractal dimension to the desired exact value for a specific choice of $N$ and $r$ if such choices are made.

\subsection{The ${\rm SU}(3)$ ferromagnetic Heisenberg model}

 For the ${\rm SU}(3)$ ferromagnetic Heisenberg model, we only consider a linear combination on a fractal decomposable into a set of the Cantor sets on the circles
 by setting $\theta_1$ and $\theta_2$ to be certain values in the interval $(0,\pi)$, whereas $\phi_1$ and $\phi_2$ vary from 0 to $2\pi$.
 Specifically,  we may consider a linear combination on a fractal consisting of two Cantor sets $C[N_1;r_1;k]$ and  $C[N_2;r_2;k]$ located on
 the two circles with fixed  $\theta_1$ and $\theta_2$,
 \begin{align*}
 	|\Phi_C(\theta_1,\theta_{2})\rangle=&\frac{1}{Z_{C}} \sum_{\phi_{1,\beta} \in C[N_1;r_1;k],\;\phi_{2,\gamma} \in C[N_2;r_2;k]} c(\phi_{1,\beta},\;\phi_{2,\gamma}) \times \nonumber \\
 	& |\psi(\theta_1,\phi_{1,\beta};\theta_{2},\phi_{2,\gamma})\rangle, \label{lcsu3}
 \end{align*}
 where $c(\phi_{1,\beta},\;\phi_{2,\gamma})$ ($\beta =1,2,\ldots,N_1^k$, $\gamma =1,2,\ldots,N_2^k$) are complex numbers, and ${Z_C}$ is a normalization factor to ensure that $|\Phi_C(\theta_1,\theta_{2})\rangle$ has been normalized. For brevity, we have assumed that $c(\phi_{1,\beta},\;\phi_{2,\gamma})$ are factorized, i.e.,  $c(\phi_{1,\beta},\;\phi_{2,\gamma})=c(\phi_{1,\beta}) c(\phi_{2,\gamma})$.

 In addition to a linear combination with equal coefficients
 two distinct types of realizations are considered: one is that both the amplitudes and the phase angles of the coefficients $c(\phi_{1,\beta})$ and $c(\phi_{2,\gamma})$ in a given linear combination
 are randomly chosen from their respective finite sets, with equal probability; the other is that both the amplitudes and the phase angles of the coefficients $c(\phi_{1,\beta})$ and $c(\phi_{2,\gamma})$ in a given linear combination are predetermined from a (piece-wisely) continuous function. In both cases, it is allowed to set some coefficients to be zero, as long as the number of zero coefficients scales polynomially with $k$ at each step $k$.
 For the first type of realization, both the amplitudes and the phase angles of $c(\phi_{1,\beta})$ and $c(\phi_{2,\gamma})$ are chosen randomly from the set $\{0.1,\;0.2,\;0.3,\;0.4,\;0.5,\;0.6,\;0.7,\;0.8,\;0.9,\;1\}$ and the set
 $\{\pi/10,\pi/5,3\pi/10,2\pi/5,\pi/2,3\pi/5,\;7\pi/10,\;4\pi/5,\;9\pi/10,\pi\}$, respectively. For the second type of realization, we have set the amplitude of $c(\phi_{1,\beta})$ to be a function $\cos(2\phi_{1,\beta}+1)+2$ and  the phase angle of $c(\phi_{1,\beta})$ to be  a function $\phi_{1,\beta}^2/(2\pi)$, whereas the amplitude of $c(\phi_{2,\gamma})$ to be a function $\cos(\phi_{2,\gamma}+1)+2$ and  the phase angle of $c(\phi_{2,\gamma})$ to be a function $\phi_{2,\gamma}^3/(4\pi^2)$, respectively.

\begin{figure}[ht!]
 	\centering
 	\includegraphics[width=0.35\textwidth]{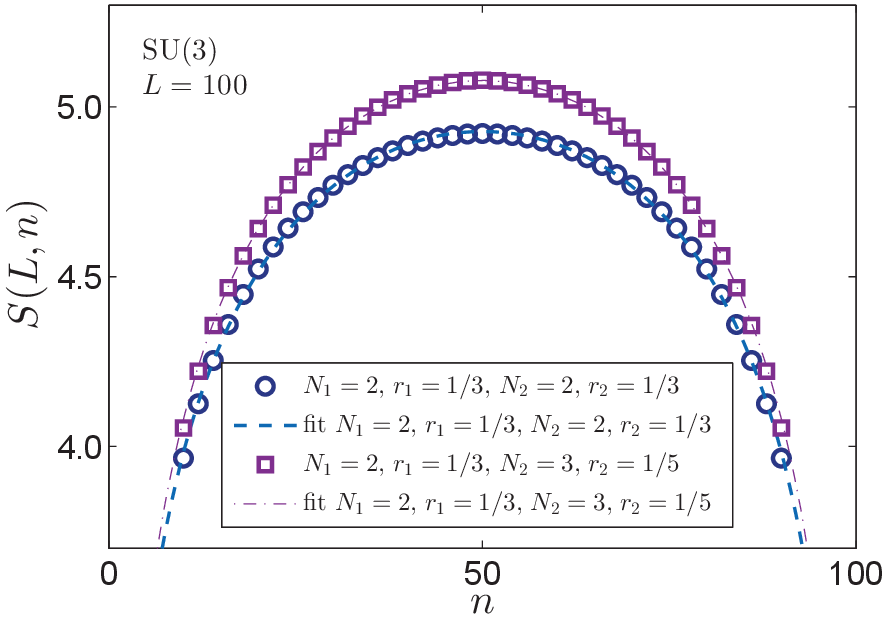}
 	\caption{The entanglement entropy $S(L,n)$ versus $n$ for a degenerate ground state $|\Phi_C(\theta_1,\theta_{2})\rangle$  on a fractal decomposable into two Cantor sets $C[N_1,r_1;\{k\}]$ and $C[N_2,r_2;\{k\}]$ with $\theta_1=\theta_2=\pi/2$, in the  ${\rm SU}(3)$ ferromagnetic Heisenberg model when $L=100$, for $N_1=N_2=2$, $r_1=r_2=1/3$ and $k=15$, and $N_1=2$,  $r_1=1/3$, $N_2=3$, $r_2=1/5$ and $k=11$, respectively.
 	The prefactor is half the sum of the fractal dimensions $d_{f1}$ and $d_{f2}$: $d_{f1} = -\ln N_1/\ln r_1$ and $d_{f2} = -\ln N_2/\ln r_2$, with a relative error being less than 3$\%$.}
 	\label{spinsu3g3g5}
 \end{figure}

 In Fig.~\ref{spinsu3g3g5}, we plot the entanglement entropy $S(L,n)$  versus $n$  for a degenerate ground state  $|\Phi_C(\theta_1,\theta_{2})\rangle$, as a linear combination with equal coefficients, on a fractal decomposable into two Cantor sets, with $\theta_1=\theta_2=\pi/2$, in the ${\rm SU}(3)$ ferromagnetic Heisenberg model when $L=100$. The Cantor sets are taken to be $C[N_1,r_1;\{k\}]$ and $C[N_2,r_2;\{k\}]$, with  $N_1=N_2=2$, $r_1=r_2=1/3$ at the step $k=15$, and $N_1=2$,  $r_1=1/3$, $N_2=3$, $r_1=1/5$  at the step $k=11$.

\begin{figure}[ht!]
	\centering
	\includegraphics[height=4.3cm]{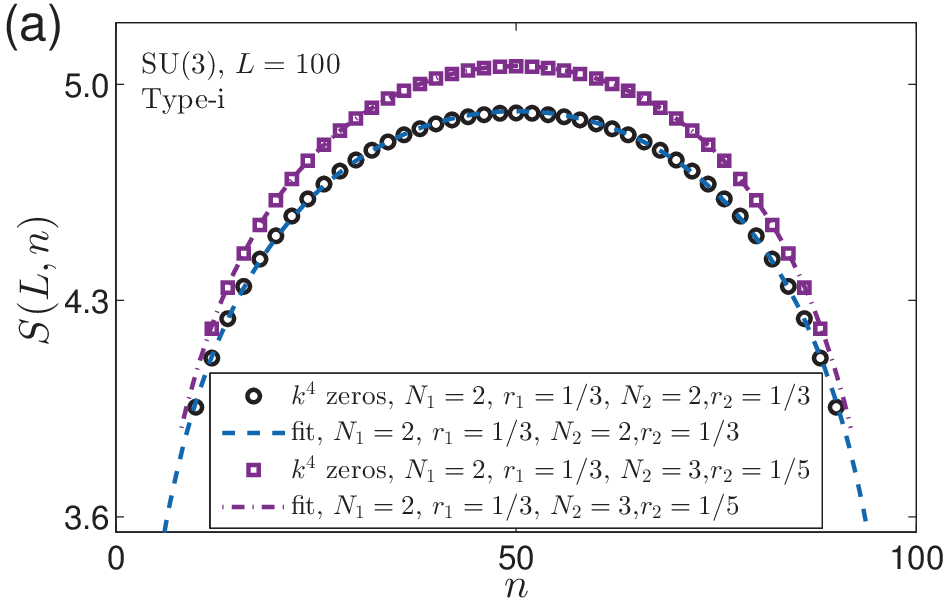}
	\includegraphics[height=4.3cm]{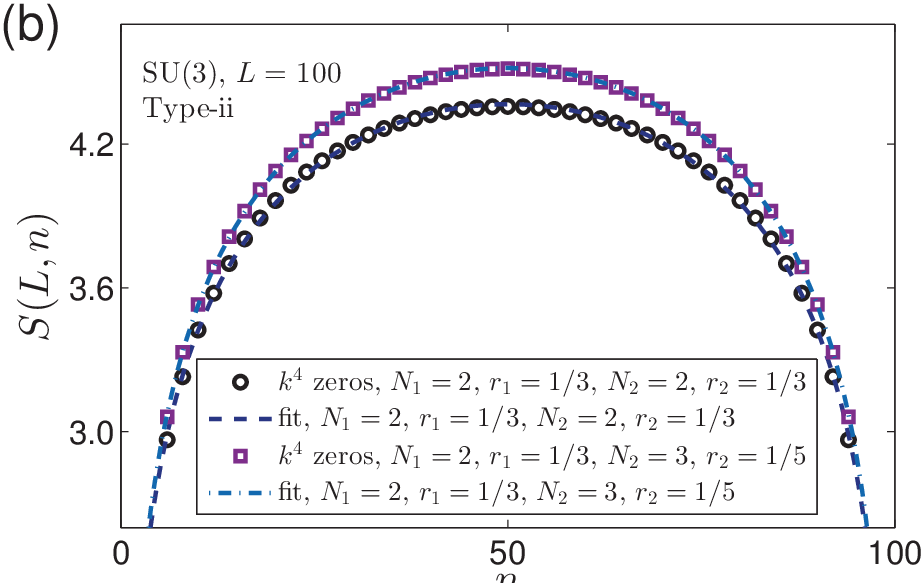}
	\caption{The entanglement entropy $S(L,n)$ versus $n$ for a degenerate ground state $|\Phi_C(\theta_1,\theta_{2})\rangle$ on a fractal decomposable into two Cantor sets $C[N_1,r_1;\{k\}]$ and $C[N_2,r_2;\{k\}]$ with $\theta_1=\theta_2=\pi/2$ in the ${\rm SU}(3)$ ferromagnetic Heisenberg model when $L=100$, for $N_1=N_2=2$, $r_1=r_2=1/3$ and $k=15$, and $N_1=2$,  $r_1=1/3$, $N_2=3$, $r_1=1/5$ and $k=11$, respectively: (a) the first type of realization (labeled as type-i); (b) the second type of realization (labeled as type-ii). The prefactor is half the sum of the fractal dimensions $d_{f1}$ and $d_{f2}$: $d_{f1} = -\ln N_1/\ln r_1$ and $d_{f2} = -\ln N_2/\ln r_2$, with a relative error being less than 3$\%$.
	}
	\label{spinsu3g3g5typeItypeII}
\end{figure}

In Fig.~\ref{spinsu3g3g5typeItypeII}, we plot the entanglement entropy $S(L,n)$  versus $n$  for a degenerate ground state $|\Phi_C(\theta_1,\theta_{2})\rangle$  on a fractal decomposable into two Cantor sets $C[N_1,r_1;\{k\}]$ and $C[N_2,r_2;\{k\}]$ with $\theta_1=\theta_2=\pi/2$, when $L=100$, for $N_1=N_2=2$, $r_1=r_2=1/3$ and $k=15$, and $N_1=2$,  $r_1=1/3$, $N_2=3$, $r_2=1/5$ and $k=11$, respectively.
We have implemented the first type of realization (labeled as type-i) in Fig.~\ref{spinsu3g3g5typeItypeII} (a)  and the second type of realization (labeled as type-ii) in Fig.~\ref{spinsu3g3g5typeItypeII} (b).
Here the number of zero coefficients has been chosen to be $k^4$.

\begin{figure}[ht!]
	\centering
	\includegraphics[height=4.3cm]{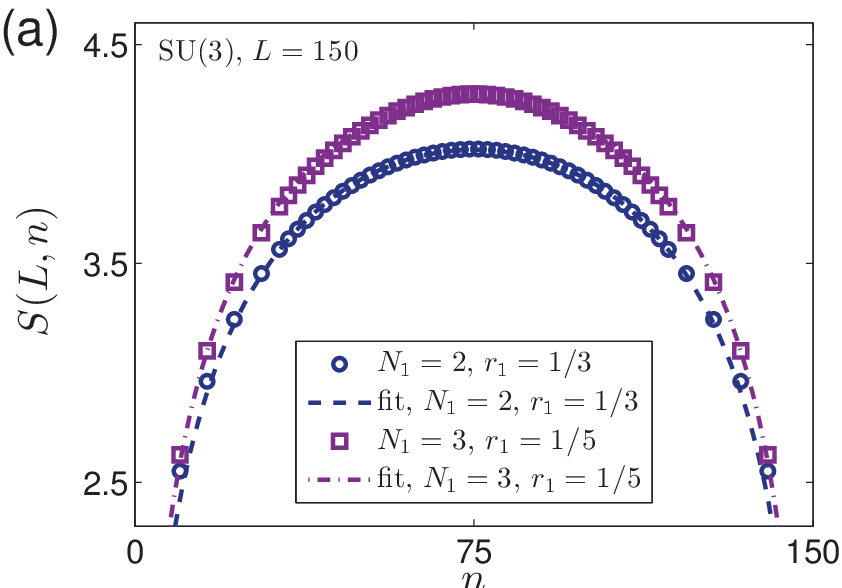}\vspace{3mm}
	\includegraphics[height=4.3cm]{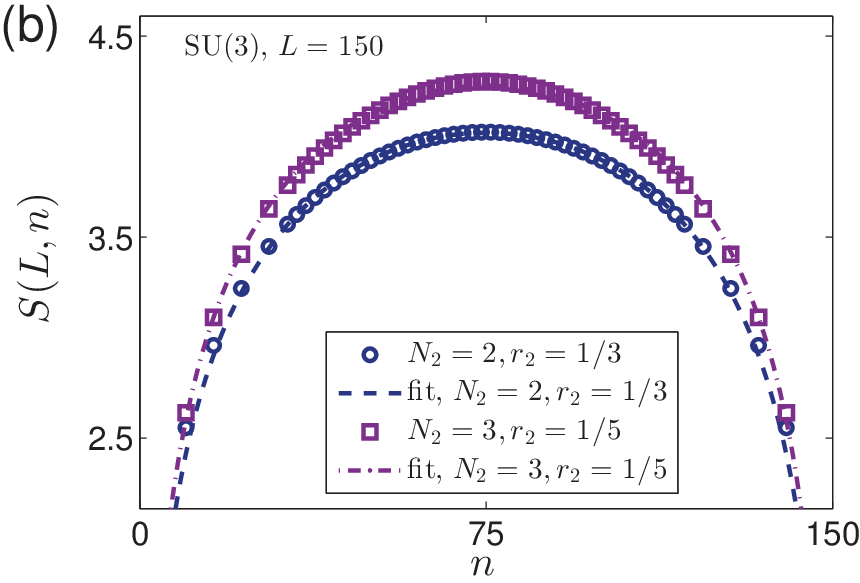}
	\caption{The entanglement entropy $S(L,n)$ versus $n$  for  a degenerate ground state in the ${\rm SU}(3)$ ferromagnetic Heisenberg model: (a) on a Cantor teepee  $C_{\rm tp}[N_1, r_1;\{k\}]$ for  $\phi_1 \in C[N_1, r_1;\{k\}]$ and $0\le\theta_1\le \pi/2$, with $\phi_2=\theta_2=0$, and (b) on a Cantor teepee  $C_{\rm tp}[N_2,r_2;\{k\}]$ for $\phi_2 \in C[N_2, r_2;\{k\}]$ and $0\le\theta_2\le \pi/2$, with $\phi_1=\theta_1=0$, when $L=150$.  Here, $N_1=2$ and $r_1=1/3$ with $k=20$, and $N_1=3$ and $r_1=1/5$ with $k=18$ have been chosen for $C_{\rm tp}[N_1, r_1;\{k\}]$, whereas $N_2=2$ and $r_2=1/3$ with $k=20$, and $N_2=3$ and $r_2=1/5$ with $k=18$ have been chosen for $C_{\rm tp}[N_2,r_2;|{k}]$. The prefactor is half the fractal dimension $d_f$ of a Cantor teepee for an indicated value of $k$, with a relative error being less than 3$\%$.
	}
	\label{su3teepee}
\end{figure}

In Fig.~\ref{su3teepee}, we plot the entanglement entropy $S(L,n)$ versus $n$  for  a degenerate ground state in the ${\rm SU}(3)$ ferromagnetic Heisenberg model: (a) on a Cantor teepee  $C_{\rm tp}[N_1, r_1;\{k\}]$ for  $\phi_1 \in C[N_1, r_1;\{k\}]$ and $0\le\theta_1\le \pi/2$, with $\phi_2=\theta_2=0$, and (b) on a Cantor teepee  $C_{\rm tp}[N_2,r_2;\{k\}]$ for $\phi_2 \in C[N_2, r_2;\{k\}]$ and $0\le\theta_2\le \pi/2$, with $\phi_1=\theta_1=0$, when $L=150$.
In Fig.~\ref{su3teepee}(a), the Cantor teepees are taken to be  $C_{\rm tp}[N_1, r_1;\{k\}]$, with $N_1=2$ and $r_1=1/3$ at the step $k=18$, and $N_1=3$ and $r_1=1/5$ at the step $k=20$, respectively. In  Fig.~\ref{su3teepee}(b), the Cantor teepees are taken to be $C_{\rm tp}[N_2,r_2;\{k\}]$, with
 $N_2=2$ and $r_2=1/3$ at the step $k=18$, and $N_2=3$ and $r_2=1/5$ at the step $k=20$, respectively.
In particular, for $N_1=N_2$ and $r_1=r_2$, the entanglement entropy $S(L,n)$ are the same with an error less than $10^{-13}$, as a result of the symmetry under the simultaneous exchange of the (local) orthonormal basis states $|1\rangle_j, |2\rangle_j$, and $|3\rangle_j$ ($j=1,\ldots,L$).

In addition,  we demonstrate how the logarithmic scaling behavior of the entanglement entropy emerges in the thermodynamic limit from a numerical perspective in Appendix C, where
two plots are shown for the spin-$1/2$ ${\rm SU}(2)$ ferromagnetic Heisenberg model,  when the Cantor set $C(N,r;\{k\})$ is chosen, with  (a) $N=2$, $r=1/3$, and $k=20$, and (b) $N=3$, $r=1/5$, and $k=18$, respectively. It is concluded that the logarithmic scaling behavior of the entanglement entropy in the thermodynamic limit emerges if $1 \ll n \ll L/2$. Here, we remark that this has already been established for the orthonormal basis states $|L,M\rangle$ in Ref.~\cite{FMGM}, but it is now extended to a generic degenerate ground state.

Our numerical tests confirm the theoretical predictions for the scaling behaviors of the entanglement entropy for linear combinations on the Cantor sets and the Cantor teepees in the spin-$s$ ${\rm SU}(2)$ ferromagnetic Heisenberg model, with $s$ up to $s=2$, and on fractals decomposable into two Cantor sets and the Cantor teepees in the ${\rm SU}(3)$ ferromagnetic Heisenberg model, with the two specific realizations of the restriction on the coefficients being implemented.
In particular, the prefactor in front of the logarithm in the scaling relation (\ref{srfs}) is confirmed numerically to be half the fractal dimension $d_f$ of a decomposable fractal, which in turn is the sum of the Cantor sets contained in the decomposition of the fractal.  For a linear combination on a decomposable fractal,  our numerical tests have been implemented not only for the identical coefficients, but also for the coefficients chosen randomly from a prescribed set or determined from a preset continuous function. In addition, the fractal dimension extracted from a finite system-size scaling analysis for the entanglement entropy of a linear combination on the Cantor teepees match the exact values, with high precision in both the spin-$s$ ${\rm SU}(2)$ ferromagnetic Heisenberg model  and the ${\rm SU}(3)$ ferromagnetic Heisenberg model, both of which  are able to accommodate only one Cantor teepee.  Our numerical results also demonstrate that some of the coefficients are allowed to be zero, as long as the number of zero coefficients scales polynomially with the step number $k$.

\section{Concluding remarks}

We have performed a systematic investigation into the entanglement entropy for highly degenerate ground states of quantum many-body systems undergoing SSB with type-B GMs, with the spin-$s$ ${\rm SU}(2)$ ferromagnetic Heisenberg model, the ${\rm SO}(4)$  ferromagnetic spin-orbital model and the ${\rm SU}(2s+1)$ ferromagnetic Heisenberg model as illustrative examples. The models  exhibit an intrinsic abstract fractal underlying the ground-state subspace. This {\it intrinsic} fractal has been revealed by introducing an {\it extrinsic} fractal  in the coset space $G/H$ and performing an investigation into the entanglement entropy for a linear combination of a set of the overcomplete basis states on the fractal.  Generically, the set of the overcomplete basis states consists of factorized (unentangled) ground states, which are the spin coherent states and their generalizations. It has been demonstrated that the fractal dimensions for all the Cantor sets form a {\it dense} subset in the interval $[0,1]$. As a result, one may restrict, in principle, to fractals decomposable into a set of the Cantor sets, although other types of fractals, such as the Cantor teepees, the Sierpinski carpet and the Sierpinski triangle, are practically useful  for a numerical implementation to evaluate the entanglement entropy.

It has been established that the entanglement entropy scales logarithmically with the block size, with the prefactor being half the fractal dimension of the support of the linear combinations.
We have achieved a full characterization of  the  coefficients in a linear combination, which demands to separate the ground-state subspace into a disjoint union of countably infinitely many regions, each of which is associated with a fractal decomposable to a set of the  Cantor sets. A restriction imposed on the coefficients is that
the norm for the linear combination scales as the square root of the number of the self-similar building blocks kept at each step $k$ for a fractal,
under an assumption that the maximum absolute value of the coefficients in the linear combination is chosen to be around one, and  the coefficients in the linear combination are almost constants within the building blocks. This restriction leads to two specific realizations: the ratio between any two nonzero coefficients either is a random constant at each step $k$ or converges to any nonzero value, as the step number $k$ tends to infinity,  subject to the condition that the number of zero coefficients scales polynomially with $k$. Indeed, the two realizations also satisfy the requirement for the coefficients in a linear combination to ensure that degenerate ground states at different steps are self-similar.
In particular, the set of all the regions  is  dense in the ground-state subspace, in the sense that there is no clear-cut boundary between any two regions. Instead, there is always a region in between, no matter how close they are. Here, the closeness is measured in terms of the fractal dimensions of the fractals.

To achieve our goal, a conceptual framework has been developed,  including an equivalence class in the set of all the supports, an approximation of a fractal to a support, and a decomposition of a fractal into a set of the Cantor sets. In particular,
the orthonormal basis states may be expressed as
a linear combination of a set of the overcomplete basis states  in a subspace of the coset space, which in turn may be regarded as a limit of a sequence of decomposable fractals.
As a consequence, we have been led to the identification of the fractal dimension $d_f$ with the number of type-B GMs $N_B$ for the orthonormal basis states $|L,M\rangle$, $|L,M_s,M_t\rangle$ and $|L,M_1,\ldots,M_{2s}\rangle$: $d_f=N_B$. We stress that the orthonormal basis states $|L,M\rangle$, $|L,M_s,M_t\rangle$ and $|L,M_1,\ldots,M_{2s}\rangle$ are {\it unique}, in the sense that they are the {\it only} degenerate ground states exhibiting the self-similarities in the real space, up to a unitary symmetry transformation. Physically, this is relevant to the fact that,  for each model, the orthonormal basis states constitute the simultaneous eigenvectors of the Hamiltonian and the Cartan generator(s) of the symmetry group, respectively.

Our argument may be extended to any quantum many-body systems undergoing SSB with type-B GMs, with a SSB pattern from $G$ to $H$, as long as $G$ is a semisimple Lie group. Generically, the number of type-B GMs is equal to the rank of the symmetry group $G$~\cite{FMGM}. If $G$ is a simple Lie group, then the type-B GMs are not independent to each other, in contrast with the case with a semisimple Lie group as a symmetry group. This has been explained in the context of the ${\rm SO}(4)$  spin-orbital model as an illustrative example for a semisimple Lie subgroup of the symmetry group $G$. For this model, as it evolves from deep inside the ferromagnetic regime to the ${\rm SU(4)}$ symmetric point at $\zeta=1/4$, the symmetry group ${\rm SO}(4)$ in the ferromagnetic regime, isomorphic to  ${\rm SU}(2) \times {\rm SU}(2)$, appears to be a semisimple Lie subgroup of the symmetry group  ${\rm SU(4)}$ at $\zeta=1/4$.

In addition, the ground-state subspace at least contains a permutation-invariant sector for a quantum many-body system undergoing SSB with type-B GMs. Indeed, the entire ground-state subspace is permutation-invariant for each of the models under investigation. However, this is not necessary, as demonstrated~\cite{TypeBtasaki}  for the flat-band Tasaki model~\cite{tasaki} at quarter filling.

Last but not least, we have performed a systematic finite system-size scaling analysis to confirm the theoretical predictions for the scaling behaviors of the entanglement entropy for linear combinations on the Cantor sets
and the Cantor teepees in the spin-$s$ ${\rm SU}(2)$ ferromagnetic Heisenberg model, with $s$ up to $s=2$, and on fractals decomposable into two Cantor sets and the Cantor teepees in the ${\rm SU}(3)$ ferromagnetic Heisenberg  model, with the two specific realizations of the restriction on the coefficients being implemented.

{\it Acknowledgment.} - We thank Murray Batchelor for his helpful comments and suggestions during the preparation of the paper. J.O.F. is grateful to the Center for Modern Physics, Chongqing University for their hospitality and financial support during his stay in the spring of 2024.
I. P. M. acknowledges funding from the National Science and Technology Council (NSTC) Grant No. 112-2811-M-007-044 and 113-2112-M-007-MY2.
%%%%%%%%%%%%%%%%%%%%%%%%%%%%%%%%%%%%Appendix%%%%%%%%%%%%%%%%%%%%%%%

\section*{Appendices}
\twocolumngrid
\setcounter{section}{0}
\setcounter{equation}{0}
\setcounter{figure}{0}
\renewcommand{\theequation}{A\arabic{equation}}
\renewcommand{\thefigure}{A\arabic{figure}}
\subsection{ The Cantor sets and the Cantor teepees}\label{AA}
Taking the interval $[0,1]$, denoted as $C[2,1/3;0]$ at step $0$,  and dividing it into three equal subintervals and removing the  middle subinterval, we have $C[2,1/3;1]$ at step 1.
Repeating the same  procedure for the two remaining subintervals, and  removing the middle subsubintervals, we have $C[2,1/3;2]$ at step 2.
Keep repeating  the same  procedure $k$ times,  we have $C[2,1/3;k]$ at step $k$. This procedure yields the Cantor set, denoted as $C[2,1/3;\{k\}]$, where $\{k\}$ denotes a set of the natural numbers, as $k$ tends to infinity. Note that the fractal dimension for the Cantor set $C[2,1/3;\{k\}]$ is $d_f=\ln 2/\ln 3$.
An extension to a (generalized) Cantor set is straightforward.
Specifically, it is created by dividing the interval $[0,\;  1]$ into $1/r$ parts, and removing  $1/r -N$ subintervals, we have $C[N,r;1]$ at step 1. Here, we have assumed that $1/r$ is always a positive integer and $1/r > N$. Repeating the same  procedure for the $N$ remaining subintervals, and  removing  the subsubintervals in the same way as done for $C[2,1/3;\{k\}]$, we have $C[N,r;2]$ at step 2. Keep repeating the same  procedure $k$ times,   we have $C[N,r;k]$ at step $k$, thus yielding a (generalized) Cantor set, denoted as $C[N,r;\{k\}]$.  Indeed, the number of subintervals in a Cantor set $C[N,r;\{k\}]$ is $N^k$ at the step $k$, and the fractal dimension $d_f$ for the Cantor set $C[N,r;\{k\}]$ is $d_f= -\ln N/\ln r$.

Strictly speaking, the notation  $C[N,r;\{k\}]$ does not fix a Cantor set uniquely for generic $N$ and $r$, because we have not yet specified the way to keep $N$ subintervals among $1/r$ subintervals.  However, the geometric information encoded in the Cantor set $C[N,r;\{k\}]$ {\it only} depends on the values of $N$ and $r$, as far as the scaling behaviors of the entanglement entropy are concerned for highly degenerate ground states arising from SSB with type-B GMs.   Mathematically, this is due to the fact that a linear combination on the Cantor set $C[N,r;\{k\}]$ at the step $k$ may be expanded in terms of the orthonormal basis states $|L,M\rangle$, with the number of the coefficients being $2sL+1$. We are thus led to a set of equations from equating the coefficients in the expansions of the two linear combinations of the overcomplete basis states on the two Cantor sets with the same $N$ and $r$, which are expressed in terms of  the orthonormal basis states $|L,M\rangle$. As long as $k$ is large enough, the number of equations, i.e., $2sL+1$, is much less than the number of the coefficients in the linear combinations of the overcomplete basis states on the two Cantor sets. As a result, if one set of the coefficients is fixed, then there is always a solution to the set of equations to yield the other set of the coefficients, and vice versa.
In addition, the Cantor set $C[N^q,r^q;\{k\}]$ at the step $k$ is identical to the Cantor set $C[N,r;\{qk\}]$
at the step $q\; k$, with $q$ being a positive integer. That is, $C[N^q,r^q;\{k\}]$ is identical to $C[N,r;\{qk\}]$, both of which share the same fractal dimension.

A Cantor teepee~\cite{cantorset,cantorteepee}, known as the Knaster-Kuratowski fan and Cantor's leaky tent, is a fractal  in a two-dimensional setting, defined as an extension of a Cantor set in a one-dimensional setting.
A Cantor teepee is constructed from the Cantor set and an apex point.
Suppose a Cantor set $C[N,r;\{k\}]$ is located on $[0,\;  1]$, with the two ends at the point $(0,\; 0)$ and $(1,\;0)$, whereas the apex point $p$ is located at (1/2, 1/2).
For $c\in C[N,r;\{k\}]$, the line segment connecting ($c$, 0) to $p$ is denoted as $L(c)$. If $c\in C[N,r;\{k\}]$ is an endpoint of an interval discarded in the construction of $C[N,r;\{k\}]$, then we define $X_c=\{ (x,y)\in L(c):y\in Q\}$; for all other points in $C[N,r;\{k\}]$, we define $X_c=\{(x,y)\in L(c): y\notin Q\}$. Then a Cantor teepee is defined as $\bigcup_{c\in C[N,r;\{k\}]}X_c$, equipped with the subspace topology inherited from the standard topology on the two-dimensional Euclidean space. For our purpose, it is convenient to denote
a Cantor teepee as $C_{\rm {tp}}[N,r;\{k\}]$, with $k$ being the step number, which tends to infinity: $k\rightarrow \infty$.
Note that, the Cantor teepee $C_{\rm {tp}}[N,r;\{k\}]$ is connected, but becomes totally disconnected upon the removal of the apex point $p$.
The fractal dimension for the Cantor teepee $C_{\rm {tp}}[N,r;\{k\}]$ is $d_f= 1-\ln N/\ln r$.

\subsection{A fractal decomposable into a set of the Cantor sets}\label{AB}

For our purpose, one may restrict to a  fractal that admits a decomposition into a set of subfractals, e.g., the Cantor sets. We stress that not all fractals admit such a decomposition, with the Cantor teepees, the Sierpinski carpet and the Sierpinski triangle being examples.  However,  as a support, they may be well approximated in terms of two Cantor sets $C[N_1,r_1; \{k\}]]$ and $C[N_2,r_2; \{k\}]$. In this context, a decomposable fractal in a two-dimensional setting, represented as  $C[N_1,r_1; \{k\}] \times C[N_2,r_2; \{k\}]$, is called a variant of the Sierpinski carpet. Actually, this may be extended to a decomposable fractal in a $4s$-dimensional setting. It is decomposed into a set of the Cantor sets, with the number of the Cantor sets being $4s$ and the fractal dimension of the fractal being  the sum of the fractal dimensions of the Cantor sets contained in the decomposition. Hence, an analog of the variants of the  Sierpinski carpet in a higher dimensional setting may be defined accordingly.

It is conceptually satisfying to restrict to the set of all the Cantor sets, which may be regarded as a minimal set of fractals to investigate the scaling behaviors of the entanglement entropy for degenerate ground states arising from SSB with type-B GMs. In practice, it is still helpful to keep fractals that do not admit such a decomposition. In particular, the Cantor teepees have been kept as supports to form linear combinations,  and the entanglement entropy for degenerate ground states thus constructed have been evaluated.

\subsection{Logarithmic scaling of the entanglement entropy in the thermodynamic limit: a finite-size numerical approach}\label{AC}

	\begin{figure}
		\centering
		\includegraphics[width=0.35\textwidth]{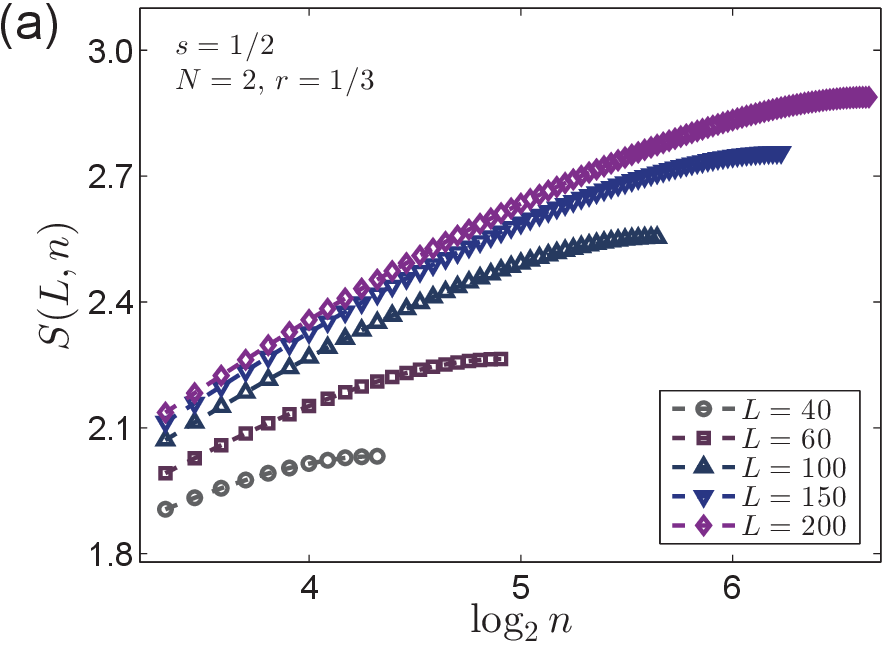}
		\includegraphics[width=0.35\textwidth]{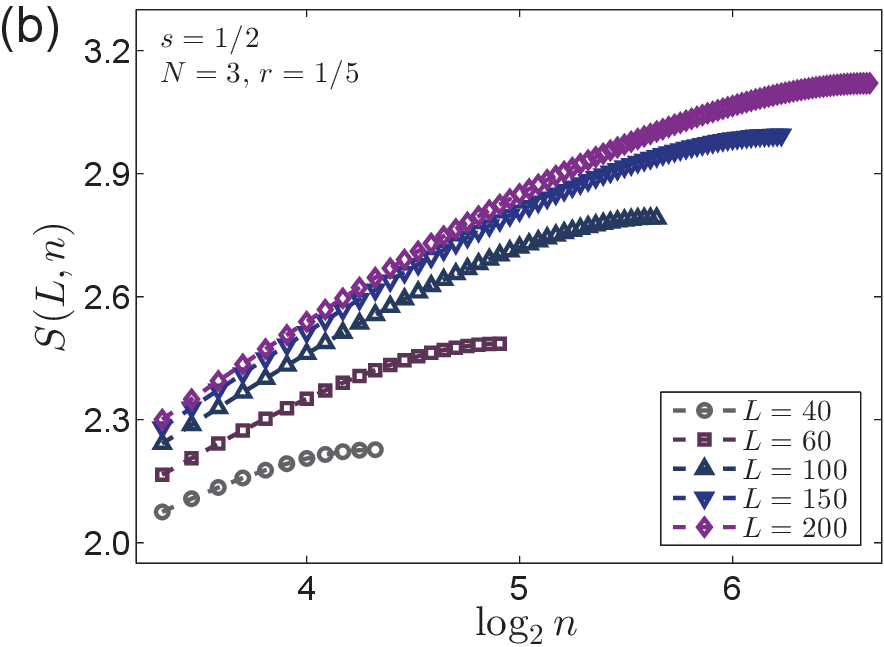}
		\caption{ The entanglement entropy $S(L,n)$ versus $\log_2 n$ for the spin-$1/2$ ${\rm SU}(2)$ ferromagnetic Heisenberg model. A degenerate ground state $|\Phi_C(\theta)\rangle$ on the Cantor set $C[N,r;\{k\}]$ [cf. Eq.~(\ref{lcfractal})], with $\theta$ being $\pi/2$, is chosen. We have fixed (a) $N=2$, $r=1/3$, and $k=20$ and (b) $N=3$, $r=1/5$, and $k=18$, as $L$ varies, respectively. Here, $L=40$, $60$, $100$, $150$ and $200$, and  $n$ ranges from 10 to $L/2$.}
		\label{Su2s12finite}
	\end{figure}

In order to understand how the finite system-size scaling relation (\ref{srfs}) reduces to the scaling relation  (\ref{srf}) as the thermodynamic limit $L \rightarrow \infty$ is approached, it is necessary to evaluate the values of the entanglement entropy for different chosen values of $L$. This offers an intuitive way to demonstrate the emergence of the logarithmic scaling behavior of the entanglement entropy in the thermodynamic limit from a numerical perspective.

Here we show two plots in Fig.~\ref{Su2s12finite} for the spin-$1/2$ ${\rm SU}(2)$ ferromagnetic Heisenberg model, where the Cantor set $C[N,r;\{k\}]$ is chosen, with  (a) $N=2$, $r=1/3$, and $k=20$, and (b) $N=3$, $r=1/5$, and $k=18$. As one may see, a significant deviation from the logarithmic scaling behavior is observed for any $L$, but tends to vanish  as $L$ increases, as long as  $n$ is relatively small, if it is compared to $L$, but large enough. More precisely,  the logarithmic scaling behavior of the entanglement entropy in the thermodynamic limit emerges if $1 \ll n \ll L/2$, as already shown for the orthonormal basis states in Ref.~\cite{FMGM}.


\begin{thebibliography}{10}
	
\bibitem{goldstone} J. Goldstone, {\it Field theories with superconductor solutions}, Nuovo Cimento \textbf{19}, 154 (1961); 
J. Goldstone, A. Salam, and S. Weinberg, {\it Broken symmetries}, Phys. Rev. \textbf{127}, 965 (1962); 
Y. Nambu and G. Jona-Lasinio, {\it Dynamical model of elementary particles based on an analogy with superconductivity}, Phys. Rev. \textbf{122}, 345 (1961).
		
\bibitem{anderson} P. W. Anderson, {\it Broken symmetry can't compare with ferromagnets}, Phys. Today {\textbf 43}, 117 (1990); R. Peierls, {\it Spontaneously broken symmetries}, J. Phys. A: Math. Gen. \textbf{24}, 5273 (1991); R. Peierls, T. A. Kaplan, and P. W. Anderson, {\it Reflections-on-broken-symmetry}, Phys. Today {\textbf 44},   13 (1991).
	
\bibitem{watanabe} H. Watanabe and H. Murayama, {\it Unified description of Nambu-Goldstone bosons without Lorentz invariance}, Phys. Rev. Lett. \textbf{108}, 251602 (2012); H. Watanabe and H. Murayama, {\it Effective Lagrangian for nonrelativistic systems}, Phys. Rev. X \textbf{4}, 031057 (2014).
	
\bibitem{nambu} Y. Nambu, {\it Spontaneous breaking of Lie and current algebras}, J. Stat. Phys. \textbf{115}, 7 (2004).
	
\bibitem{nielsen} H. B. Nielsen and S. Chadha,  {\it On how to count Goldstone bosons}, Nucl. Phys. B \textbf{105}, 445 (1976).
	
\bibitem{schafer}  T. Schafer, D. T. Son, M. A. Stephanov, D. Toublan, and J. J. M. Verbaarschot, {\it Kaon condensation and Goldstone's theorem}, Phys. Lett. B \textbf{522}, 67 (2001).
	
\bibitem{miransky} V. A. Miransky and I. A. Shovkovy,  {\it Spontaneous symmetry breaking with abnormal number of Nambu-Goldstone bosons and Kaon condensate},  Phys. Rev. Lett. \textbf{88}, 111601 (2002).
	
\bibitem{nicolis}  A. Nicolis and F. Piazza,  {\it Implications of relativity on nonrelativistic Goldstone theorems: gapped excitations at finite charge density}, Phys. Rev. Lett. \textbf{110}, 011602 (2013); 
H. Watanabe, T. Brauner, and H. Murayama, {\it Massive Nambu-Goldstone bosons}, Phys. Rev. Lett. \textbf{111}, 021601 (2013).
	
\bibitem{brauner-watanabe} H. Watanabe and T. Brauner, {\it Number of Nambu-Goldstone bosons and its relation to charge densities}, Phys. Rev. D \textbf{84}, 125013 (2011).
	
\bibitem{NG} Y. Hidaka, {\it Counting rule for Nambu-Goldstone modes in nonrelativistic systems}, Phys. Rev. Lett. \textbf{110}, 091601 (2013); T. Hayata and Y. Hidaka, {\it Dispersion relations of Nambu-Goldstone modes at finite temperature and density}, Phys. Rev. D \textbf{91}, 056006 (2015); D. A. Takahashi and M. Nitta,  {\it Counting rule of Nambu-Goldstone modes for internal and spacetime symmetries: Bogoliubov theory approach},  Ann. Phys. \textbf{354}, 101 (2015).
	
\bibitem{Metlitski} M. A. Metlitski and T. Grover,  {\it Entanglement entropy of systems with spontaneously broken continuous symmetry},  arXiv: 1112.5166 (2015).

\bibitem{Rademaker} L. Rademaker,  {\it Tower of states and the entanglement spectrum in a coplanar antiferromagnet}, Phys. Rev. B \textbf{92}, 144419 (2015).

\bibitem{typeAmany1} B. Kulchytskyy, C. M. Herdman, S. Inglis, and R. G. Melko, {\it Detecting Goldstone modes with entanglement entropy}, Phys. Rev. B \textbf{92}, 115146 (2015).

\bibitem{typeAmany2} D. J. Luitz, X. Plat, F. Alet, and N. Laflorencie, {\it Universal logarithmic corrections to entanglement entropies in two dimensions with spontaneously broken continuous symmetries}, Phys. Rev B \textbf{91}, 155145 (2015).

\bibitem{typeAmany3} N. Laflorencie, D. J. Luitz, and F. Alet, {\it Spin-wave approach for entanglement entropies of the $J_1$-$J_2$ Heisenberg antiferromagnet on the square lattice}, Phys. Rev. B \textbf{92}, 115126 (2015).

\bibitem{typeAmany4}  D.-V. Bauer and J. O. Fjaerestad, {\it  Spin-wave study of entanglement and Rényi entropy for coplanar and collinear magnetic orders in two-dimensional quantum Heisenberg antiferromagnets}, Phys. Rev. B \textbf{101}, 195124 (2020).

\bibitem{FMGM}  Q.-Q. Shi, Y.-W. Dai, H.-Q. Zhou, and I. P. McCulloch, {\it Fractal dimension and the counting rule of the Goldstone modes}, J. Phys. A: Math. Theor. \textbf{58}, 05LT01 (2025).

\bibitem{goldensu3} H.-Q. Zhou, Q.-Q. Shi, I. P. McCulloch, and M. T. Batchelor, {\it Goldstone modes and the golden spiral in the ferromagnetic spin-1 biquadratic model}, J. Phys. A: Math. Theor. \textbf{58}, 39LT01 (2025). 


\bibitem{spinorbitalsu4} Q.-Q. Shi, H.-Q. Zhou, I. P. McCulloch, and  M. T. Batchelor, {\it  Type-B Goldstone modes and a logarithmic spiral in the staggered SU(4) ferromagnetic spin-orbital model},  arXiv: 2309.04973 (2023).
	
\bibitem{finitesize} H.-Q. Zhou, Q.-Q. Shi, I. P. McCulloch, and M. T. Batchelor, {\it Entanglement entropy for scale-invariant states: universal finite-size scaling}, arXiv: 2304.11339 (2023).
	
\bibitem{mwc} N. D. Mermin and H. Wagner, {\it Absence of ferromagnetism or antiferromagnetism in one- or two-dimensional isotropic Heisenberg models}, Phys. Rev. Lett. \textbf{17}, 1133 (1966); 	S. R. Coleman, {\it There are no Goldstone bosons in two dimensions}, Commun. Math. Phys. \textbf{31}, 259 (1973).
	
\bibitem{2dtypeb}  H.-Q. Zhou, Q.-Q Shi, I. P. McCulloch, and M. T. Batchelor, {\it Entanglement entropy for a type of scale-invariant states in two spatial dimensions and beyond: universal finite-size scaling}, arXiv: 2412.06396 (2024).
	
\bibitem{tasakibook}  H. Tasaki, {\it Physics and Mathematics of Quantum Many-Body Systems} (Springer, Berlin, 2020).
	
\bibitem{doyon} O. A. Castro-Alvaredo and B. Doyon, {\it Entanglement entropy of highly degenerate states and fractal dimensions}, Phys. Rev. Lett. \textbf{108}, 120401 (2012); O. A. Castro-Alvaredo and B. Doyon, {\it Entanglement in permutation symmetric states, fractal dimensions, and geometric quantum mechanics}, J. Stat. Mech. P02016 (2013).
	
\bibitem{TypeBtasaki} H.-Q. Zhou, Q.-Q Shi, I. P. McCulloch, and J. O. Fj{\ae}restad,  {\it Entanglement entropy for the one-dimensional flat-band ferromagnetic Tasaki model: spontaneous symmetry breaking with a type-B Goldstone mode}, arXiv: 2412.00739 (2024).
	
\bibitem{tasaki} H. Tasaki, {\it Ferromagnetism in the Hubbard models with degenerate single-electron ground states}, Phys. Rev. Lett. \textbf{69}, 1608 (1992).	
	
\bibitem{faddeev} L. A. Takhtadzhan and L. D. Faddeev, {\it The quantum method of the inverse problem and the Heisenberg XYZ model}, Russ. Math. Surv. \textbf{34}, 11 (1979).
	
\bibitem{baxterbook} R. J. Baxter, {\it Exactly Solved Models in Statistical Mechanics} (Academic Press, London, 1982).
	
\bibitem{So4}  K. I. Kugel and D. I. Khomskii, {\it The Jahn-Teller effect and magnetism: transition metal compounds}, Sov. Phys. Usp. \textbf{25}, 231 (1982).
	
\bibitem{barber} A. Kl\"{u}mper, {\it New results for q-state vertex models and the pure biquadratic spin-1 Hamiltonian}, Europhys. Lett. \textbf{9}, 815 (1989); M. T. Batchelor and M. N. Barber, {\it Spin-s quantum chains and Temperley-Lieb algebras}, J. Phys. A: Math. Gen. \textbf{23}, L15 (1990).
	
\bibitem{tla} H. N. V. Temperley and E. Lieb, {\it Spin-s quantum chains and Temperley-Lieb algebras}, Proc. R. Soc. Lond. \textbf{A322}, 251 (1971).
	
\bibitem{martin} P. Martin, {\it Potts models and related problems in statistical mechanics} (World Scientific, Singapore, 1991).
	
\bibitem{sutherland} B. Sutherland, {\it Model for a multicomponent quantum system}, Phys. Rev. B \textbf{12}, 3795 (1975).
		
\bibitem{mila} P. Nataf and F. Mila, {\it Exact diagonalization of Heisenberg SU(N) models}, Phys. Rev. Lett.  \textbf{113}, 127204 (2014).
		
\bibitem{popkov} V. Popkov and M. Salerno, {\it Logarithmic divergence of the block entanglement entropy for the ferromagnetic Heisenberg model}, Phys. Rev. A \textbf{71}, 012301 (2005); V. Popkov, M. Salerno, and G. $\&$ Sch\"{u}tz,  {\it Entangling power of permutation-invariant quantum states},  Phys. Rev. A \textbf{72}, 032327 (2005).
	
\bibitem{klich} I. Klich and L. Levitov, {\it Quantum Noise as an Entanglement Meter}, Phys. Rev. Lett. \textbf{102}, 100502 (2009).
	
\bibitem{shankar} W.-M. Zhang, D. H. Feng, and R. Gilmore, {\it Coherent states: Theory and some applications}, Rev. Mod. Phys. \textbf{62}, 867 (1990).
	
\bibitem{settheory} A. A. Fraenkel, Y. Bar-Hillel, and A. Levy, {\it Foundations of Set Theory} (North-Holland, Amsterdam, 1973).

\bibitem{perelomov} A. M. Perelomov, {\it Coherent states for arbitrary Lie groups}, Communications in Mathematical Physics, \textbf{26}, 222 (1972); M. Mathur, I. Raychowdhury, J. Phys. A \textbf{44}, 035203 (2011);
 S. T. Ali, J-P. Antoine, and J-P. Gazeau, {\it Coherent states, wavelets and their generalizations} (Springer-Verlag, New York, 2000).
 
\bibitem{Glauber} R. J. Glauber, {\it Coherent and incoherent states of the radiation field}, Physical Review, \textbf{131}, 2766(1963).

\bibitem{cantorset} B. Knaster and C. Kuratowski, {\it Sur les ensembles connexes}, Fundamenta Mathematicae, \textbf{2},  206 (1921).

\bibitem{cantorteepee} L. A. Steen and J. A. Jr.  Seebach, {\it Counterexamples in topology} (Springer New-York, New York-Heidelberg, 1978).


\end{thebibliography}
\end{document}